\documentclass[a4paper,11pt,final]{article}

\usepackage{graphicx,amsmath,amssymb} % Add all your packages here
\usepackage{epsfig}

\begin{document}

% paper title: Must keep \ \\ \LARGE\bf in it to leave enough margin.
\title{\ \\ \LARGE\bf  Dynamic landscape models of coevolutionary games}

\author{Hendrik Richter \\
HTWK Leipzig University of Applied Sciences \\ Faculty of
Electrical Engineering and Information Technology\\
        Postfach 301166, D--04251 Leipzig, Germany. \\ Email: 
hendrik.richter@htwk-leipzig.de. }

\maketitle

\begin{abstract}
%% Text of abstract
Players of coevolutionary games may update not only their strategies but also their  networks of interaction. Based on interpreting the  payoff of players as fitness, dynamic landscape models are proposed. The modeling procedure is carried out for  Prisoner's Dilemma (PD) and  Snowdrift (SD) games that both use either birth--death (BD) or death--birth (DB) strategy updating. The main focus is on using dynamic fitness landscapes as a mathematical model of coevolutionary game dynamics. Hence, an alternative tool for analyzing coevolutionary games becomes available, and landscape measures such as modality, ruggedness and information content can be computed and analyzed.  In addition, fixation properties of the games and quantifiers characterizing the interaction networks  are calculated numerically. 
Relations are established between landscape properties expressed by landscape measures and  quantifiers of coevolutionary game dynamics such as fixation probabilities, fixation times and network properties.

\end{abstract}

\section{Introduction}
For describing evolutionary dynamics the  framework of fitness landscapes has been introduced,~\cite{kauffm91,richengel14,stad03}.   A fitness landscape formulates relationships between genetic specifications,  individual instantiations,  and their fitness. Together with postulating differences in fitness over all possible genetic specifications and a moving bias towards higher fitness, the setup suggests the picture of an evolving  population that is moving directedly on the landscape. On a conceptual level, this picture is based on the notion of evolutionary paths that are created by the topological features of the fitness landscape. Evolutionary paths are a succession of moves on the landscape with   ascending fitness values. The existence and abundance of such evolutionary paths gives rise to estimates about how likely a transition from low--fitness regions to high--fitness regions in the landscape is. 

Apart from fitness landscapes, another approach for specifying evolutionary dynamics is evolutionary games,~\cite{szabo07,nowak04,nowak06,maysmit91}. 
Evolutionary games are mathematical models of  dynamic interactions between individuals in a population and explain how  their behavioral strategies (for instance cooperation or competition) spread in a population. The main question is how  adoption of the strategies contributes to payoff allocation and consequently to the fitness characterizing the success of each individual.  An evolutionary game becomes dynamic if it is played iteratively over several rounds and the individuals are allowed to change their strategies and/or to recast the network describing with whom they are  interacting.  
Such an iterated evolutionary game comprises of an evolving population of individuals acting as players and can be seen as an expression of evolutionary dynamics.

Given the fact  that there are two frameworks for addressing evolutionary dynamics, it is natural to ask about their relationships. Unfortunately, both frameworks are not immediately compatible.
Although it is acknowledged that evolutionary games cast fitness landscapes, it has become  clear that such game landscapes change with an evolving population of players,~\cite{nowak04}. This is attributed to frequency--dependent selection. In other words, game landscapes are dynamic.  Based on 
some earlier results on dynamic fitness landscapes, e.g.~\cite{foster13,rich08,
rich14b}, there are some first attempts  at applying these ideas to games, for instance,~\cite{rich15}. In this paper  dynamic landscapes are employed for analyzing coevolutionary games by using and extending a framework introduced  recently,~\cite{rich16}. Games are considered where the players may update their strategies (evolutionary games), see e.g.~\cite{allen14,green12,szabo07}, but also games where the players may additionally change their interaction network (coevolutionary games), see e.g.~\cite{perc10,szol09,tani07}. In particular, it is shown that
the proposed method makes it possible to model and analyze evolutionary games and coevolutionary games within the same framework.

The paper is structured as follows. In Sec.  \ref{sec:desc}, some basic definitions are given, and evolutionary and coevolutionary games are briefly recalled. Sec.  \ref{sec:gamedynamic} reviews game dynamics, particularly the processes to update strategies and interaction networks. Dynamic landscape models of coevolutionary games are introduced and discussed in  Sec. \ref{sec:land}.  It is shown that evolutionary games can be modeled by either player landscapes or  strategy landscapes. As a player landscape is dynamic due to frequency--dependence, it is difficult to extend such a model to coevolutionary games. This paper introduces a method by which player--wise strategy landscapes can be aggregated to obtain a game landscape, which is static for evolutionary games and dynamic for coevolutionary games.  The modeling procedure is demonstrated for  Prisoner's Dilemma (PD) and  Snowdrift (SD) games that both use either birth--death (BD) or death--birth (DB) strategy updating. It is further shown that BD and DB updating yield landscapes with symmetry properties, and that replacement restrictions entail symmetry breaking. Moreover,   the local topological features of absorbing configurations of the games are interpreted as absorption structure.   It is  described how landscape properties may be linked to fixation via the absorption structure.    Sec.  \ref{sec:num} reports numerical  experiments on landscape measures such as modality, ruggedness and information content. Fixation probabilities and fixation times are calculated as well as network measures characterizing the interaction networks of the coevolutionary games. It is shown and discussed how the landscape measures relate to both  fixation properties and network measures.     Sec. \ref{sec:con} closes the paper with a summary and conclusions.

\section{Definitions and game description}  \label{sec:desc}
The coevolutionary  dynamics of the games considered in this paper stems from  three levels of activity: (i) game playing, (ii) updating  the strategy,  and (iii) updating  the interaction network.  The game playing is done by a finite population of $N$  players $\mathcal{I}$ that use one of two strategies $\pi$ during each round $k=\{0,1,2,\ldots \}$. A player $\mathcal{I}_i \in \mathcal{I}$, $i=1,2,\ldots,N$, can either cooperate ($C_i$) or defect ($D_i$). A pairwise interaction between two players $\mathcal{I}_i$ and $\mathcal{I}_j$ (which can be seen as player and coplayer) yields  rewards in form of  payoff $(p_i,p_j)$ as given by the payoff matrix      
\begin{equation} 
\bordermatrix{~ & C_j & D_j \cr
                  C_i & R & S \cr
                  D_i & T & P \cr}. \label{eq:payoff}
\end{equation}
Here, $T$ stands for temptation to defect, $R$ is reward for mutual cooperation, $P$ is punishment for mutual defection, and $S$ the sucker payoff for cooperating with a defector. 
Depending on  the numerical values of $(R,P,S,T)$ and their order, particular examples of the game are obtained, which have become metaphors for studying social dilemmas and discussing strategy selection along with the effect on short-- and long--term success in accumulating payoff,~\cite{maysmit91,nowak06}. Most prominently, there are Prisoner's Dilemma (PD) games, where $T>R>P>S$, and  Snowdrift (SD) games, where $T>R>S>P$.

The payoff $p_i(k)$ of a player $\mathcal{I}_i$ in round $k$ depends not only on  the player's  strategy $\pi_i(k)$ and the values of the payoff matrix (\ref{eq:payoff}), but also on who its coplayer is (or more precisely as to what the coplayer's strategy is) and how many coplayers there are. The question of who--plays--whom in a given round of the game is addressed by the interaction network. A convenient way of expressing and visualizing the interaction network is by using elements from evolutionary graph theory,~\cite{allen14,lieb05,ohts07,sha12}. Evolutionary graph theory places each player of the population on a vertex of an (undirected) graph. This graph describes the interaction network and consequently it can be called an interaction graph.  As there are no empty vertices and a vertex can only be occupied by one player, the number of vertices of the graph equals the number of players $N$. For each player, its coplayers are indicated by edges that connect the vertex of the player with the vertices of the coplayers. Such an edge defines the connected players to be adjacent. Each vertex can have up to $N-1$ edges (self--play is not allowed). As the degree $d$ is the number of edges incident with a vertex, the degrees of the interaction graph equal the number of coplayers that are engaged with each player in a single round. A graph is called regular if the degree is the same for all vertices. Hence, a regular interaction graph means that all players have the same number of coplayers.   

The interaction graph can be described algebraically by its (interaction) adjacency matrix $A_I$, which is called an interaction matrix.  The matrix $A_I \in [0,1]^{N \times N}$ is a symmetric $N \times N$ matrix with  entries $a_{ij}=1$ indicating an edge between vertex $i$ and $j$ and $a_{ij}=0$  showing that players $\mathcal{I}_i$ and $\mathcal{I}_j$ are not adjacent. The diagonal elements $a_{ii}=0$ because there is no self--play.    The symmetry of $A_I$ reflects the fact that two players $\mathcal{I}_i$ and $\mathcal{I}_j$ mutually engage in the game.  From the perspective of player $\mathcal{I}_i$, the player $\mathcal{I}_j$ may be the coplayer. If so, then from the perspective of player $\mathcal{I}_j$, the player $\mathcal{I}_i$ is the coplayer. Consequently, $a_{ij}=a_{ji}$ in the adjacency matrix $A_I$. If all $a_{ij}=1$ (except $a_{ii}=0$), the graph is complete, meaning that every player has all other players as coplayers and the evolutionary game is understood to be well--mixed,~\cite{szabo07}.   To summarize, for describing completely and deterministically  the game and the allocation of payoff, apart from the payoff matrix  (\ref{eq:payoff}) only two other entities are necessary: the strategy vector $\pi=(\pi_1 \pi_2 \ldots \pi_N)$ with $\pi_i \in [C_i, D_i]$ and the adjacency matrix $A_I$. This setting deterministically fixes the payoff $p=(p_1,p_2,\ldots,p_N)$ for each player.
Hence, the distribution of payoff $p(k)$ amongst the players remains the same if the
players  were to engage in the game with the same entities for a second
time in round $k+1$. Put another way for these entities being constant, the game  can be seen as static.  
Consequently, making the evolutionary game dynamic  requires  updating either the players'
strategies or the  interaction network, or both.

\section{Coevolutionary game dynamics} \label{sec:gamedynamic}

\subsection{Updating strategies}  \label{sec:statupdate}
There is a huge amount of work devoted to the modes of updating the player strategies in evolutionary games,~\cite{allen14,ohts07,patt15}. Most models use versions of  stochastic strategy updating based on a Moran process, but there are also works emphasizing  limiting the effect of randomness and  including the self--interest of players, e.g.~\cite{green14}.  According to a Moran process, in each round a player $\mathcal{I}_i$ (or more precisely its strategy) is replaced by (the strategy of) a player $\mathcal{I}_j$. The players $\mathcal{I}_i$ and $\mathcal{I}_j$ are selected at random, but the probabilities of the selection may not be uniform, for instance  depending on the players' fitness, which may vary. Versions of stochastic updating rules differ in several respects. Differences are, for example,  the actual probabilities that given players $\mathcal{I}_i$ and $\mathcal{I}_j$ are selected 
or whether or not 
there is an order between selecting the player providing the strategy (the source) and selecting the player receiving the strategy (the target).  
Finally, there may be general restrictions as to which players are allowed to be a possible source and/or target of another player.  Such predetermined restrictions imply a replacement structure,~\cite{ohts07}.  Conceptually similar to interaction,  the question of who--may--replace--whom can be described by a network of replacement. This network is expressible by a replacement graph and consequently by a (replacement) adjacency matrix $W_R$, which is  called a replacement matrix. The 
 matrix $W_R \in \mathbb{R}^{N \times N}$  has  entries $w_{ij}\geq 0$, and $w_{ij}>0$ indicates that player $\mathcal{I}_i$ may provide its strategy for player $\mathcal{I}_j$ to receive.   The values of $w_{ij}>0$ contribute to the probabilities that player $\mathcal{I}_i$ is source and player $j$ is target.  If all $w_{ij}=\bar{w}$  for a constant $\bar{w}\neq0$,  every player $\mathcal{I}_i$ may be the source to every target player $\mathcal{I}_j$ with equal probability. Consequently, if there are no restrictions, the replacement graph is fully connected with evenly weighted  edges. 

 Amongst  strategy updating, the following replacement rules are frequently studied: birth--death (BD), death--birth (DB),  imitation (IM), and pair--wise comparison (PC),~\cite{allen14,patt15,sha12}. All these  Moran--based updating rules share that they
depend only on random (and possibly on players' fitness and replacement restrictions), but not on details of the interaction  (for instance who the source or target are actually interacting with and what those strategies are). Therefore, they do not account for self--interested players,~\cite{green14}.  This property  makes  it possible to disentangle player and strategy in the sense that it makes no difference from which source the target receives its strategy updating. 
In other words, for all these updating rules it is possible to specify probabilities that the strategy of a source $\mathcal{I}_i$ replaces the strategy of a target $\mathcal{I}_j$ depending only on replacement matrix and fitness,~\cite{patt15}.

\subsection{Updating interaction networks} \label{sec:netwupdate}

If, in addition to the strategies, also the  interaction network can be updated in evolutionary games, the game is called coevolutionary.  However, the players $\mathcal{I}$ of the coevolutionary game are functionally alike and can hence be thought as  belonging to the same species. Therefore, 
 coevolution takes place within a single population of players and is between different features of the players' function. The focus of this paper is on coevolving of game strategies and interaction networks. There are alternative settings, such as coevolution between game strategies and other features or parameters of the players, for instance their ability to promote their strategies, which is known as teaching,~\cite{szol08,szol09b}, or their temptation to defect, which affects their  perception of the social dilemma  and leads to multigames,~\cite{szol14}. Apart from the network structure, coevolution can also act on network interdependence,~\cite{wang14}.
All these coevolutionary settings are methodologically different from an alternative understanding of coevolution, which is between different ecological functions (and hence different species), for instance between predator and prey, or between host and parasite, see e.g.~\cite{thomp95}.
 
According to the focus of this paper, coevolution in evolutionary games  is in essence  considering the  interaction network as dynamic, from which follows that the interaction matrix $A_I$ must be time--dependent. There is a substantial variety of schemes and rules for coevolution,~\cite{pach06,perc10,szolet09,szol09,tani07}.
Unfortunately, the topic of dynamic network updating has not yet matured as far as
to express for a given coevolutionary rule the transitions from one interaction network to another as a probabilistic function. Whereas for strategy updating, there are replacement probabilities for different updating rules,~\cite{patt15}, the same is not known for network updating. 
However, it might be reasonable to assume that network updating involves creating an interaction matrix $A_I(\kappa+1)$ at point in time $\kappa+1$ from a matrix $A_I(\kappa)$ at the previous point $\kappa$, for an integer time variable $\kappa=\{0,1,2,\dots\}$. 

Such a succession of interaction networks can be modeled by instances of an Erd{\"o}s--R{\'e}nyi graph. In this paper, the discussion is restricted to the case where the number of coplayers is the same for all players and constant for all updating instances. Employing such a model precludes situations where a more highly connected player possesses high fitness  due to its connectedness, but not necessarily due to the effectiveness of its strategy. For $d$ coplayers, such an interaction graph has degree $d$ and belongs to a special class of Erd{\"o}s--R{\'e}nyi graphs, namely  random $d$--regular graphs. Modeling the interaction network by  random $d$--regular graphs makes it possible to systematically carry out  numerical experiments because recently efficient algorithms for generating such graphs became available,~\cite{bay10,blitz11,kim06}. Moreover, for  random $d$--regular graphs, some analytic results about the number of different graphs are known. This, in turn, corresponds to the number of possible player--coplayer combinations. As a $d$--regular graph with $N$ vertices has $\frac{dN}{2}$ edges, the number $dN$ needs to be even. Thus, employing such an interaction network model implies that we cannot have an odd number of players with an odd number of coplayers.   

For a small number of edges ($=$ coplayers)  $d$, the number $\mathcal{L}_d(N)$ of different   $d$--regular graphs on $N$ vertices ($=$ players) can be found by enumeration, see for instance the entries in the Sloane
encyclopedia of integer sequences,~\cite{slon15}. Thus, $\mathcal{L}_2(4)=3$ and $\mathcal{L}_3(4)=1$, while $\mathcal{L}_2(6)=70$,  $\mathcal{L}_3(6)=70$,  $\mathcal{L}_4(6)=15$ and  $\mathcal{L}_5(6)=1$, and $\mathcal{L}_2(8)=3507$, $\mathcal{L}_2(10)=286884$. Note that $\mathcal{L}_{N-1}(N)=1$ for all $N$, which means that a complete  interaction network  representing  a well--mixed population   holds only one instance of the matrix $A_I$.   
Thus, for a complete network graph the game cannot be coevolutionary. It  is always static with respect to interaction because no dynamic changes can be cast out of a single instance of $A_I$.  
 Further note that $\mathcal{L}_d(N)$ grows rapidly.  For interaction networks with $d=2$ coplayers, the number of possible player--coplayer combinations $\mathcal{L}_2(N)$ can be calculated exactly, as there is a recursive formula for the number of $2$--regular graphs,~\cite{boll01}, p.56: \begin{equation} \mathcal{L}_2(N)=(N-1)\cdot \mathcal{L}_2(N-1)+{N-1 \choose 2}\cdot \mathcal{L}_2(N-3)   \label{eq:2regular} \end{equation}  
valid for $N \geq3$, with $\mathcal{L}_2(0)=1$, $\mathcal{L}_2(1)=0$ and $\mathcal{L}_2(2)=0$. For $d>2$,
 no formula is known to compute exactly the total number $\mathcal{L}_d(N)$  of $d$--regular graphs on $N$ vertices, but asymptotic expressions have been found,~\cite{worm99}. Asymptotically, and for $d=\hbox{o}(\sqrt{N})$ and $dN$ even, the number is   \begin{equation} \mathcal{L}_d(N)=\frac{ (dN)! \: \cdot \:  \exp{\left(\frac{1-d^2}{4}-\frac{d^3}{12N}+\mathcal{O}\left (\frac{d^2}{N}\right )\right)}   }{\left(\frac{dN}{2}\right)!\: 2^{\frac{dN}{2}} \: (d!)^N}. \label{eq:LDN} \end{equation}
Based on a collection of  random $d$--regular graphs the effect of different networks of interaction on payoff collecting and fitness can be analyzed, for which a landscape approach is proposed in the next section.    

\section{Landscape models  
of 
game dynamics
} \label{sec:land}

\subsection{Static and dynamic fitness landscapes}
A general definition of a (static) fitness landscape $\Lambda_S$ is the triple $\Lambda_S=(\mathbb{X},n,f)$,
where $\mathbb{X}$ is a configuration space, $n(x)$ is a
neighborhood structure that assigns to every $x \in \mathbb{X}$ a
set of direct neighbors, and $f(x):  \mathbb{X}
\rightarrow \mathbb{R}$ is a fitness function that provides every
$x \in \mathbb{X}$ with a proprietary quantity to be interpreted as a
'quality' information or fitness,~\cite{richengel14,stad03}. In this definition, the configuration space together with the neighborhood structure expresses a (multi--dimensional) 'location', while fitness is a property of this location. 
The configuration space itself can be seen as  an unordered (finite or infinite) list of configurations that genetic specifications of biological systems can have. The neighborhood structure defines a  (possibly multi--dimensional) order of this list by establishing what is directly next to each element of the configuration space. As direct neighbors of an element have a neighborhood structure themselves, this naturally establishes distant neighbors of the element as well. 

The definition of a (static) landscape has the consequence of each configuration possessing a constant fitness value.  For several reasons this might not realistically  reflect the evolutionary scenario to be described and  may generally restrict the descriptive power and versatility of the landscape model. Hence,
assuming that fitness may change over time, while configuration space and neighborhood structure remain constant, the  definition above can be extended to a dynamic fitness landscape, which can be expressed as the quintuple $\Lambda_D=(\mathbb{X},n,\mathcal{K},F,\phi)$,~\cite{rich14a}. In addition to the elements of the static landscape, there is the time set $\mathcal{K}$, the set of all fitness functions $F$ in time $\kappa \in \mathcal{K}$, and the transition map $\phi$ defining how the fitness function changes over time. For a discrete time set $\mathcal{K}$, for instance for the integer numbers $\mathcal{K}= \{0,1,2,\ldots \}$, the notion of a dynamic landscape coincides with the notion of a series of static landscapes. Hence, two static landscapes $\Lambda_S^{(1)}=(\mathbb{X},n,f^{(1)})$ and $\Lambda_S^{(2)}=(\mathbb{X},n,f^{(2)})$ can be reformulated as one dynamic landscape $\Lambda_D$  with $(f^{(1)}, f^{(2)}) \in F$ and $\phi$ describing how $f^{(1)}$ changes into $f^{(2)}$. Such a dynamic landscape model implies the time variable $\kappa \in \mathcal{K}$ to act as an integer counting and ordering scale for dynamic instances of a static landscape. Hence, $\kappa \in \mathcal{K}$ is numerically tantamount  to yet conceptually different from counting the rounds of an coevolutionary game by  $k=\{0,1,2,\ldots \}$. 

Applying a landscape approach for describing evolutionary dynamics requires addressing what may constitute a configuration $x \in \mathbb{X}$ and its neighborhood $n(x)$, and also what defines fitness $f(x)$. 
For the coevolutionary games described in the previous sections, there are several modeling options, which are discussed in the following. The actual modeling choice of $\mathbb{X}$, $n$ and $f$ and their interdependencies may either result in a static landscape or entail a  landscape that is dynamic and additionally requires  $\mathcal{K}$, $F$ and $\phi$ to be specified.

Evolutionary as well as coevolutionary games allocate payoff to players according to the payoff matrix (\ref{eq:payoff}). 
 For making the payoff $p$  interpretable as reproduction rate or survival probability (and lastly as fitness $f$), it has been suggested to rescale $p$ by a positive, increasing, differentiable function,~\cite{allen14,sha12}. In the following the linear function $f=1+\delta  p$ is used with  the intensity of selection $\delta\geq 0$. 

\subsection{Player landscapes}
The simplest landscape model of an evolutionary game  arises from equating configurations with players $\mathcal{I}$, which for $N$ players leads to a   player configuration space $\mathbb{X}=\mathcal{I}$ with $N$ elements.  The neighborhood structure follows from the $d$ coplayers that each player has, which can be $1 \leq d \leq N-1$.
 Thus, the neighborhood of a player consists of all the other players with which it is mutually engaged in a game according to the interaction matrix $A_I$. Hence, assuming that the players $\mathcal{I}$ can be attributed with a fitness $f$, such a player landscape  $\Lambda_\mathcal{I}$ could be specified by $\Lambda_\mathcal{I}=(\mathcal{I},A_I,f)$. 
 A popular form of such player landscapes is to place the players on a two--dimensional square lattice and define the coplayers to be Von Neumann (or Moore) neighborhoods, which consists of the lattices cells orthogonally (or additionally diagonally--adjacent) surrounding a central cell,~\cite{nowak93}. Admittedly, such an arrangement fixes the number of direct neighbors to $d=4$ (or $d=8)$, but yields a convenient way of visualizing the quality information over the resulting two--dimensional structure, which might be one reason for the popularity of these neighborhoods. 
  The most obvious choice of the quality information is payoff $p$ or quantities directly derived from it such as the linear fitness $f=1+\delta  p$ introduced earlier. 
This has led to label such landscapes as payoff landscapes,~\cite{brede11}.

 There are, however, several problems with such a player landscape model. The main problem is that the configuration is  the player, not its strategy, nor the strategies of its coplayers. Hence, with the player's and coplayers' strategies, two quantities decisive for the amount of payoff are not directly attached to the configuration. Strategies can be seen as ambiguous and polyvalent properties of the configuration of players. This means that the payoff attributable to a configuration depends on both the  player's strategy and also on the  strategies of its neighboring coplayers. This aspect is known as frequency--dependence, as the payoff can be seen as to depend on how frequent the strategy that the player adopts also occurs in the coplayers. Consequently, frequency--dependent fitness refutes the assumption that each player $\mathcal{I}_i$ can be attributed with a unique and static fitness. In short, fitness derived from payoff can be seen as dynamic so that the real player landscape cannot be static,  but should be dynamic: $\Lambda_\mathcal{I}=(\mathcal{I},A_I,\mathcal{K},f(\kappa),\phi)$. However, the dynamics of $f(\kappa)$ is caused not only by frequency--dependence, but also by strategy updating for which the player landscape model does not directly account and  both these causes can hardly be separated from each other. Hence, the transition map $\phi$ describing how the fitness $f(\kappa+1)$ relates to $f(\kappa)$ is not straightforwardly definable. 
In addition, modeling configurations of a landscape by players means that the neighborhood structure is expressed by the adjacency matrix $A_I$. A variable interaction network, as in coevolutionary games, therefore implies a changing neighborhood structure. To conclude a player landscape of a coevolutionary game would involve changing neighborhood structure as well as  dynamic fitness. This may make analyzing such a landscape rather complicated.
 
There is another reason for the difficulties to deduce meaningful conclusions from 
 payoff--based fitness over a  player landscape. Topological features of the landscape can be used as a starting point for estimating how likely transitions from low--fitness configurations to high--fitness configuration are and also which configurations are most likely to be a  steady state of evolutionary dynamics.   However, which player in a symmetric game as specified by the payoff matrix (\ref{eq:payoff}) exactly is  a likely high--fitness  outcome of an evolutionary process is  not very relevant. 
A much more important question is what fraction of the players in the long run settles stably to one of the possible strategies. In pursuing this question, there are several works that define the quality information of the landscape to be the strategy to which a player temporarily or finally settles. This means the 'fitness' is expressed by the strategy vector $\pi(k)$. The results have been visualized by coloring the players according to their strategy,~\cite{nowak93,nowak04}. Such a model has the advantage that the spatial and temporal distribution of the strategy preferences can be visualized with respect to the player--coplayer structure. However, payoff--based fitness as the main drive of evolutionary game dynamics is not an explicit component of such a landscape 
 and the number of coplayers is defined by the restrictions of the adjacency of the lattice grid.

\subsection{Strategy landscapes} \label{sec:stratland}

An alternative landscape model arises from
equating configurations with all possible combinations of strategies that each player and its coplayers can have. An element $\pi \in \Pi$ of the strategy configuration space $\Pi$ is comprised of the strategies of any player $\mathcal{I}_i$, $i=1,2,\ldots,N$, and its up to $N-1$ coplayers: $\pi=(\pi_1\pi_2 \ldots \pi_N)$. The strategy configuration space  $\mathbb{X}=\Pi$ generalizes the time--dependent strategy vector $\pi(k)$ towards all of its possible instances. 
  Hence, for $N$ players with two possible strategies,  $\Pi$ contains $\ell=2^N$ elements. If we binary code the strategies cooperation and defection  (for instance $C_i=1$, $D_i=0$), an element  $\pi \in \Pi$ appears as binary string of length $N$. Note that for this case the bit--count of $\pi$, $\text{bc}(\pi)$, provides a simple way of expressing the fraction of cooperators $\text{bc}(\pi)/N$. 
It is assumed that only one player or coplayer  can change its strategy at a given point in time. This implies
a
 neighborhood structure where  each element $\pi$ has  $N$ direct neighbors which are   distanced by 
Hamming distance  of $1$, which is denoted by $\mathcal{H}_d^1(\pi)$. For instance,   $\mathcal{H}_d^1(0000)=\{1000,0100,0010,0001 \}$ With such a model we obtain for payoff--based fitness $f$ 
a unique and static landscape $\Lambda_{\Pi}^i=(\Pi,\mathcal{H}_d^1,f_i)$ for each player $\mathcal{I}_i$ and each interaction network. As the game specified by the payoff matrix (\ref{eq:payoff})
is symmetric,  the strategy landscapes $\Lambda_{\Pi}^i$ are topologically alike for all players $\mathcal{I}_i$. The landscapes $\Lambda_{\Pi}^i$ can be transformed into each other by shifting and reshuffling.  For a landscape interpretation this topological likeness implies that landscape quantifiers such as the number of maxima, or correlation structure, or information content do not vary over the $\Lambda_\Pi^i$.

For $N=4$, the landscapes   can be visualized in two dimensions, see Fig. \ref{fig:n4}. It shows  $\Lambda_{\Pi}^i$,  $i=1,2,3,4$, for  the payoff matrix
$\bordermatrix{~ & C_j & D_j \cr
                  C_i & 3 & 0 \cr
                  D_i & 5 & 1 \cr}$  and  the adjacency matrix $A_I=\left(\begin{smallmatrix} 0 & 1 & 1 & 1 \\ 1 &  0 & 1 & 1 \\ 1 & 1 & 0 & 1\\ 1 & 1 & 1 & 0 \end{smallmatrix} \right)$  specifying a PD game with  a complete  interaction network and $d=3$ coplayers for each player $\mathcal{I}_i$. 

We find that $\mathcal{L}_3(4)=1$ and hence the game is static with respect to updating the interaction network. Observe that for each player  there is only one maximum fitness value (the player is defecting, while all coplayers cooperate) and one minimum fitness value (the player cooperates, while all coplayers defect). Apart from the single maximum and the single minimum, there are several configurations that have the same fitness value. Interestingly, these configurations do not form a neutral network,~\cite{richengel14}, as they have Hamming distance of $2$ and hence are not direct neighbors.   From the strategy landscape $\Lambda_{\Pi}^i$ it can be concluded which strategy for the player $\mathcal{I}_i$  (with respect to the strategies of the coplayers) yields the highest fitness and is therefore most desirable from the perspective of $\mathcal{I}_i$. Nonetheless, the evolutionary path from a given initial configuration may depend on, and be influenced by, the strategies provided to and/or received from the coplayers. In addition, from the perspective of another player, another strategy configuration is best. Best configurations for respective players, however, are mutually exclusive, which is a defining feature of social dilemma games such as the PD.   Consequently, each strategy landscape $\Lambda_{\Pi}^i$ can be seen as a building block that constructs a strategy landscape of the game.
Such a game landscape would allow conclusions as to what strategy configurations are adopted if  all players and their interactions are taken into account. In other words, a game landscape   may model the dynamics caused by the strategy updating processes discussed in Sec.    \ref{sec:statupdate}.

\subsection{Game landscapes} \label{sec:gameland}

Reconsider the game with $N=4$ players, for which Fig. \ref{fig:n4} depicts the player--wise strategy landscapes  $\Lambda_{\Pi}^i$. At every point in time $k$, the game can be seen as occupying one of its $2^4=16$ configurations. Put another way, the actual strategy vector $\pi(k)$ specifies an actual configuration on the landscapes $\Lambda_{\Pi}^i$. For each player $\mathcal{I}_i$, its landscape     $\Lambda_{\Pi}^i$ gives its actual fitness $f_i(k)$.  The strategy updating process means that one player provides its strategy for another player to receive. The receiving player changes its strategy. According to the landscape view this process corresponds with changing the actual configuration $\pi(k)$ to a neighboring configuration $\pi(k+1)$. As the change of configuration affects all players (and consequently all player--wise strategy landscapes), it entails that all players may experience a change of fitness as well. No player can hold onto its configuration as long as the strategy updating process is underway unless one of the two absorbing configurations $\pi_\infty$ are reached, namely $\pi_\infty=(0000)$ or $\pi_\infty=(1111)$. 

In the following, the strategy updating processes birth--death (BD) and death--birth (DB) are discussed. For these processes transition probabilities can be derived,~\cite{patt15}, which can now be employed to define game landscapes. Therefore,  it will be convenient to rewrite the landscape $\Lambda_{\Pi}^i$ as its decomposition $\Lambda_{\Pi}^i= \{ \lambda_\ell^i\}$, $\ell=1,2,\ldots,2^N$, where each $\lambda_\ell^i$ contains the fitness and preserves the neighborhood of configuration $\ell$. 
Assume that the game is in configuration $\pi(k)=(1101)$, which means that player $\mathcal{I}_3$ is defecting, while the three other players are cooperating. According to the PD game,  the fitness of $\mathcal{I}_3$  is highest, the three other players have the same (albeit lower) fitness. 
To start with, let us consider a BD strategy updating, which selects source before target,~\cite{patt15,sha12}. A player's strategy is chosen at random with a probability proportional to fitness to be a source (hence birth). The birth probability of a configuration $\ell$ of player $\mathcal{I}_i$ therefore scales to \begin{equation} b_\ell^i= \frac{\lambda_\ell^i}{ \sum_{\ell=1}^{2^N} \lambda_\ell^i}, \label{eq:birth} \end{equation} where the  $\lambda_\ell^i$ are the decompositions of the landscape $\Lambda_{\Pi}^i$ containing the fitness.  The player with the highest fitness is most likely to be a source, 
 which is presumably $\mathcal{I}_3$ with strategy $\pi_3(k)=0$. 
Which one of the three players is the target  to receive the strategy (hence death) is due to chance but influenced by possible restrictions regarding the replacement. Hence, the death probability of a player  $\mathcal{I}_i$  is \begin{equation}d_i=\frac{1}{N}\displaystyle \sum_{j=1}^{N} \frac{w_{ji}}{\sum_{i=1}^{N} w_{ji} }, \label{eq:deathbd}\end{equation} where the  $w_{ji}$ are the elements of the  replacement matrix $W_R$ possibly restricting replacements of strategies as discussed in Sec.  \ref{sec:statupdate}. Note that the death probability is independent of fitness and hence the same for all configurations of each player.  A  BD (and also a DB) updating does not envisage self--replacement and hence the replacement matrix $W_R$ must have diagonal elements $w_{ii}=0$. If, on the other hand, there are no replacement restrictions, then the death probability is invariant over players: $d_i=\frac{1}{N}$ for all players using a BD updating.
 Assume that all players can be a target and  $\mathcal{I}_2$ is chosen. Hence, the strategy configuration after the  strategy updating is $\pi(k+1)=(1001)$. The players $\mathcal{I}_2$ and $\mathcal{I}_3$ have leveled their fitnesses, while the fitness of both the other players is fallen even more.

 Now consider a DB strategy updating, which selects target before source,~\cite{patt15,sha12}. Here, a player's strategy is chosen at random with a probability proportional to the inverse of its fitness to be a target (hence death).  Therefore, the death probability of  a configuration $\ell$ of player $\mathcal{I}_i$ can be expressed as scaling to \begin{equation}d_\ell^i= 1-\frac{\lambda_\ell^i}{ \sum_{\ell=1}^{2^N} \lambda_\ell^i}. \label{eq:death} \end{equation} Still assume that the game is in configuration $\pi(k)=(1101)$ and as the players $\mathcal{I}_1$, $\mathcal{I}_2$, and $\mathcal{I}_4$ have the same (low) fitness values, one of them is most likely to be the target. Suppose  $\mathcal{I}_1$ is chosen. Which one of the three players provides its strategy  as a source (hence birth), depends on chance and possible replacement restrictions. We get the birth probability  \begin{equation} b_i=\frac{1}{N}\displaystyle \sum_{j=1}^{N} \frac{w_{ij}}{\sum_{i=1}^{N} w_{ij} }, \end{equation} which is the same as the death probability (\ref{eq:deathbd}) in BD, but the target and the source are switched in the elements of the replacement matrix. Note that only if the player $\mathcal{I}_3$ is chosen, a change in configuration takes place, that is the strategy configuration after the  strategy updating is $\pi(k+1)=(0101)$. Put differently, the outcome of both a BD and a DB updating may be the same, but the probabilities to reach it may be different.

For a sufficiently large number of  strategy updating events (and therefore changes of configuration), there may be some configurations that the game occupies more often than others.   These may, for instance, be the absorbing configurations $\pi_\infty$ with a bit count $\text{bc}(\pi_\infty)=0$ and $\text{bc}(\pi_\infty)=N$. Analyzing whether or not these absorbing configurations are reached and how long this takes, gives rise to fixation probabilities and  fixation times, which  will be discussed in Sec. \ref{sec:fix}. Before, however, we focus on the question of how frequent any configuration is  over strategy updating events. 
The frequency of reaching a configuration
 scales to the probabilities of  birth and death discussed so far. Hence, for a BD updating the game landscape
\begin{equation}\Lambda_{\Pi}^{BD} =  \{ \lambda_\ell^{BD}\}=\left \{  \frac{1} {\frac{1}{2} \left(1+\exp{\left(  \frac{ \alpha}{N} \sum_{i=1}^{N}    b_\ell^i d_i  \right) }  \right) }\right \}, \label{eq:bd} \end{equation}
can be defined, while for a DB updating, we set
\begin{equation} \Lambda_{\Pi}^{DB} = \{ \lambda_\ell^{DB}\}=\left \{  \frac{1} {\frac{1}{2} \left(1+\exp{\left(  \frac{ \alpha}{N} \sum_{i=1}^{N}    d_\ell^i b_i  \right) } \right)}\right \}, \label{eq:db} \end{equation}
both with $\ell=1,2,\ldots,2^N$ and $\alpha$ being a sensitivity weight.  Both game landscapes retain the configuration space and the neighborhood structure of the player--wise strategy landscapes $\Lambda_{\Pi}^i$, hence using them as building blocks. The fitness of each configuration $\ell$ summarizes via a Fermi function the probabilities to reach the configuration according to the birth and death events of the strategy updating process. The fitness of the game landscape therefore depends on the fitness of each player--wise landscape and possible replacement restrictions. 
Moreover, 
different updating processes cast different game landscapes $\Lambda_{\Pi}^{BD}$ and  $\Lambda_{\Pi}^{DB}$ out of the same strategy landscapes  $\Lambda_{\Pi}^i$ of the players $\mathcal{I}_i$. Given that the $\Lambda_{\Pi}^i$  are topological alike, and hence might be seen as possessing symmetry properties, different strategy updating rules break the symmetry of the player--wise strategy landscapes. 
At the same time, the BD and DB updating processes themselves possess symmetry properties via the birth and death probabilities (\ref{eq:birth}) and (\ref{eq:death}). Consequently (and in the absence of replacement restrictions), the game landscapes $\Lambda_{\Pi}^{BD}$ and  $\Lambda_{\Pi}^{DB}$ also retain symmetry properties. Replacement restrictions induced by different $W_R$, however,  yield another symmetry breaking. These symmetries (and broken symmetries) are reflected by the landscape properties discussed in the next section.

The discussion so far has been for a constant interaction network, that is for a specific matrix $A_I$. As pointed out in Sec.  \ref{sec:netwupdate}, network updating can be described by a series of adjacency matrices $A_I(\kappa)$.
Hence, as the genetic description  of the coevolutionary game comprises of the strategy vector {\em and} the interaction network,  the strategy configurations made up by the space $\Pi$ could be augmented by interaction configurations built by all possible networks of interaction. 
Consequently, the $\mathcal{L}_d(N)$ different interaction graphs enumerated approximately by Eq. (\ref{eq:LDN}) could be seen as  configurations according to the landscape definitions discussed above. However, in view of the rather large number of possible graphs for given $N$ and $1<d<N-2$ (a rough estimate of Eq. (\ref{eq:LDN}) for $d \ll N$ yields $\mathcal{L}_d(N) = \mathcal{O}( N^N)$), an alternative model is to understand different interaction graphs as dynamic instances of the strategy  landscape.  Put differently, the dynamics of the strategy landscape is the results of its variability with respect to the  interaction network. A consequence of such a modeling is that the timely order of  varying interaction networks  could be interpreted as  temporal neighborhoods according to the neighborhood structure inherent in landscapes.  
With network updating expressed as dynamic instances of player--wise strategy landscapes, we get a  dynamic landscape
$\Lambda_{\Pi}^i=(\Pi,\mathcal{H}_d^1,\mathcal{K},f_i(\kappa),\{ A_I(\kappa), A_I(\kappa+1)\})$ for describing a coevolutionary game. Apart from the strategy configuration $\Pi$ with neighborhood $ \mathcal{H}_d^1$ and the integer time set  $\mathcal{K}$, the quantity $f_i(\kappa)$ gives payoff--based fitness for each configuration, each player $\mathcal{I}_i$, and the $\kappa$--th interaction network. The matrix pair $\{ A_I(\kappa), A_I(\kappa+1)\}$ of subsequent adjacency matrices specifies how the fitness $f_i(\kappa+1)$ relates to $f_i(\kappa)$, thus constructing the transition map $\phi$. 

Taking up the example of $N=4$ with the same values of the payoff matrix as in Sec. \ref{sec:stratland}, but $d=2$ coplayers, we get  $\mathcal{L}_2(4)=3$ and hence a game that is dynamic with respect to updating the interaction network. The three dynamic instances are shown in Fig. \ref{fig:n5}, where Fig. \ref{fig:n5}a is for the adjacency matrix $A_I(0)=\left( \begin{smallmatrix} 0 & 0 & 1 & 1 \\ 0 &  0 & 1 & 1 \\ 1 & 1 & 0 & 0\\ 1 & 1 & 0 & 0 \end{smallmatrix} \right)$, Fig. \ref{fig:n5}b is for $A_I(1)=\left(\begin{smallmatrix} 0 & 1 & 0 & 1 \\ 1 &  0 & 1 & 0 \\ 0 & 1 & 0 & 1\\ 1 & 0 & 1 & 0 \end{smallmatrix} \right)$, and  Fig. \ref{fig:n5}c is for $A_I(2)=\left(\begin{smallmatrix} 0 & 1 & 1 & 0 \\ 1 &  0 & 0 & 1 \\ 1 & 0 & 0 & 1\\ 0 & 1 & 1 & 0 \end{smallmatrix} \right)$. It can be seen that the three different networks produce three different player--wise strategy landscapes for each player, which means that we  indeed obtain dynamic changes over the three  instances of $2$--regular graphs on $N=4$ vertices. 

Comparing these strategy landscapes with those for the complete interaction network (see Fig. \ref{fig:n4})   reveals  differences. A first is that for each player, there are now two maxima and two minima. Each player retains a maximum (minimum) if this player itself defects (cooperates), while its two coplayers cooperate (defect). The two maxima (minima) come about as it makes no difference for the player's payoff whether the fourth player (who is no coplayer as $d=2$) cooperates or defects.  A second difference is  that two neighboring configurations may build a block of equal fitness in connection with every configuration belonging to such a same--fitness block.  Consequently, there is neutrality in these fitness landscapes. Moreover, these results suggest that the number of maxima and the degree of neutrality scales to the number of coplayers, which can be confirmed by numerical experiments for landscapes with more than $N=4$ players. 

Within the given modeling framework of coevolutionary games, the timely order of the adjacency matrices is not associated with a particular updating process of the interaction network. The main reason is the general lack of established algebraic descriptions of network updating. The dynamic landscapes proposed may offer such an algebraic description as the transition map $\phi$ can be formulated uniquely for regular graphs, for instance for the transient between $A_I(0)$ and $A_I(1)$ of the example considered above as $\{ A_I(0), A_I(1)\}=\phi_{01}=A_I(1)-A_I(0)=\left( \begin{smallmatrix} \: \: \: 0 & \: \: \:1 & -1 & \: \: \:0 \\ \: \: \:1 &  \: \: \:0 & \: \: \:0 & -1 \\ -1 & \: \: \:0 & \: \: \:0 & \: \: \:1\\ \: \: \:0 & -1 & \: \: \:1 & \: \: \:0 \end{smallmatrix} \right)$.
For  dynamic player--wise strategy landscapes $\Lambda_{\Pi}^i=(\Pi,\mathcal{H}_d^1,\mathcal{K},f_i(\kappa),  \{ A_I(\kappa),  A_I(\kappa+1)\})$, game landscapes for BD and DB updating can be defined according to the probabilities of birth/death and expressed as dynamic counterparts of Eqn. (\ref{eq:bd}) and (\ref{eq:db}).  As the fitness $f_i(\kappa)$ of each player $\mathcal{I}_i$ now depends on the time variable $\kappa$ specifying dynamic instances of the adjancency matrix, the death and birth probabilities $d_i(\kappa)$, $b_i(\kappa)$,  $d_\ell^i(\kappa)$, $b_\ell^i(\kappa)$ are also dynamic. Consequently, the static games landscapes  (\ref{eq:bd}) and (\ref{eq:db}) become dynamic game landscapes:  $\Lambda_{\Pi}^{BD}(\kappa) =  \{ \lambda_\ell^{BD}(\kappa) \}$ and $ \Lambda_{\Pi}^{DB} (\kappa) = \{ \lambda_\ell^{DB}(\kappa) \}$. These dynamic BD and DB landscapes are the main topic of the numerical experiments reported in  Sec. \ref{sec:num}.  

\subsection{Landscapes and fixation}  \label{sec:fix}
The game specified by the payoff matrix (\ref{eq:payoff}) and the updating processes described in Sec.  \ref{sec:gamedynamic} instantiate evolutionary dynamics describable by the game landscapes  (\ref{eq:bd}) and (\ref{eq:db}) introduced above. As updating processes such as BD and DB  depend on random processes, the resulting game dynamics can also be seen as a stochastic process. Consequently, stochastic properties such as fixation probability and fixation time have been suggested for evaluating and comparing the long--term outcome of evolutionary  game dynamics, and studied widely in theory and numerical experiment, see, for instance,~\cite{lieb05,nowak06,patt15,sha12}. These fixation properties particularly account for whether or not the game dynamics settles on a steady state, and if so, how long this  takes on average, and how frequent it is.

The fixation probability quantifies how likely it is that  one of the two strategies that a player can use (cooperate or defect) spreads to the whole population of players, given that only one player started using this strategy. According to the landscape interpretation, this corresponds to reaching one 
 of  the two absorbing configurations  $\pi_\infty$ with  bit count $\text{bc}(\pi_\infty)=0$ and $\text{bc}(\pi_\infty)=N$, given 
that the initial configuration $\pi(0)=\pi_0$ had bit count $\text{bc}(\pi_0)=N-1$ and $\text{bc}(\pi_0)=1$, respectively. For each absorbing configuration, there are $N$ different configurations that can possibly serve as  an initial configuration.  Hence, as $\frac{N}{2^N}$ tends to zeros for $N$ getting larger, initial configurations getting scarce in the overall topological structure of the game landscape for a sufficiently large number of players. The same also applies to absorbing configurations. This is in agreement with the observation that fixation probabilities generally decrease while $N$ increases (see also the results of numerical experiments in Sec. \ref{sec:results}). As there are two absorbing configurations, two distinct fixation probabilities can be defined, one for complete cooperation and another for complete defection. The probability to reach the configuration where all player cooperate, $\text{bc}(\pi_\infty)=N$, is denoted by $\varrho_c$, while the probability of all players defecting, $\text{bc}(\pi_\infty)=0$, is named $\varrho_d$.

      Fixation probabilities can be analytically calculated for Moran processes based on properties of Markov chains for well--mixed populations,~\cite{nowak06,hinder15} and numerically for games on graphs,~\cite{hinder16}. For games on graphs with replacement restrictions, estimates of the fixation probabilities using diffusion theory have been reported,~\cite{ohts07}. For coevolutionary games considering dynamic networks of interaction of varying degree, numerical experiments can be carried out. Following previous experimental  works, the 
 fixation probabilities are approximated by the relative frequency of fixation. The fixation time quantifies how many changes in configuration it takes on average to reach an absorbing configuration $\pi_\infty$. This corresponds with the average amount of time needed to achieve fixation. The notion of fixation time can be refined by distinguishing which one of the two absorbing configuration is reached, which gives rise to conditional  fixation times,~\cite{traul09}.  The fixation times for the cooperative and defective absorbing configurations are denoted by  $\tau_c$ and  $\tau_d$, respectively. 

As fixation probability and fixation time are the most important quantities in stochastic game dynamics, 
these quantities are discussed next  with respect to the landscape interpretation proposed in the previous sections. 
The fitness of the landscapes (\ref{eq:bd}) and (\ref{eq:db}) derives from payoffs of each player and summarizes the probabilities that a particular configuration is occupied by the game. Hence, possible differences in fitness across neighborhoods generate topological features of the landscape.
These topological features, in turn, create  evolutionary paths on the landscape, which any evolutionary dynamics has to observe. However, the evolutionary dynamics is governed by a move bias towards higher fitness, which is not a move imperative. In other words, the landscape view implies that there are probabilities that the maxima are reached, but no certainties. Moreover, these probabilities depend on what exactly the topological features of the landscape are, for instance, on the number of maxima, their distribution and their accessibility. For evaluating the effect of landscape features, just focusing on the maxima is not sufficient. Therefore, different types of landscape measures have been proposed which aim at reflecting, in a more general sense, the impact that landscape features have on evolutionary dynamics, see also the discussion in Sec. \ref{sec:meas}.

Fixation occurs if a succession of changes in configuration leads from prescribed initial configurations to the absorbing configurations. Fixation probabilities $\varrho_c>0$, $\varrho_d>0$  require evolutionary paths  connecting initial configurations with respective absorbing configurations.   The values of $\varrho_c>0$, $\varrho_d>0$ scale to how easy or how difficult it is that these evolutionary paths can be accessed and maintained. The fixation time, on the other hand, varies with the length of the evolutionary path. Consequently, by analyzing the topological structure of the game landscape, it may be feasible to infer fixation properties. This kind of analysis, however, is impeded by the fact that 
 absorbing configurations in game landscapes are, topologically interpreted,  non--passable points in the landscape. However, non--passable points are not a standard concept of landscapes.  Perhaps most similar are steady states of a landscape, but there is the difference that the evolutionary dynamics can, under certain conditions, leave a steady state and there is the even more fundamental difference that steady states are by definition maxima of the landscape.  
Absorbing configurations may or may not be maxima of the game landscape. In the same way, the initial configurations marking the origin of the evolutionary path may or may not be minima of the landscape. The numerical experiments discussed in Sec. \ref{sec:results} confirm such a characteristics for the game landscapes  (\ref{eq:bd}) and (\ref{eq:db}). 

This line of reasoning suggests that a landscape analysis should take into account that fixation properties  are likely to be related to landscapes via the local (and possibly also the global) topological features of absorbing and initial configurations. In analogy to the landscape structure, which describes globally the topological features of the entire landscape, these  topological features and their interdependences with fixation we may call absorption structure. 
 The numerical  experiments reported next section address not only topological features of the landscapes, but also the absorption structure, where the focus is on the local structures.

\section{Numerical experiments} \label{sec:num}
\subsection{Landscape measures} \label{sec:meas}
Game landscapes can only be visualized as two--dimen\-sional projections up to $N=4$ players. For analyzing landscapes with more players, we need to resort to  landscape measures. A first measure we look at is modality expressed by the number of local maxima $\#_{LM}$.  Local maxima are potential steady states on the evolutionary path. Hence, the number of local maxima relates to the variety of possible evolutionary paths and consequently to the complexity of the evolutionary dynamics displayed. If there is just one (smoothly accessible) maximum, then all evolutionary paths converge to it and the evolutionary dynamics displayed is rather simple.  If, on the other hand, there is a large number of maxima, then the possible evolutionary paths may differ from each other massively resulting in more complex evolutionary dynamics.
For a  landscape $\Lambda_\Pi=(\Pi,\mathcal{H}_d^1,f)$ a configuration $\pi$ is a local maximum $\hat \pi$, if $\forall \pi \in \mathcal{H}_d^1(\hat \pi)$, the fitness of this strategy configuration satisfies $f(\hat \pi) \geq f(\pi)$. Moreover,  if this condition  holds  $\forall \pi \in \Pi$, then  $\hat \pi$  is a  global maximum.

 For evaluating the local absorption structure, we need to consider three further local topological features: minimum, neutrality, and saddles. A local minimum $\check \pi$ has at least one neighbor that has a smaller fitness and no neighbors that have larger fitness than itself. A neutral configuration $\bar \pi$ has only neighbors with the same fitness, which means that a landscape area containing $\bar \pi$  and its neighbors is flat. Lastly, a saddle $\acute \pi$ has some neighbors that are larger and some other neighbors that are smaller. In numerical experiments, the number of local maxima  $\#_{LM}$ can be computed for the game landscapes (\ref{eq:bd}) and (\ref{eq:db}). The same applies to whether the absorbing configuration $\pi_\infty$ and its initial configurations $\pi_0$ are maxima,  minima, neutral or saddles. Consequently, for the dynamic instances of these landscapes, a time series containing the numbers of local maxima $\#_{LM}(\kappa)$ is obtained. The same applies to all other measures of dynamic landscapes. 

There are two immediate problems with analyzing landscapes by modality expressed by the number of local maxima $\#_{LM}$. First, on a practical level, we find that $\#_{LM}$ can only be calculated by enumeration, which entails the proverbial curse of dimensionality. Second, on a conceptual level, there is the fact that the pure number of local maxima  is a decisive (and arguable the most important) factor defining evolutionary paths, but the distribution of the maxima and their accessibility is also profoundly influential. To overcome these issues, other types of measures have been proposed for quantifying smoothness, ruggedness, or neutrality of the landscape. Two of them are studied here, correlation length $\lambda$ and information content $h_{IC}$. 

The correlation length
 evaluates  across  the landscape how strongly the fitness of any configuration relates to the fitness of neighboring configurations and hence is a measure of the landscape's ruggedness,~\cite{stad96,richengel14}. For calculating the correlation length $\lambda$, a random walk on the landscape of length $T_w$ is used. The fitness values for each step on the walk are recorded by the sequence \begin{equation} \mathcal{S}=(f(0),f(1),\ldots,f(T_w-1) ) \label{eq:seq} \end{equation} and thus a series of neighboring fitness relations is obtained. Assuming that the landscape is isotropic, these neighboring fitness relations account for general changes in fitness across the landscape. Hence, the autocorrelation of sequence (\ref{eq:seq}) with time lag $t_L$  defines the landscape's random walk correlation function
$r(t_L)=\frac{\underset{i=0}{\overset{T_w-1-t_L}{\sum}} \left(
f(i)-\mu\right) \left(f(i+t_L)-\mu
\right) }{\sigma^2}$,
where
$\mu=\frac{1}{T_w}\underset{i=0}{\overset{T_w-1}{\sum}}
f(i)$, $\sigma^2 =\frac{1}{T_w} \underset{i=0}{\overset{T_w-1}{\sum}}
\left(f(i)-\mu\right)^2$ and $T_w \gg t_L>0$.  The
function $r(t_L)$ measures the correlation between different
regions of the fitness landscape and expresses a measure of how smooth or rugged the landscape is. 
 The correlation length \begin{equation} \lambda=-1/\log(|r(1)|) \label{eq:corrle}\end{equation} derives from the autocorrelation  $r(1)$ of time lag $t_L=1$,~\cite{stad96,richengel14}. 

The information content $h_{IC}$,~\cite{mun15,vassi00}, is an entropic landscape measure, which also uses the fitness sequence (\ref{eq:seq})
generated by a random walk on the
landscape. It can be interpreted as a measure of the amount of information required to reconstruct the landscape structure. From the time series (\ref{eq:seq}), 
differences in fitness  between two consecutive walking steps are coded by
symbols $s_{i} \in \mathbb{S}$, $i=0,1,2,\ldots,T_w-1$,
taken from the set $\mathbb{S}= \{-1,0,1\}$. These symbols are
obtained by
\begin{equation}s_i(\epsilon)=\left\{ \begin{array}{rcccc} -1,
& \quad \text{if} \quad& f(i+1)-f(i) & < & \epsilon \\ 0, & \quad \text{if} \quad &  |f(i+1)-f(i)|&\leq & \epsilon\\
1, & \quad \text{if}  \quad& f(i+1)-f(i)&> & \epsilon
\end{array} \right.
\end{equation}
for a fixed $\epsilon \in [0,L]$, where $L$ is the maximum difference
between two fitness values. The obtained symbols are concatenated
to a string
\begin{equation} S_{IC}(\epsilon)=s_0s_1\ldots s_{T_w-1}. \label{eq:entstring} \end{equation}
The parameter $\epsilon$ defines the sensitivity by which the string
$S_{IC}(\epsilon)$ accounts for differences in the fitness values. For example, 
if $\epsilon=0$, the string $S_{IC}(\epsilon)$ contains the symbol zero only for the
random walk reaching a strictly flat area. Hence, $\epsilon=0$
discriminates very sensitively between increasing and decreasing
fitness values. By contrast, for $\epsilon=L$, the string only
contains the symbol zero, which makes evaluating the structure of
the landscape pointless. A fixed value of $\epsilon$ with
$0<\epsilon<L$ defines a level of detail with respect to the information gained about the
landscape structure.

For defining the information content of the landscape, the
distribution of subblocks of length two, $s_{i}
s_{i+1}$, $i=0,1,\ldots T_w-2$, within the string
(\ref{eq:entstring}) is analyzed. These subblocks stand for local patterns in
the landscape. The probabilities of the occurrence of the
pattern $ab$ with $a,b \in
\mathbb{S}$ and $a \neq b$ are denoted by $p_{ab}$. For numerical calculation,  these
probabilities are approximated by the relative frequency of the patterns within the
string (\ref{eq:entstring}). As the set $\mathbb{S}$ consists of three
elements, we find six different kinds of subblock $s_{i}
s_{i+1}=ab$ with $a \neq b$
within the string $S_{IC}(\epsilon)$. From their probabilities
and a given sensitivity level $\epsilon$  the entropic
measure
\begin{equation} h_{IC}(\epsilon)=- {\underset{a, b \in
\mathbb{S} \atop a \neq b}{\sum}} p_{ab}  \log_6  p_{ab},
\label{eq:infcont}
\end{equation}
is calculated, which is called  information content of the fitness landscape,~\cite{mun15,vassi00}.
Note that by taking the logarithm in Eq. (\ref{eq:infcont}) with
the base $6$, the information content is scaled to the interval
$[0,1]$.

\subsection{Graph--theoretical properties of interaction networks}
Networks of interaction may be described by instances of a random $d$--regular graph, as set out in Sec.  \ref{sec:netwupdate}. Based on this description,  varying  interaction networks have been interpreted in this paper as to cast dynamic instances of a  landscape characterizing the coevolutionary games. Instances of interaction networks specify who--plays--whom in the game, which means that even if for each player the number of coplayers is constant, who in fact the coplayers are is not. As different coplayers may imply different strategies and hence different allocations of payoff, different networks of interaction may result in topologically different game landscapes. Put another way, if properties of game landscapes vary over dynamic instances, the variations should be reflected by properties of interaction networks, that is graph--theoretical quantifiers of $d$--regular graphs. Spectral graph theory defines several such quantifiers, which take advantage of connections between the algebraic description of a graph and its structural properties; see for instance~\cite{biggs94,brou12,cvet09,li12,spiel09}, upon which the remainder of this section about quantification of graph--theoretical properties of interaction networks relies.  The main propose of this quantification is to map structural differences of the interaction graph to different values of graph--spectral invariants, which, in turn,  are interpretable as (graph--theoretical) network measures. For definitions of these invariants, also see~\cite{biggs94,brou12,cvet09,li12,spiel09}.

The quantities considered are based on algebraic properties of the adjacency matrix $A_I$, or the Laplacian matrix $L=dI-A_I$, which is derived from $A_I$ to include the degree $d$ explicitly. 
For the matrices $A_I$ and $L$, the spectra of eigenvalues, $eig(A_I)=\left(\alpha_1,\alpha_2,\ldots,\alpha_N \right)$ and $eig(L)=\left(\lambda_1,\lambda_2,\ldots,\lambda_N \right)$, are starting points for further consideration. For connected $d$--regular graphs, we find for  spectra of the adjacency matrix $A_I$ that all eigenvalues are real and  $-d\leq \alpha_1 \leq \alpha_2 \leq \ldots \leq \alpha_N\leq d$, while eigenvalues of the Laplacian $L$ are also all real, and non--negative as well as  sortable according to $0=\lambda_1 \leq \lambda_2 \leq\ldots \leq\lambda_N$.

A first quantity is the (normalized) energy of a graph: \begin{equation}\text{ene}= \frac{1}{N}\sum_{i=1}^{N}  |\alpha_i |,\label{eq:ener} \end{equation}
which can be interpreted as the spectral distance between a given graph and an empty graph, and can hence be seen as scaling to the degree of difference between graphs. 
A second graph--theoretical network measure based on the eigenvalues $eig(A_I)$ is the independence number \begin{equation} \text{ind}=\frac{-N \alpha_1}{d-\alpha_1}, \label{eq:ind}\end{equation}
which is an approximation of the size of the largest independent  set of vertices in a graph. An independent set is a set of vertices in a graph such that no two vertices of the set are connected by a edge.   

A network measure based on the Laplacian derives from the smallest eigenvalue of $L$ larger than zeros, $\lambda_2$,  is termed (normalized) algebraic connectivity \begin{equation} \text{acl}=\frac{\lambda_2}{\lambda_N}, \label{eq:conn} \end{equation}  and scales to how well a graph is connected.  Connectivity denotes the structural property of a graph that removal of vertices or edges disconnects the graph. The Laplacian eigenvalue $\lambda_2=0$ if the graph is not connected, and $\lambda_2=N$ if the graph is complete (that is fully connected). Larger values of $\lambda_2$  indicate graphs with a  rounder shape, and high connectivity and girth, while for smaller values of $\lambda_2$ the graph is more path--like with low connectivity and girth. 
 Also calculated from the Laplacian is the expander index
\begin{equation} \text{exi}=\max{\left(1-\frac{\lambda_2}{d},\frac{\lambda_N}{d}-1\right)} \label{eq:expan}\end{equation}
which is a measure for the $d$--regular graph possessing expander properties. The expander index has small values for all eigenvalues $\lambda_i$ being close to $d$, and larger values for the opposite. Expander graphs are marked by all small sets of vertices usually having a larger number of neighbors. Thus, they can be seen as their neighborhood expanding.

As far as possible and needed, 
the graph--theoretical quantifiers are normalized with respect to the order of the graph. Hence, they can be compared over a varying number of vertices and hence players.  In Sec. \ref{sec:results} results are given that analyze the correlation between the network measures (\ref{eq:ener})--(\ref{eq:expan}) and the landscape measures (\ref{eq:corrle}) and (\ref{eq:infcont}).

\subsection{Experimental setup} \label{sec:setup}

The numerical experiments with the dynamic fitness landscapes of coevolutionary games consider a PD game and a SD game with  
$\bordermatrix{~ & C_j & D_j \cr
                  C_i & 3 & 0 \cr
                  D_i & 5 & 1 \cr}$ and
$\bordermatrix{~ & C_j & D_j \cr
                  C_i & 3 & 1 \cr
                  D_i & 5 & 0 \cr}$, respectively, which is a parametrization as suggested by Axelrod's seminal work,~\cite{axel80}.  The dynamics of the landscape is addressed by examining the effect of varying networks of interaction. Algorithms are employed that numerically generate  adjacency matrices $A_I(\kappa)$ specifying random regular graphs with given order and degree,~\cite{bay10,blitz11,kim06}. It is checked to see if the graphs are connected. If a graph fails the check, there are isolated vertices that may bias controlling the interaction network via the graph's degree and hence such graphs are discarded. 
For the experiments, different sets  of graphs with prescribed $N$ and $d$ are generated and used. The absorption structure was analyzed with a set of $G=6000$ graphs. Some experiments studying landscape measures and fixation properties were done with a set of $G=3000$ graphs. These experiments have shown that for a considerable number of different networks, rather similar results are obtained.  For this reason and also to facilitate the numerical experiments, unless stated differently the results presented in the figures are based on a set of $G=1000$ different interaction networks. In all cases 
for $\mathcal{L}_d(N)<G$, the complete set of possible networks of interaction is taken. The experiments are conducted for $N$ even to guarantee the existence of $d$--regular graphs for all $2\leq d \leq N-1$. 

Different replacement structures are analyzed.  The experimental setup follows previous works,~\cite{ohts07}, and defines the replacement matrix $W_R$ as random regular graph with given degree and guaranteed connectivity.  Additionally, the elements $w_{ij} \neq 0$ are filled with realizations of a random variable uniformly distributed on the interval $[0,1]$. The degree of $W_R$ is set to match the degree of $A_I$. All these experiments are carried out for BD and DB landscapes. Other updating schemes such as PC or IM can be treated likewise.  For these processes transition probabilities are known,~\cite{patt15}, and hence  landscapes similarly to (\ref{eq:bd}) and (\ref{eq:db}) can be computed.  With the conventional PC--based  computational resources that were available, it was possible to experiment within a reasonable time--frame with up to $N=20$ players. All experiments employ a linear relationship $f=1+\delta p$ between payoff and fitness with $\delta=0.25$. The game landscapes are computed with a sensitivity weight $\alpha=5$. For calculating the correlation length $\lambda$ and the information content $h_{IC}$, a random walk of length $T_w=10000$ was used, and the results are averaged over $50$  independent walks. Numerical experiments have shown that the results obtained are statistically equivalent over different initial configurations that the walks starts with. Hence, it appears reasonable to assume  for the tested cases that the game landscapes are isotropic. The information content (\ref{eq:infcont}) is computed with a sensitivity level that scales to the number of players by $\epsilon=\exp{(-12-N)}$.

The numerical experiments calculating fixation properties are based on $2500$ repetitions. This is a rather small amount compared to other recent experimental works, e.g.~\cite{hinder15,zuk13}. Some auxiliary experiments with a larger amount of repetition, however, have shown that the values of the fixation properties analyzed converge well so that the numerical setup used appears sufficient for up to $N=20$ players. 

\subsection{Experimental results} \label{sec:results}

\begin{figure*}[t]

\hspace{0.8cm} PD game \hspace{5.0cm} SD game 

\vspace{0.5cm}

\includegraphics[trim = 15mm 65mm 10mm 100mm,clip, width=6cm, height=6cm]{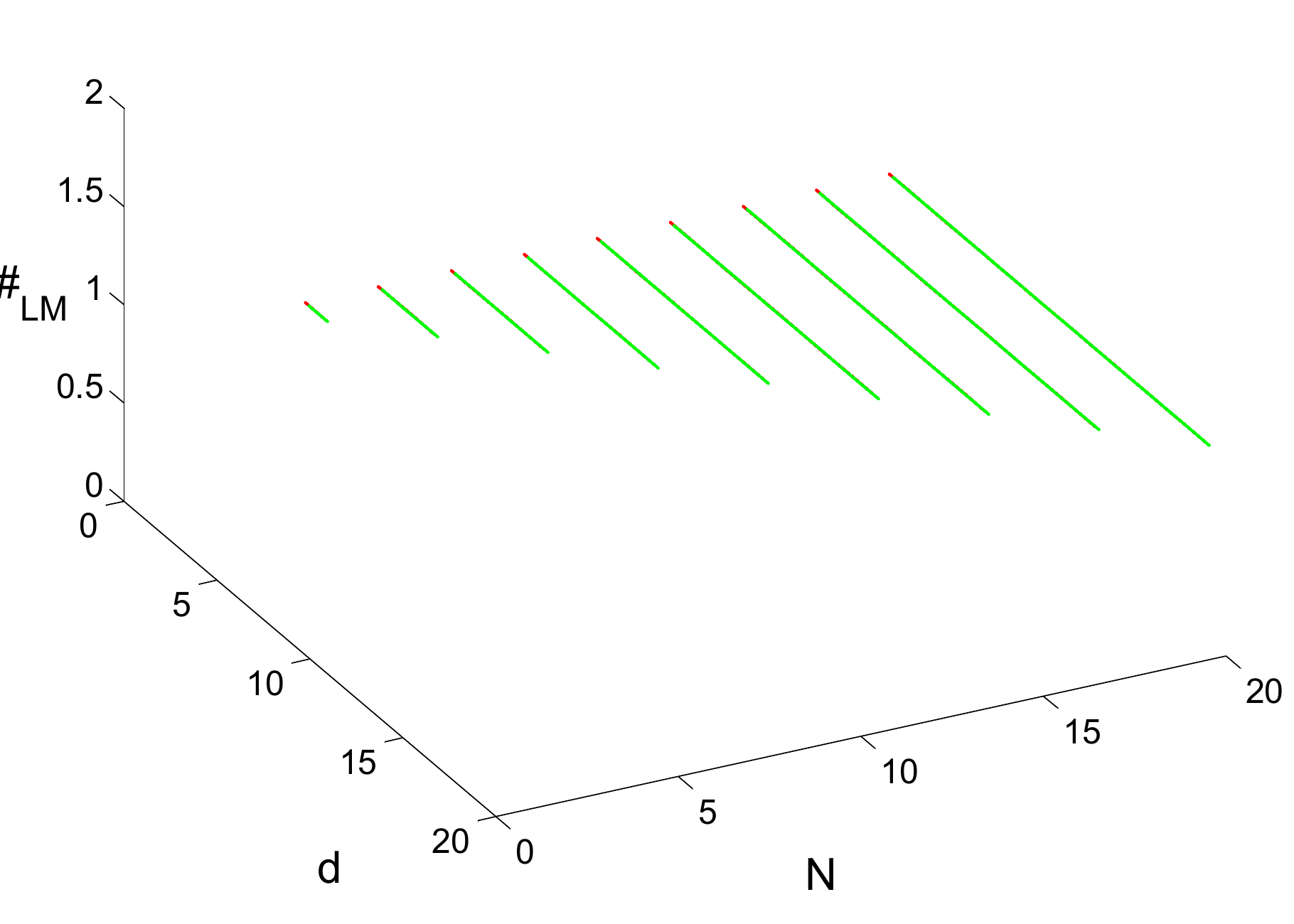} 
\includegraphics[trim = 15mm 65mm 10mm 100mm,clip, width=6cm, height=6cm]{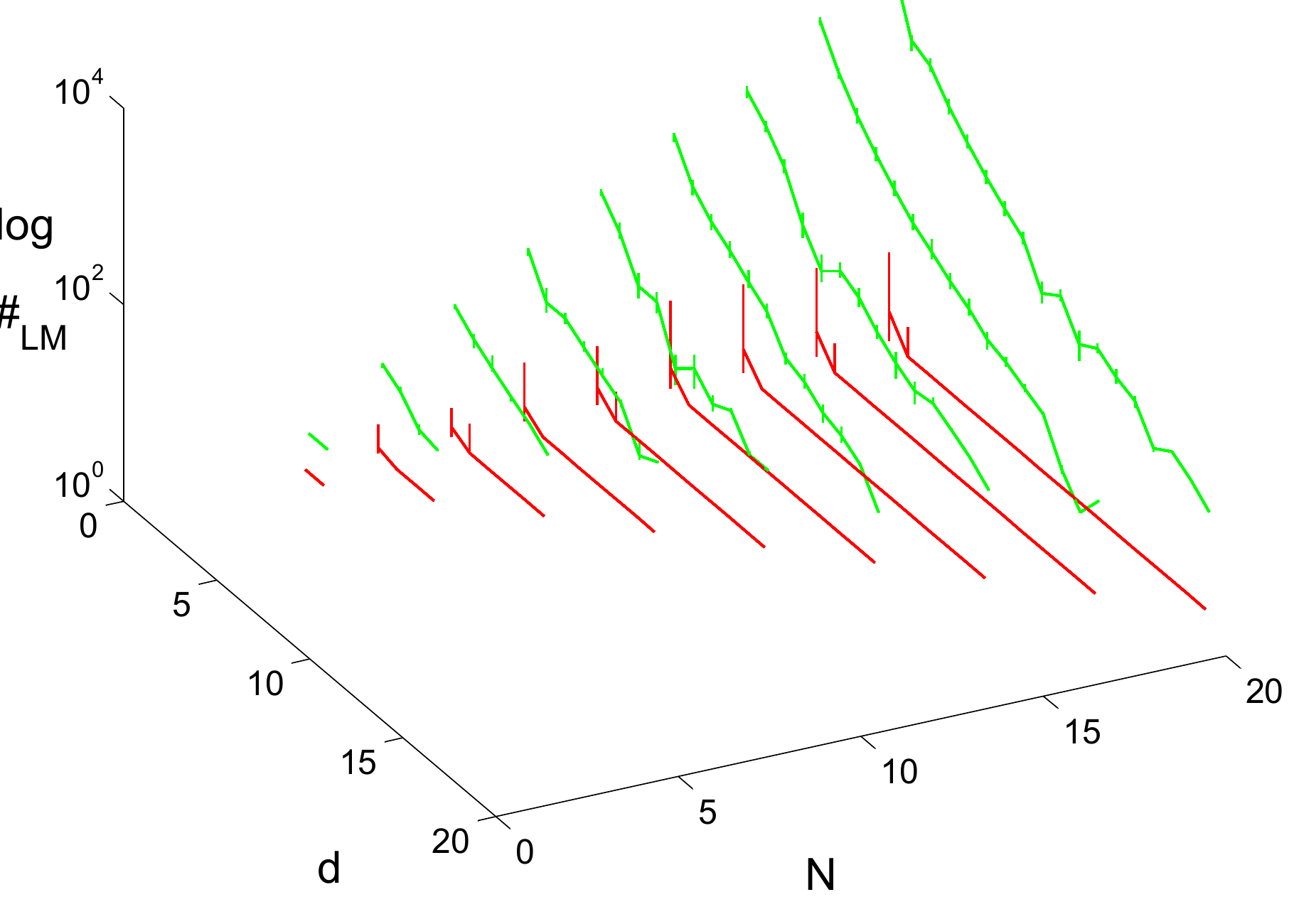} 

\hspace{0.8cm} (a)  \hspace{5.3cm} (b)

\vspace{0.5cm}

\includegraphics[trim = 15mm 65mm 10mm 100mm,clip, width=6cm, height=6cm]{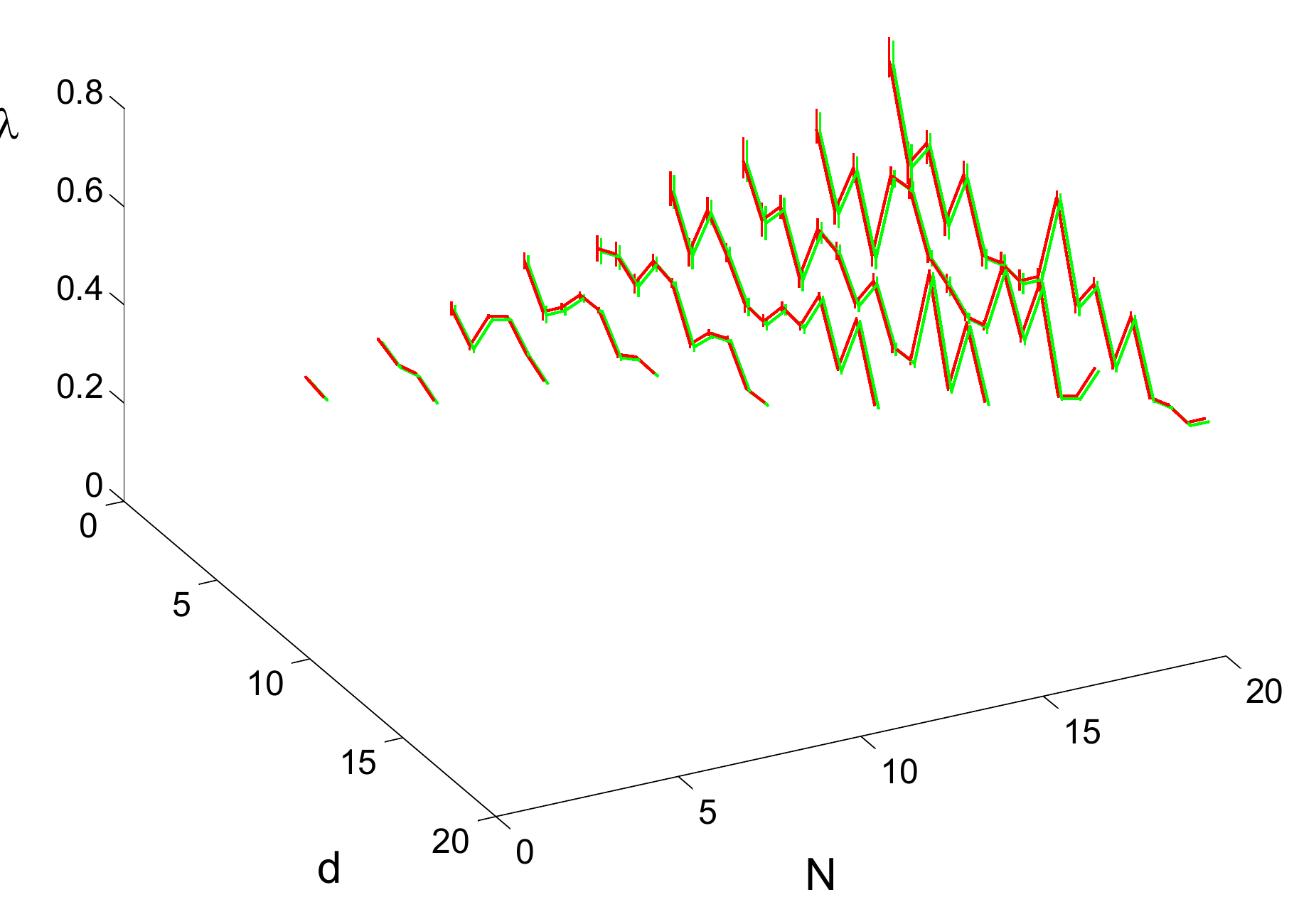} 
\includegraphics[trim = 15mm 65mm 10mm 100mm,clip, width=6cm, height=6cm]{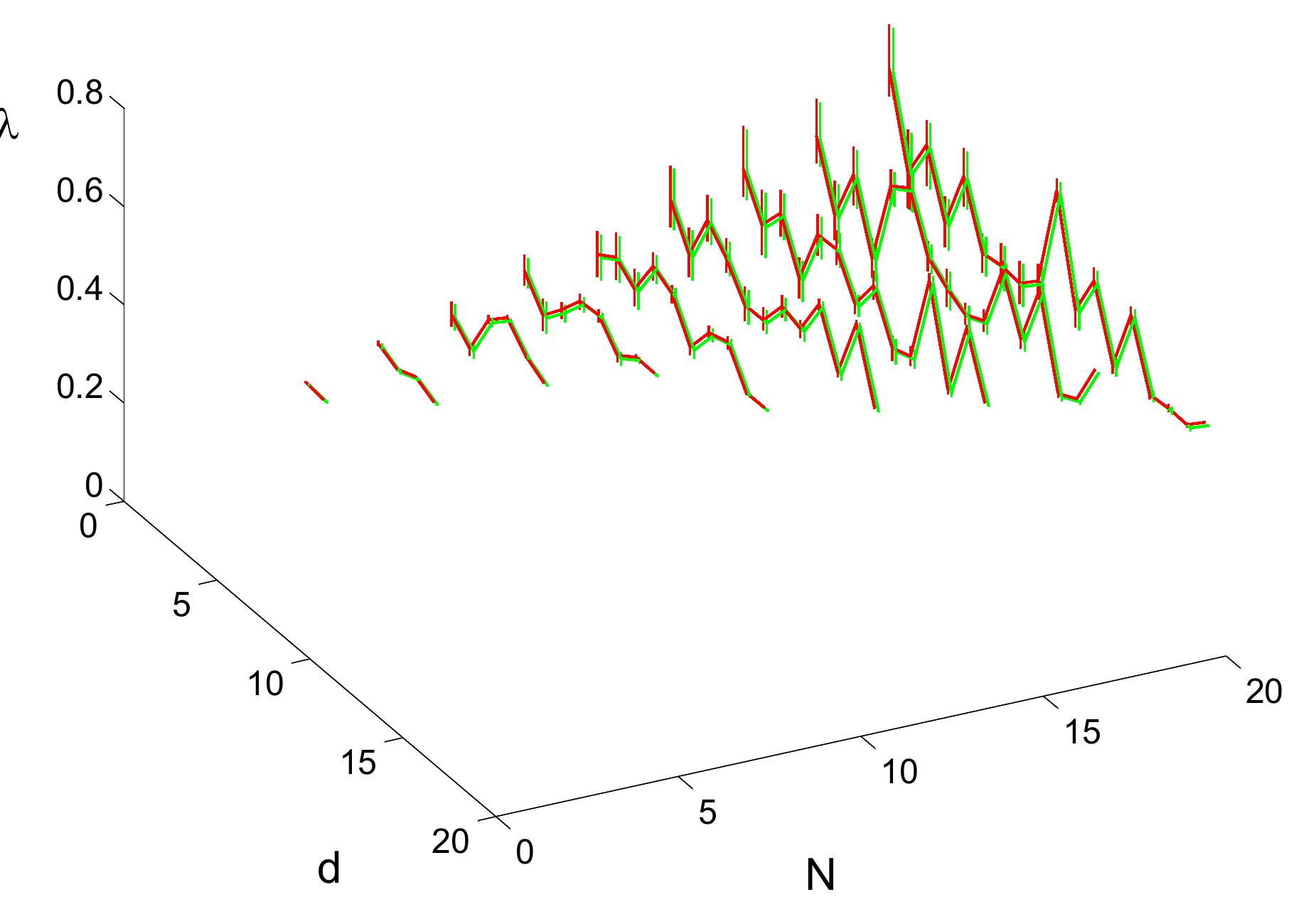} 

\hspace{0.8cm} (c)  \hspace{5.3cm} (d)

\includegraphics[trim = 15mm 65mm 10mm 100mm,clip, width=6cm, height=6cm]{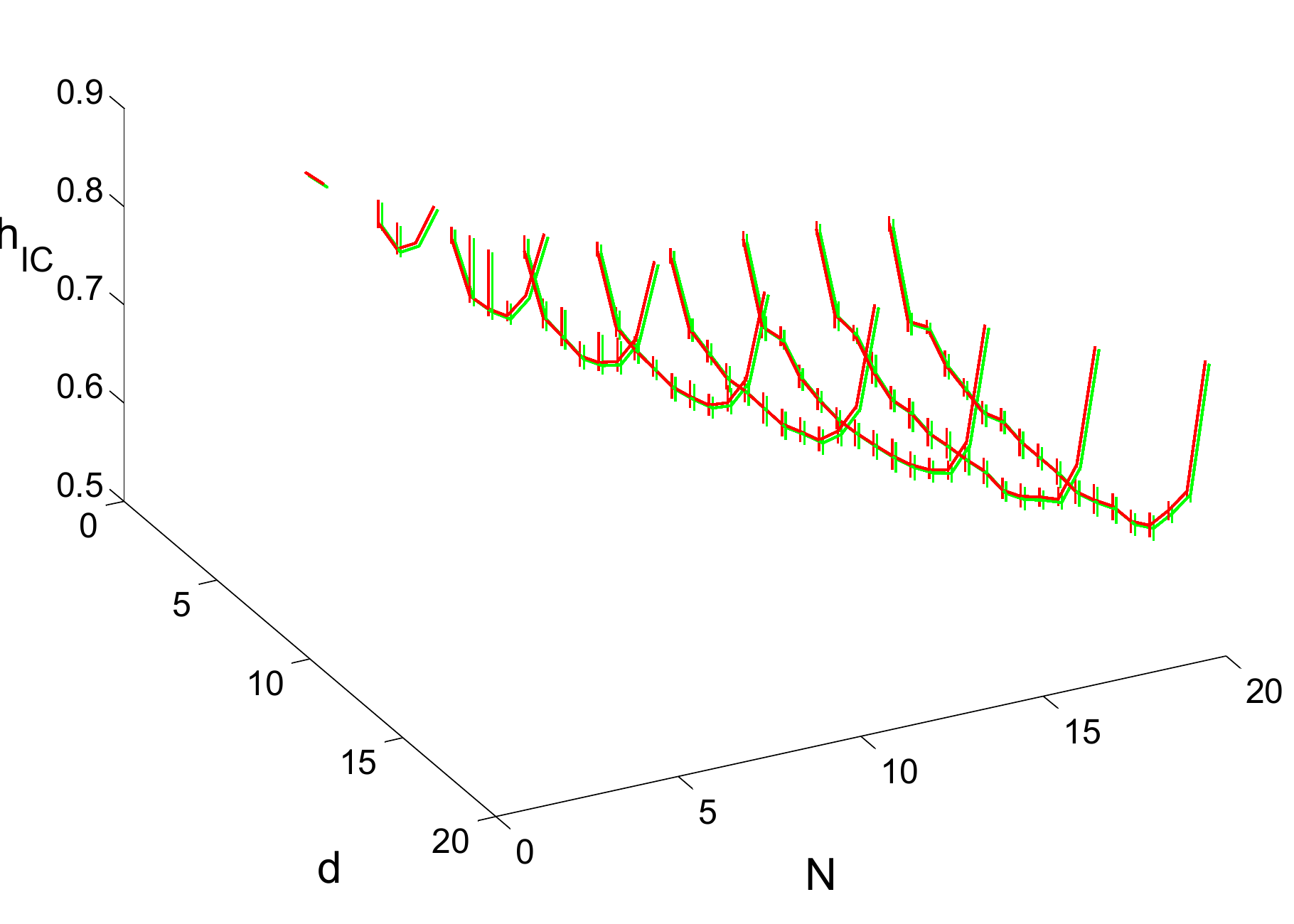} 
\includegraphics[trim = 15mm 65mm 10mm 100mm,clip, width=6cm, height=6cm]{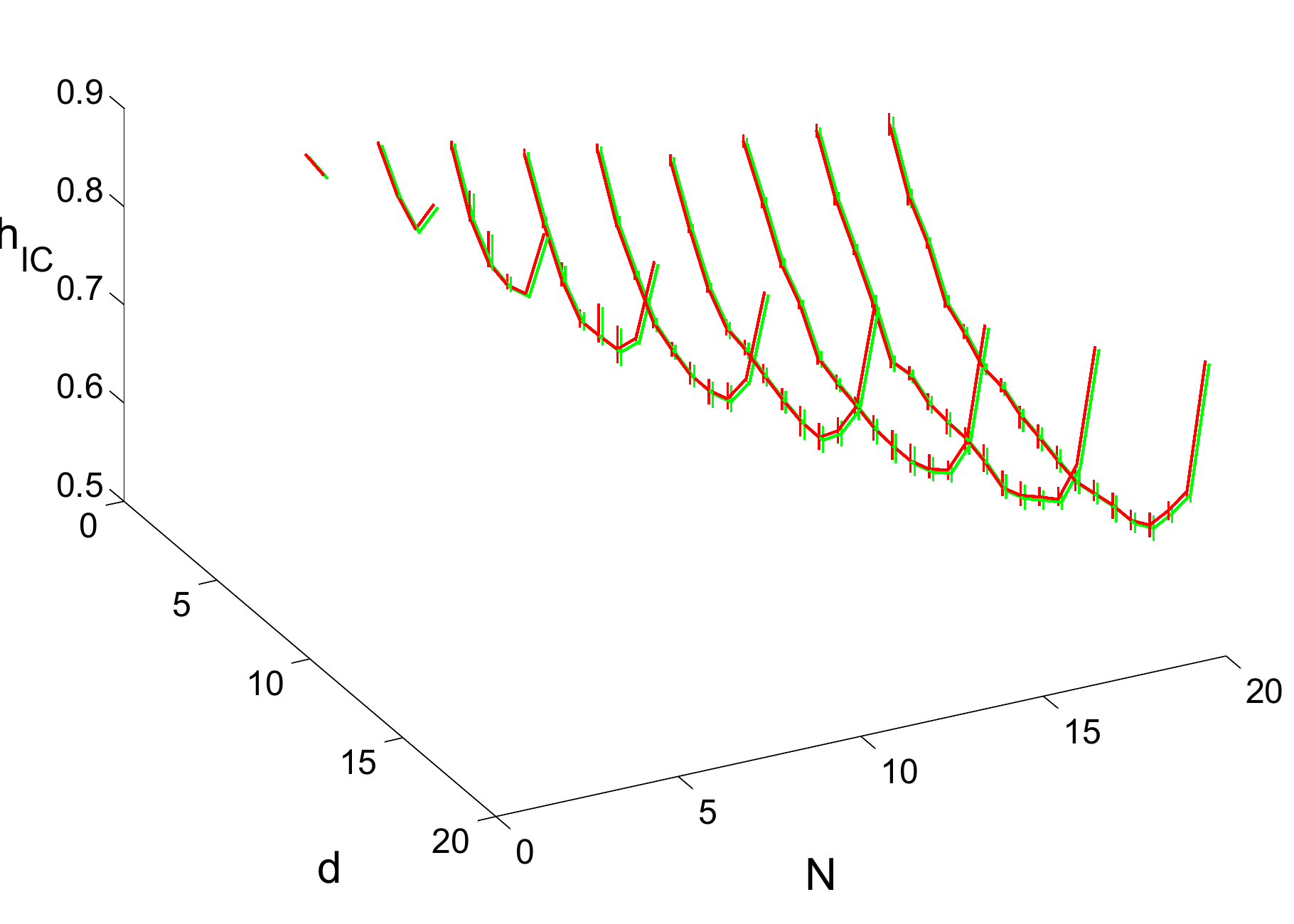} 

\hspace{0.8cm} (e)  \hspace{5.3cm} (f)

\caption{Landscape measures  as a function of the number of players $N$ and coplayers $d$, while  no replacement restriction are imposed. Red lines give the results for BD updating, green lines are for DB updating. Vertical spikes indicate the range between smallest and largest value.  (a), (b) Modality measured by the number of local maxima $\#_{LM}$, (c), (d) correlation length $\lambda$ and (e), (f) information content $h_{IC}$.  }
\label{fig:meas}

\end{figure*}

Fig. \ref{fig:meas} shows the   landscape measures number of  local maxima $\#_{LM}$,  correlations length $\lambda$ and information content $h_{IC}$ over  $N$ and $d$. The red lines indicate a BD updating, the green lines a DB updating. In addition to the quantities averaged over the up to $1000$ different interaction networks tested (horizontal lines), the vertical spikes indicate the range between the least and the largest value of $\#_{LM}$, $\lambda$ or $h_{IC}$ over these networks described by the adjacency matrices $A_I(\kappa)$.   This presentation and color code is kept for all landscape measures and fixation properties.

For the SD game, the number of  local maxima $\#_{LM}$ are given as semi--logarithmic plots, see Fig.  \ref{fig:meas}b. 
The results show that  for the PD landscape (Fig.  \ref{fig:meas}a) there is only one maximum for all tested $N$ and $d$, and both BD and DB updating. It hence is unimodal, while for the SD game the number of maxima increases with $N$  and is hence multimodal.  Moreover, for the SD game, we find for DB updating that $\#_{LM}$ decreases for a given $N$ and $d$ getting larger.  Also, for the SD game accounting for $\#_{LM}$ does not reflect the symmetry between BD and DB landscapes, which is not the case for the PD game.  Another observation is that for the SD landscape  the number of local maxima $\#_{LM}$ sometimes shows vertical spikes indicating the amount to be not constant for a given $N$ and $d$ and varying networks.  In other words, there is a certain variety in the number of local maxima over instances of interaction networks expressed as $d$--regular graphs. Some interaction graphs yield game landscapes with a larger (or smaller) number of local maxima than average.   Further note that
$\#_{LM}/2^N \rightarrow 0$ for $N$ getting larger. 
All these results support previous findings about evolutionary games, for instance that PD games and BD updating do not provide an advantage for cooperators,~\cite{ohts07,zuk13}. Thus, for PD the small number of maxima of the player--wise landscapes $\Lambda_\pi^i$ (compare to Fig. \ref{fig:n4}) corresponds with the small number of maxima in the game landscape. By contrast, for the SD game and DB updating not only configurations where the defecting player earns the largest payoff are maxima of the game landscapes. Consequently, the number $\#_{LM}$ is larger. 
For the landscape measures $\lambda$ and $h_{IC}$ in Fig. \ref{fig:meas}c--f, we find almost identical results for BD and DB landscapes, yet the different games can be distinguished, albeit not as clearly as for $\#_{LM}$. Hence,  correlation length and information content largely reflect the symmetry of the game landscapes. It can also be seen that the variety over different networks of interaction is slightly stronger for $h_{IC}$ than for $\lambda$. Also, it is interesting to see that the landscape measures are similar for PD and SD, while modality is different. A possible explanation is that for the PD  game the landscape is, apart from being unimodal, structurally similar to holey landscapes,~\cite{gav04} where neutral ridges are  mixed with holes of lower fitness.

\begin{figure*}

\hspace{0.8cm} PD game \hspace{5.0cm} SD game 

\vspace{0.5cm}

\includegraphics[trim = 15mm 65mm 10mm 100mm,clip, width=6cm, height=6cm]{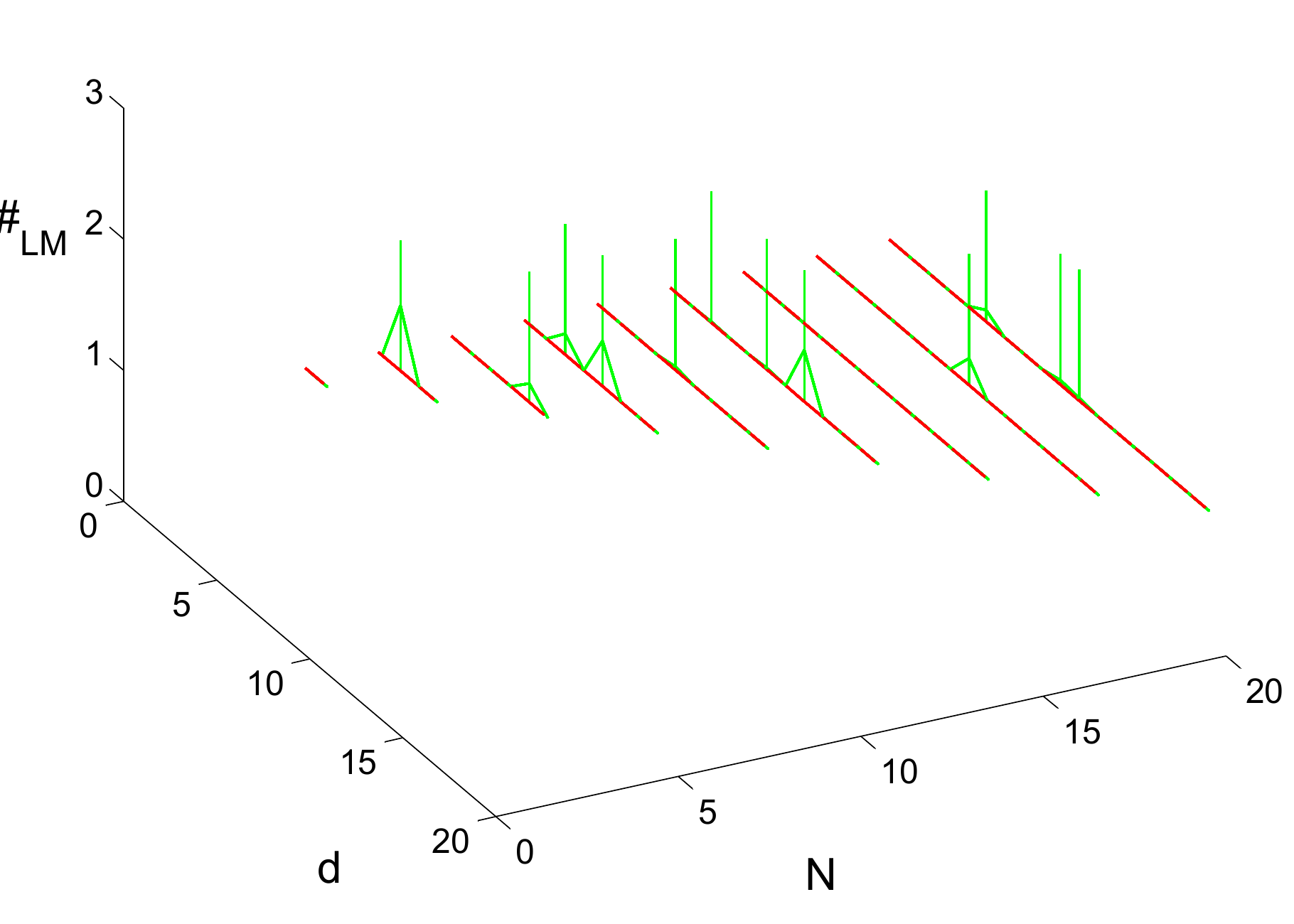} 
\includegraphics[trim = 15mm 65mm 10mm 100mm,clip, width=6cm, height=6cm]{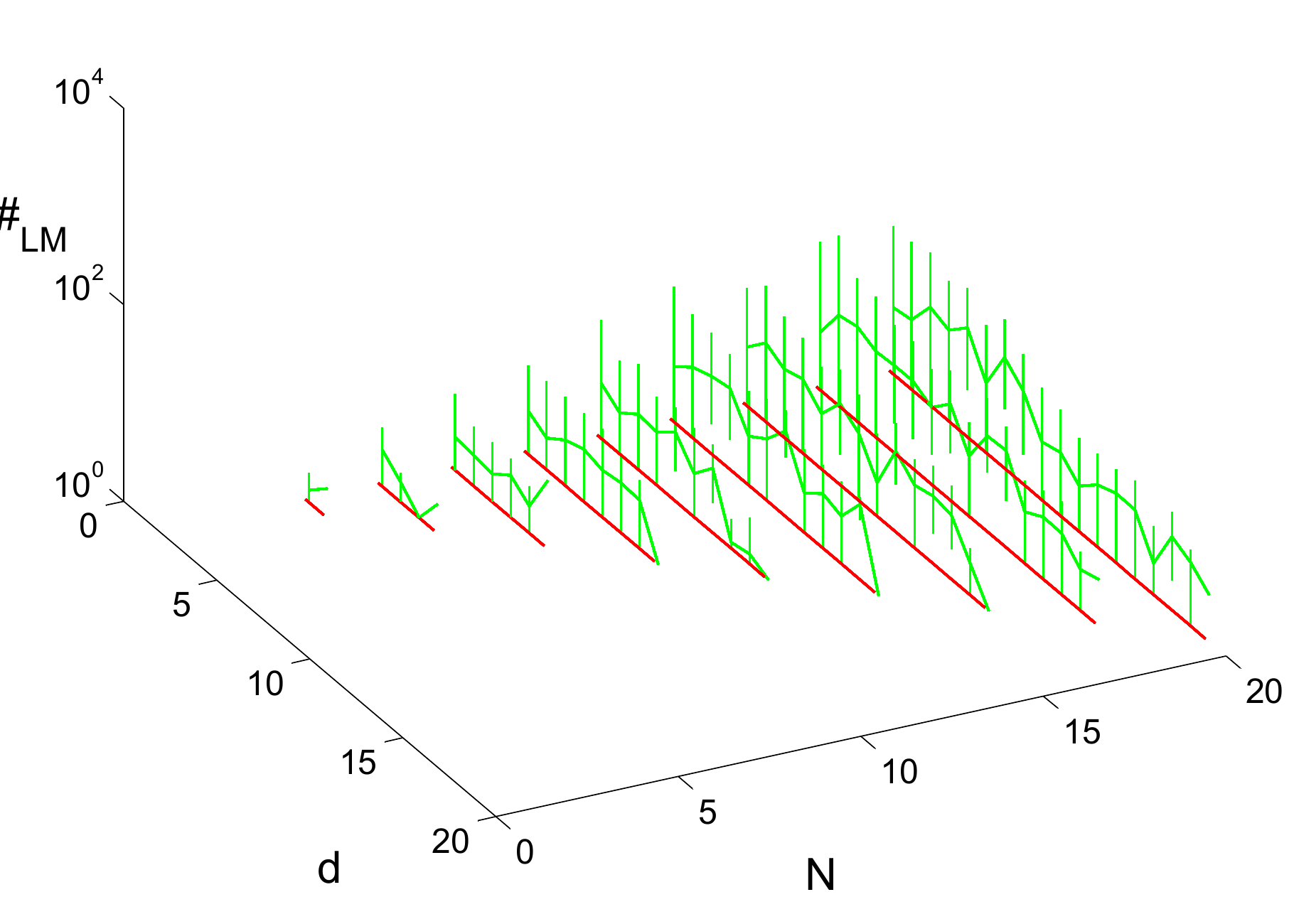} 

\hspace{0.8cm} (a)  \hspace{5.3cm} (b)

\vspace{0.5cm}

\includegraphics[trim = 15mm 65mm 10mm 100mm,clip, width=6cm, height=6cm]{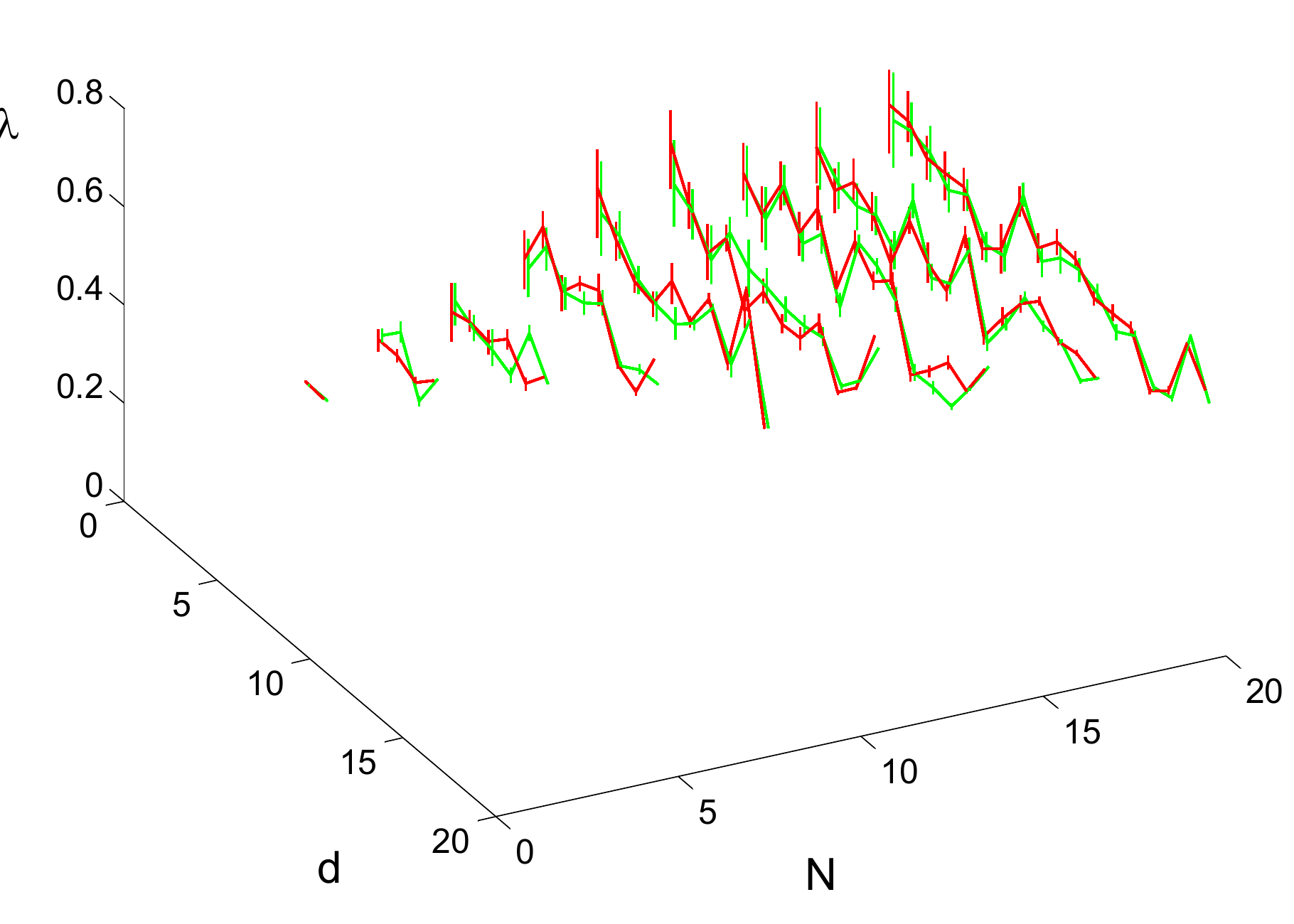} 
\includegraphics[trim = 15mm 65mm 10mm 100mm,clip, width=6cm, height=6cm]{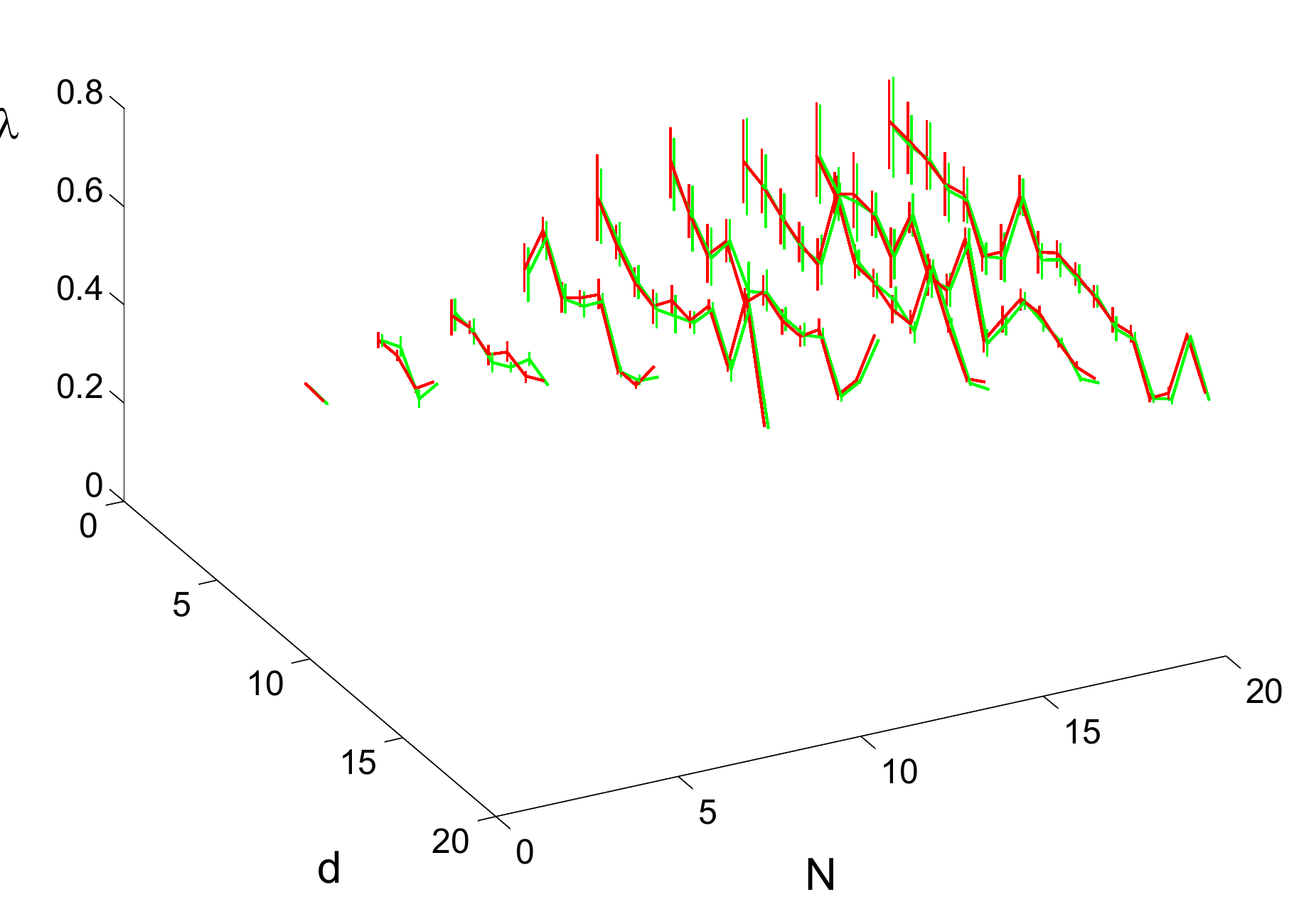} 

\hspace{0.8cm} (c)  \hspace{5.3cm} (d)

\vspace{0.5cm}

\includegraphics[trim = 15mm 65mm 10mm 100mm,clip, width=6cm, height=6cm]{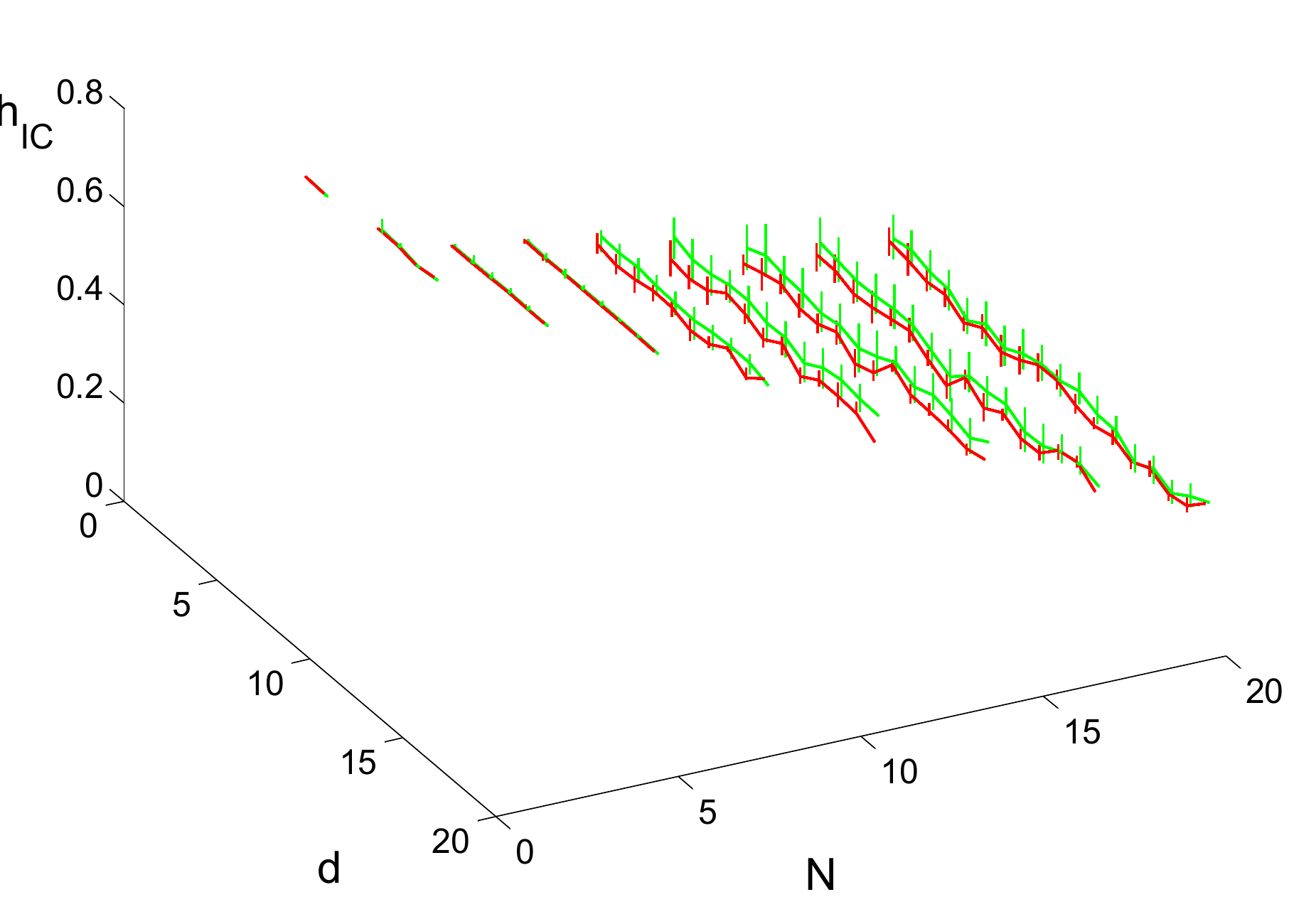} 
\includegraphics[trim = 15mm 65mm 10mm 100mm,clip, width=6cm, height=6cm]{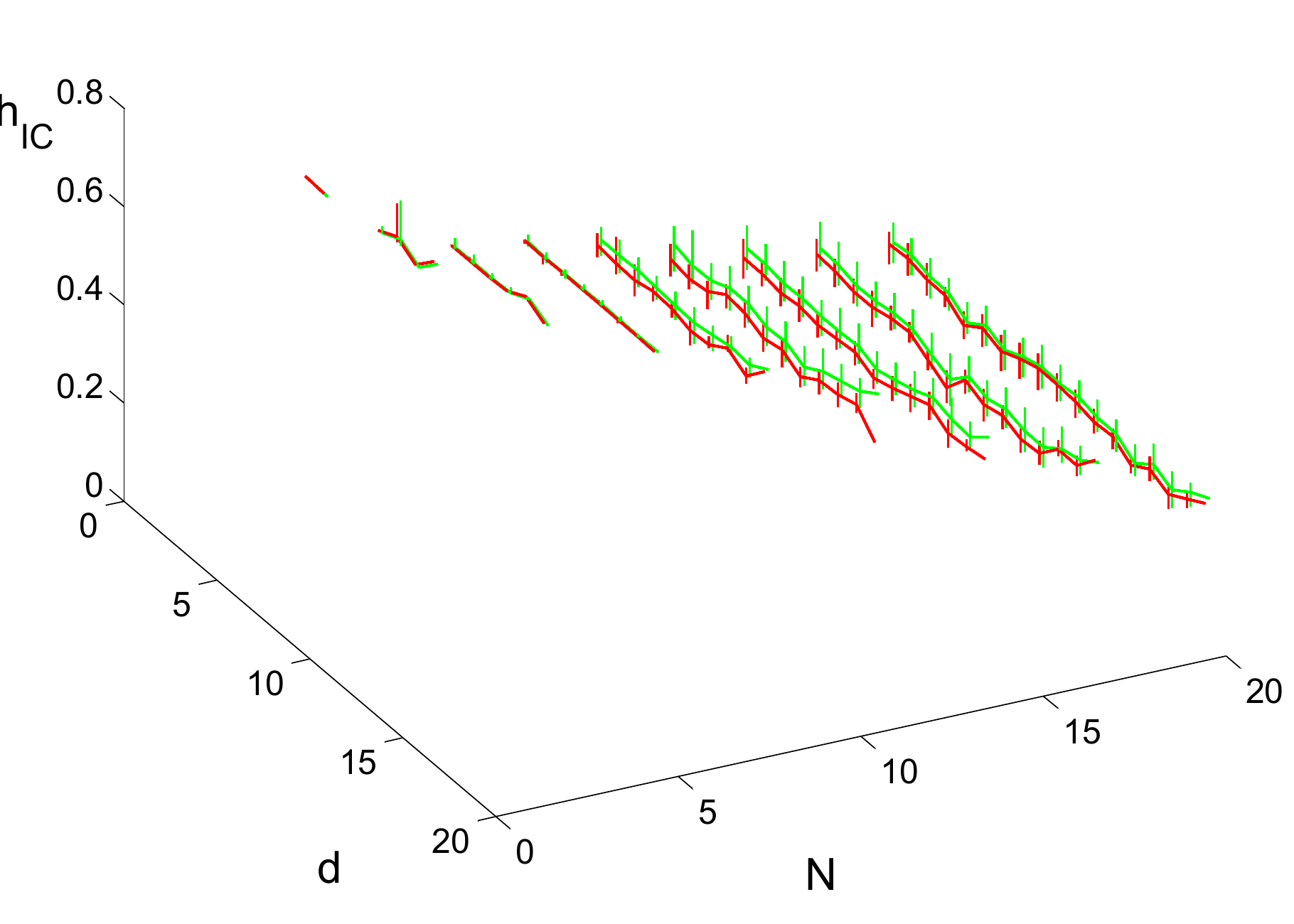} 

\hspace{0.8cm} (e)  \hspace{5.3cm} (f)

\caption{Landscape measures  as a function of the number of players $N$ and coplayers $d$ with replacement restriction specified by the replacement matrix $W_R$. Red lines give the results for BD updating, green lines are for DB updating.    Vertical spikes indicate the range between smallest and largest value. (a), (b) Modality measured by the number of local maxima $\#_{LM}$, (c), (d) correlation length $\lambda$ and (e), (f) information content $h_{IC}$.   }
\label{fig:meas_rest}

\end{figure*}

We next analyze the effect of replacement restrictions imposed by the replacement graph not being fully connected with evenly weighted edges, and focus on the differences between replacement restriction being set or not, see Fig. \ref{fig:meas_rest}. A main observation is that replacement restrictions modify the game landscapes, which is also shown by the landscape measures.  For instance, the number of local maxima $\#_{LM}$ for the PD game  and DB updating is no longer strictly unimodal (compare Figs.   \ref{fig:meas}a and \ref{fig:meas_rest}a). Interestingly, for BD updating, even replacement restrictions do not alter unimodality. These is still just one maximum for all tested $N$ and $d$, see the red lines in Fig. \ref{fig:meas_rest}a. For the SD game, the inverse proportionality between    $\#_{LM}$ and $d$ ceases, and generally the number of local maxima does vary more strongly for different networks of interaction. These characteristics can also be seen for the landscape measures correlation length $\lambda$ (see Fig. \ref{fig:meas_rest}c,d) and information content $h_{IC}$ (see  Fig. \ref{fig:meas_rest}e,f). Here, the main difference is that the measures are no longer the same (or almost the same) for BD and DB updating. This reflects the broken symmetry that replacement imposes on BD and DB game landscapes. Generally, 
 replacement restrictions imply landscapes that vary more substantially  over different networks of interaction. Furthermore, as there is an inverse relationship between  ruggedness and correlation length $\lambda$, it can be noted that   ruggedness decreases as the number of player gets larger. This effect is amplified by replacement restrictions.

\begin{figure*}[t]

\hspace{0.8cm} PD game \hspace{5.0cm} SD game 

\vspace{0.5cm}

\includegraphics[trim = 15mm 65mm 10mm 100mm,clip, width=6cm, height=6cm]{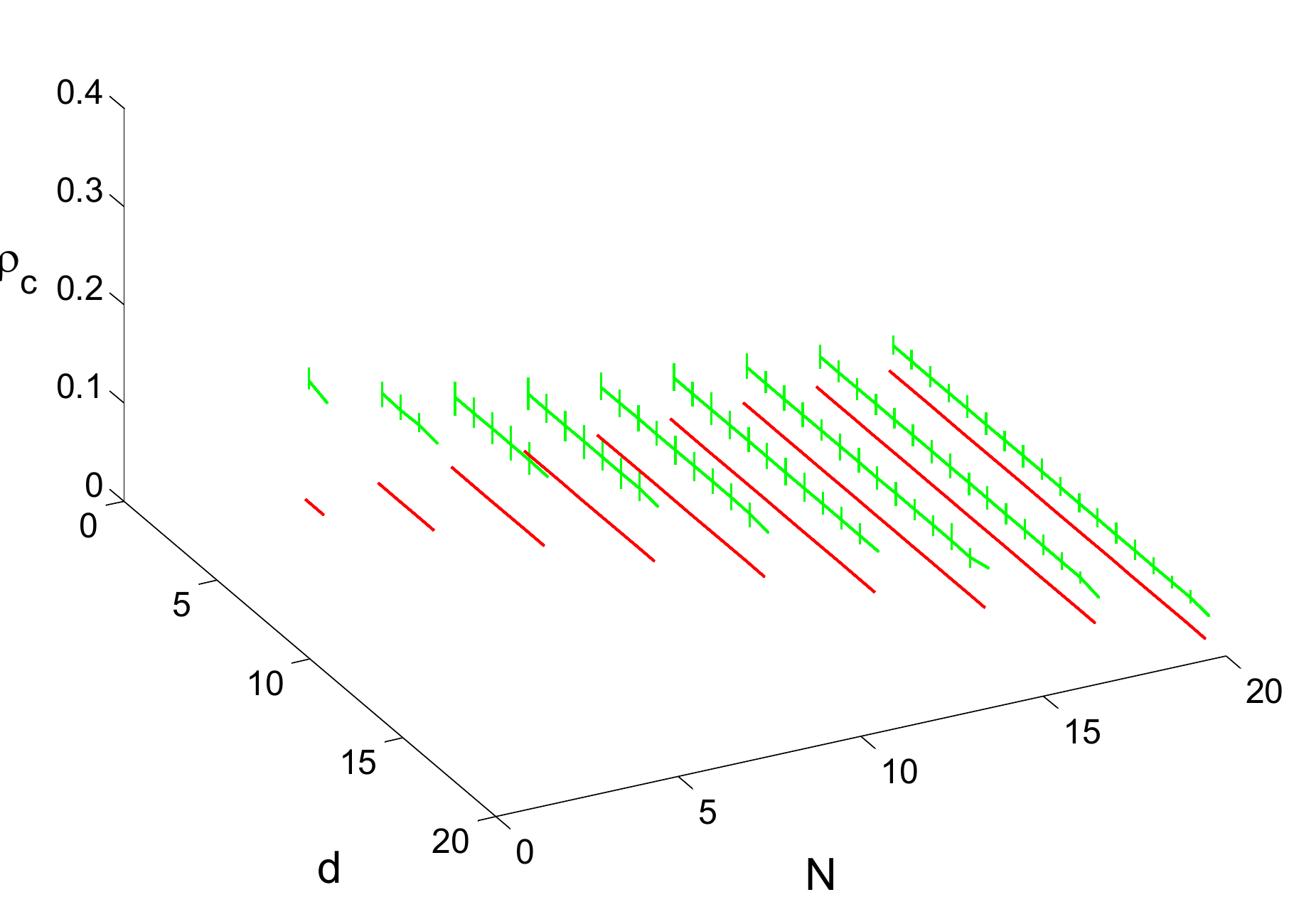} 
\includegraphics[trim = 15mm 65mm 10mm 100mm,clip, width=6cm, height=6cm]{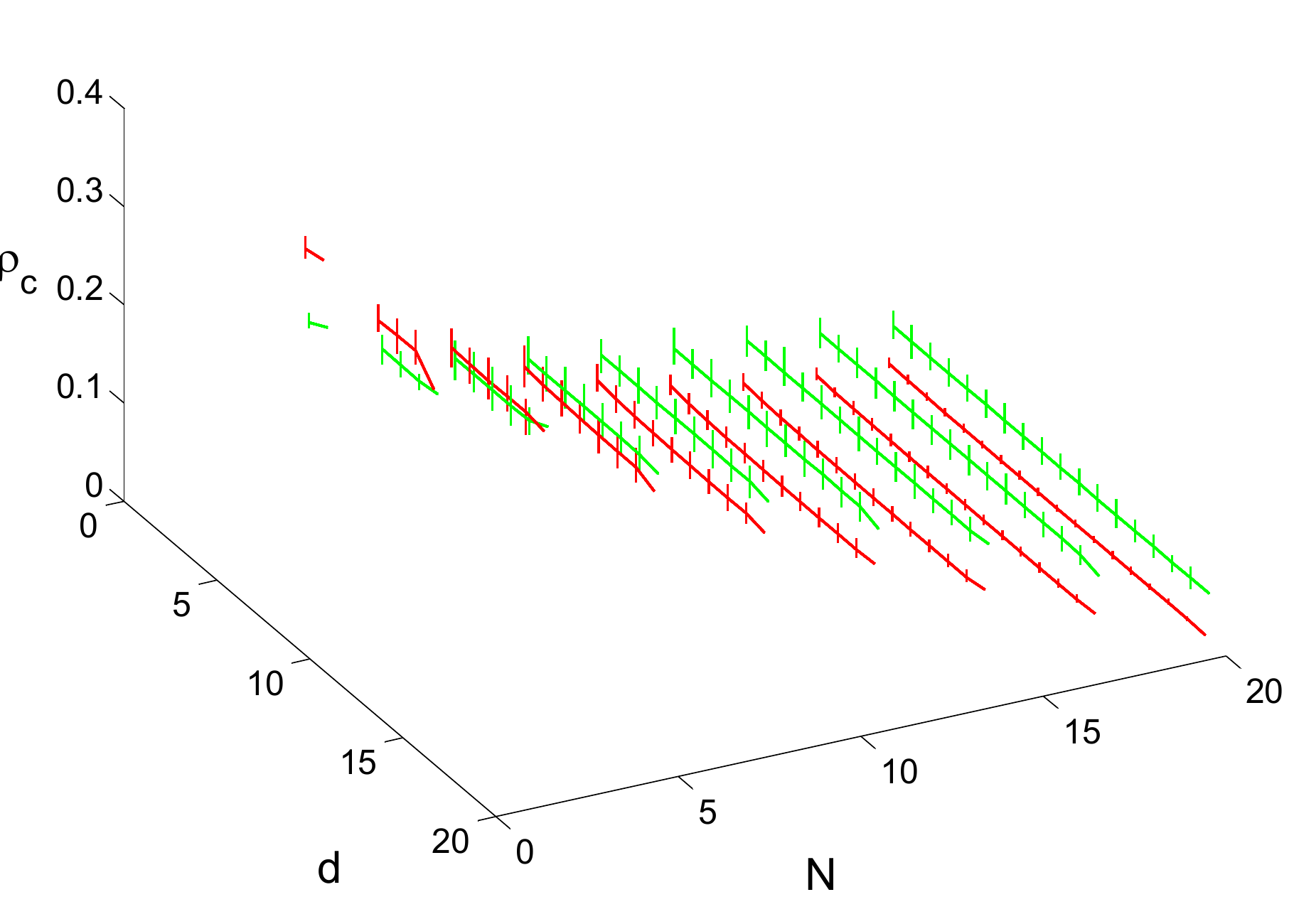} 

\hspace{0.8cm} (a)  \hspace{5.3cm} (b) 
\vspace{0.3cm}

\includegraphics[trim = 15mm 65mm 10mm 100mm,clip, width=6cm, height=6cm]{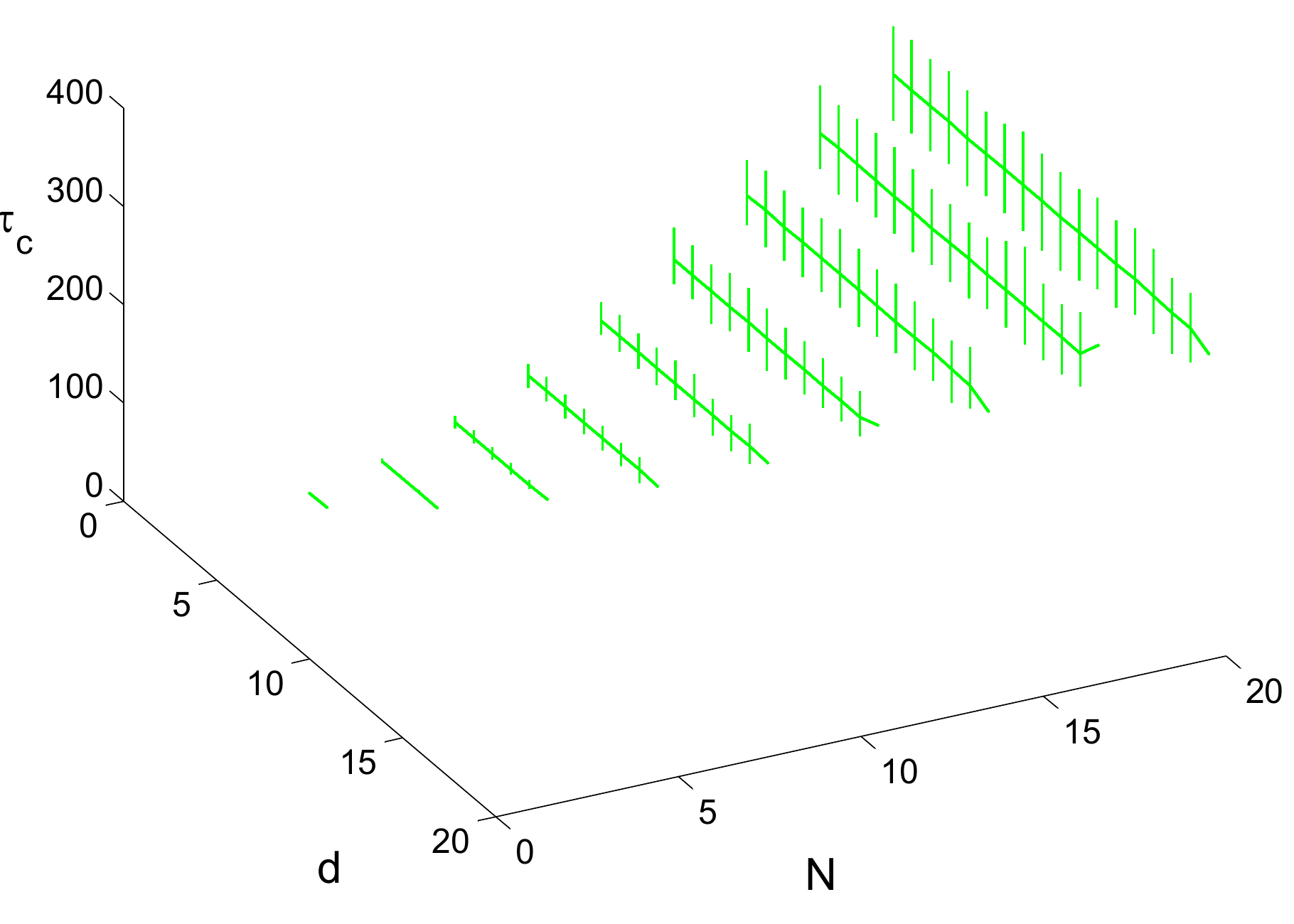} 
\includegraphics[trim = 15mm 65mm 10mm 100mm,clip, width=6cm, height=6cm]{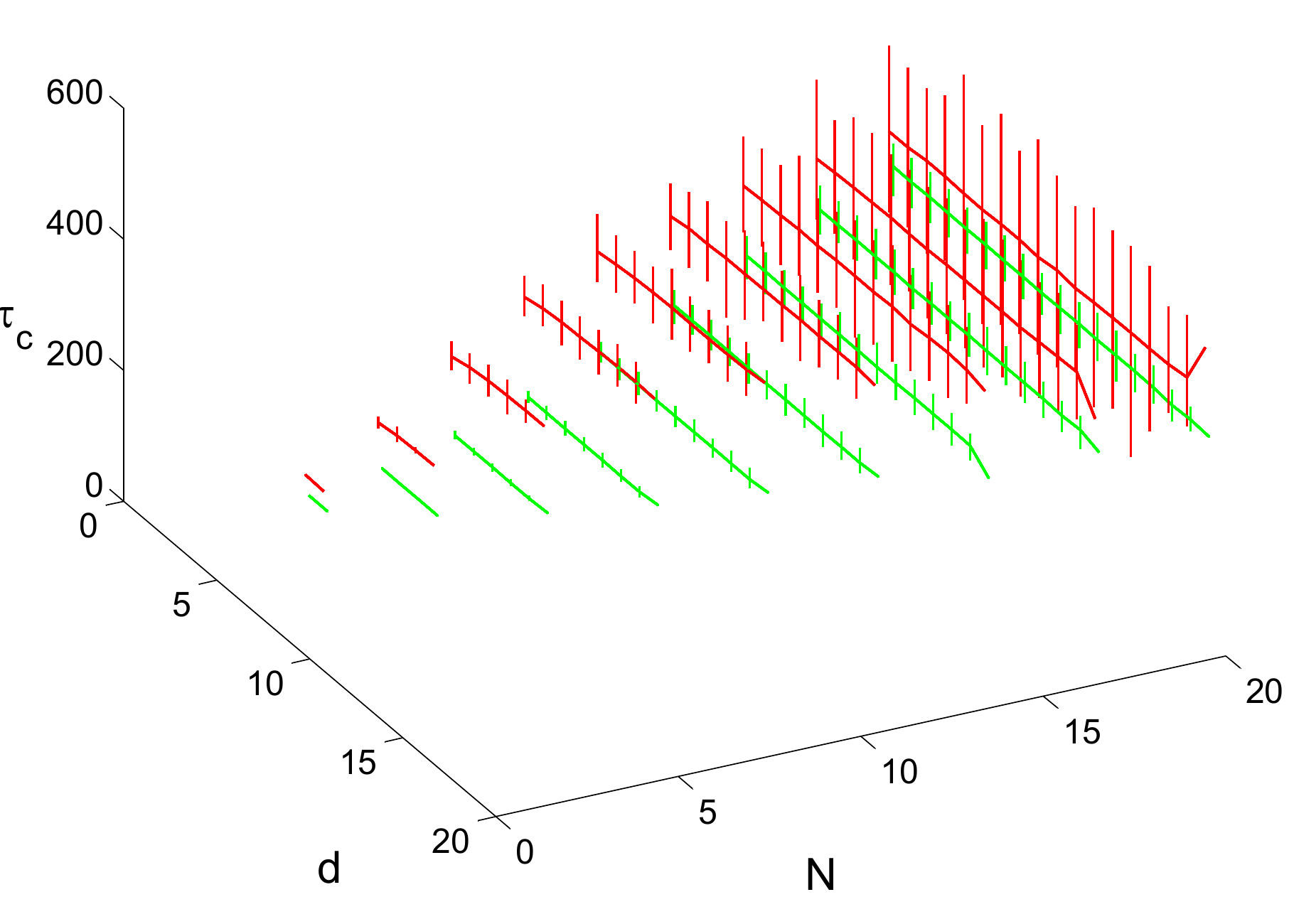} 

\hspace{0.8cm} (c)  \hspace{5.3cm} (d)

\caption{Fixation probability $\varrho_c$  and fixation time $\tau_c$ of the cooperative absorbing configuration over $N$ and $d$. Red lines give the results for BD updating, green lines are for DB updating. No replacement restriction imposed.  Vertical spikes indicate the range between smallest and largest value over the considered interaction matrices $A_I$. (a), (b): Fixation probability $\varrho_c$, (c), (d): fixation time $\tau_c$.       }
\label{fig:fix_coop}

\end{figure*}

\begin{figure*}[t]

\hspace{0.8cm} PD game \hspace{5.0cm} SD game 

\vspace{0.5cm}

\includegraphics[trim = 15mm 65mm 10mm 100mm,clip, width=6cm, height=6cm]{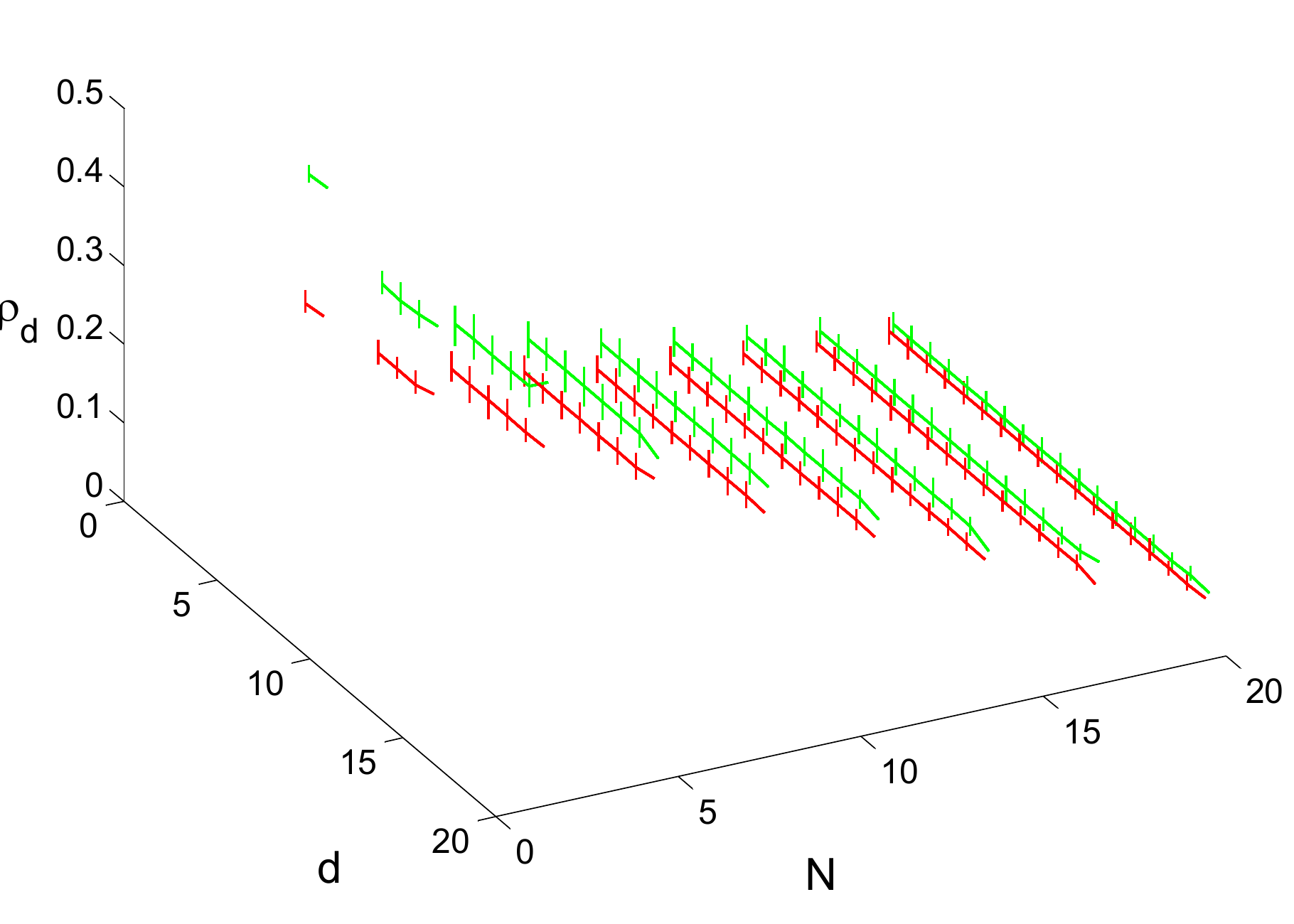} 
\includegraphics[trim = 15mm 65mm 10mm 100mm,clip, width=6cm, height=6cm]{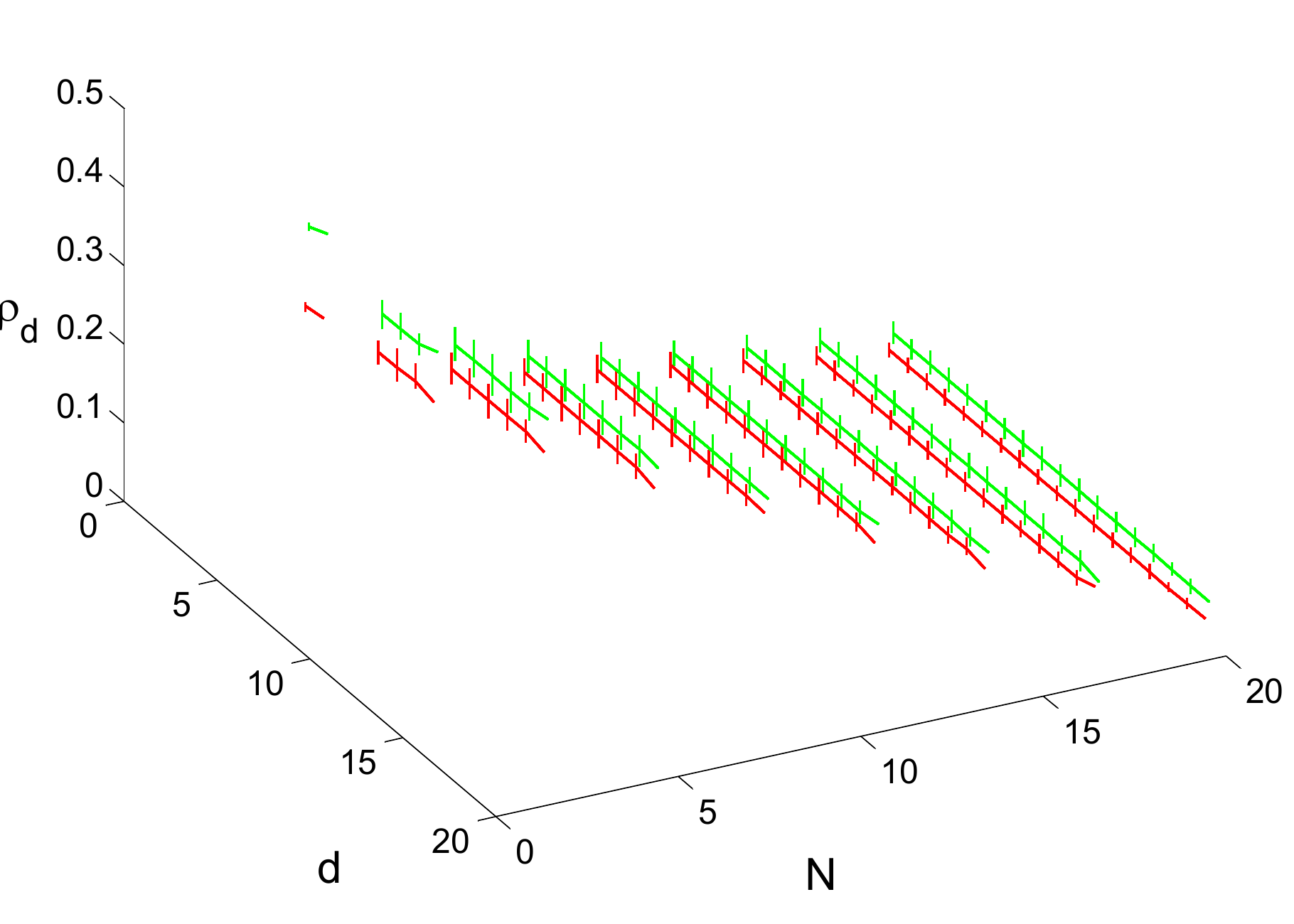} 

\hspace{0.8cm} (a)  \hspace{5.3cm} (b) 
\vspace{0.3cm}

\includegraphics[trim = 15mm 65mm 10mm 100mm,clip, width=6cm, height=6cm]{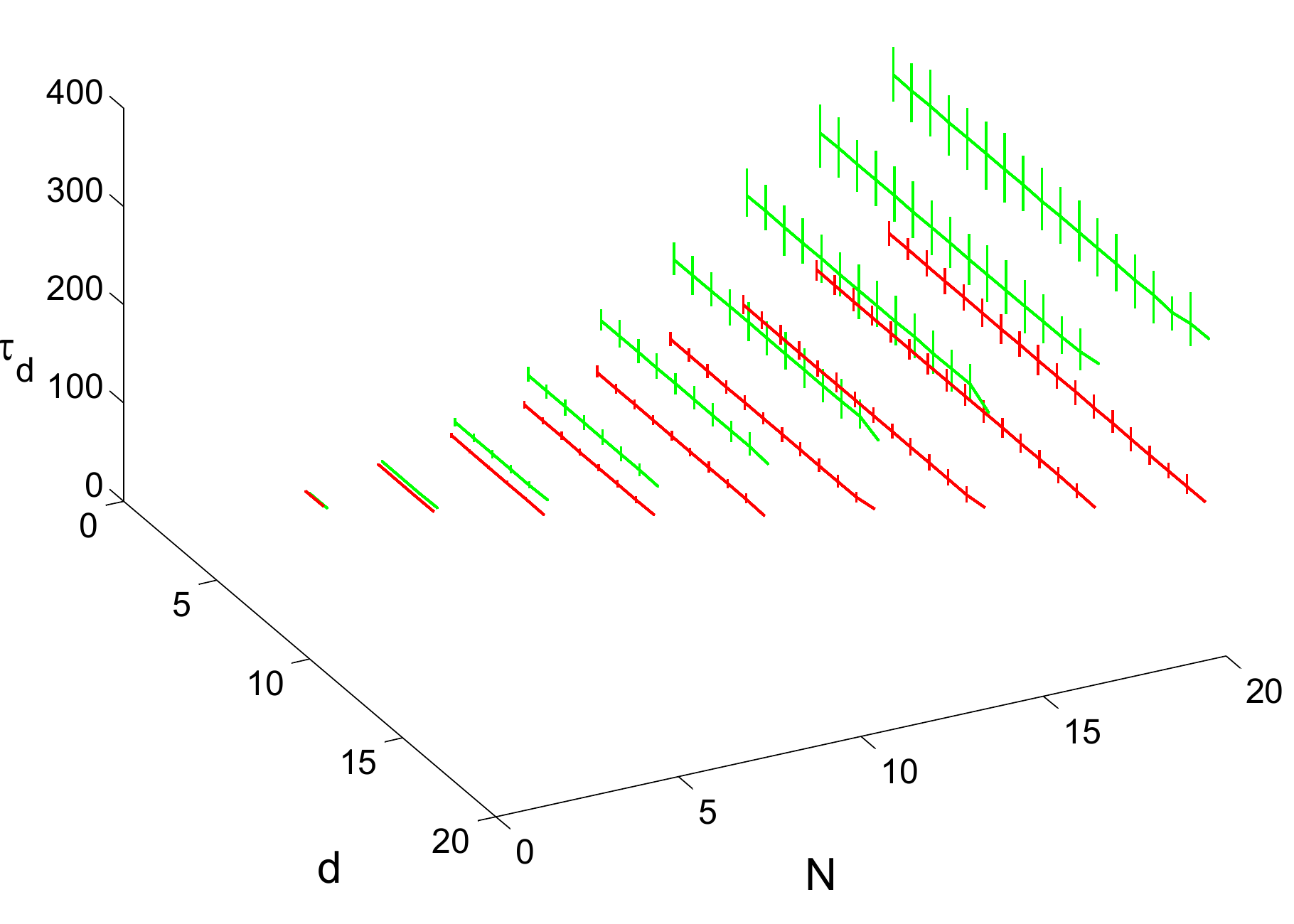} 
\includegraphics[trim = 15mm 65mm 10mm 100mm,clip, width=6cm, height=6cm]{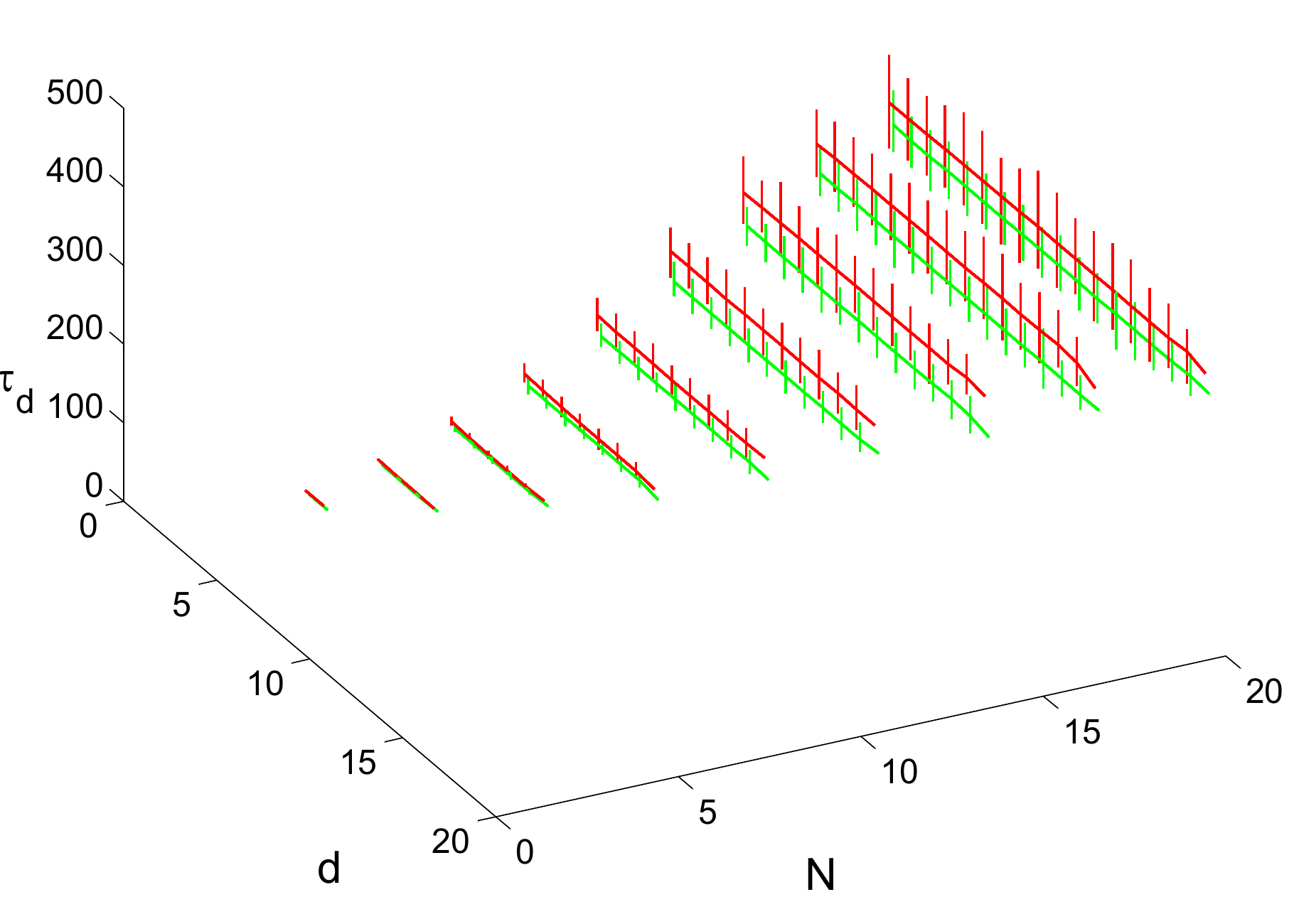} 

\hspace{0.8cm} (c)  \hspace{5.3cm} (d)

\caption{Fixation probability $\varrho_d$  and fixation time $\tau_d$ of the defective absorbing configuration over $N$ and $d$. Red lines give the results for BD updating, green lines are for DB updating. No replacement restriction imposed. Vertical spikes indicate the range between smallest and largest value over the considered interaction matrices $A_I$. (a), (b): Fixation probability  $\varrho_d$, (c), (d): fixation time  $\tau_d$.       }
\label{fig:fix_def}

\end{figure*}

In Fig.  \ref{fig:fix_coop} fixation properties of the cooperative absorbing configuration $\pi_\infty$ with  $bc(\pi_\infty)=N$ are given over $N$ and $d$. The fixation probability $\varrho_c$ is zero for the PD game and BD updating, which corresponds to previous results showing that cooperation is never favored or beneficial under such an updating,~\cite{ohts07,zuk13}. Apart from this result,   $\varrho_c$ falls with the number of players $N$ getting larger, but for a given $N$, the fixation probability is the same for a varying number of coplayers $d$. These results are in line with  finding that regular evolutionary graphs  are generally isothermal,~\cite{lieb05}.  Moreover, except for very small numbers of players ($N=4$ and partly $N=6$), the fixation probability for the well--mixed case ($d=N-1$) is the same as for a smaller number of coplayers.  This, however, is only the case for averages over interaction networks. There are for a constant $N$ and $d$ interaction networks with fixation probabilities larger or smaller than average indicated by the vertical spikes (for $d=N-1$ there cannot be a spike as only one instance of $A_I$ exists). Hence, these results suggest that the graph structure of the interaction network does matter for $2 \leq d <N-1$. Moreover, which $A_I(\kappa)$ causes the largest or smallest $\varrho_c$ varies over $N$ and $d$. Regarding the fixation times $\tau_c$, similar observations can be made. Note that for the PD game no fixation time for BD updating are given as the fixation probability is zero. For the SD game the fixation times for BD updating are much larger than for DB.
Fixation properties of the defective absorbing configuration $\pi_\infty$ with  $bc(\pi_\infty)=0$ are given in Fig.  \ref{fig:fix_def}. All game settings produce fixation probabilities $\varrho_d>0$. Apart from this, the results are similar to the general trends for the cooperative absorption, namely fixation probability differs for BD and DB updating, falls with an increasing number of players $N$ and is isothermal for given $N$ and a varying number of coplayers $d$. The fixation times $\tau_d$ in Fig. \ref{fig:fix_def}c,d also show some similarity, but also differences.  The main observation is that for the SD game the maximal fixation times are not substantially larger than for the PD game.

We next consider the local absorption structure of the game landscapes, which is based on up to $6000$ different interaction matrices $A_I(\kappa)$. The results for the  PD and SD game   with BD and DB landscapes  are given in Tab. \ref{tab:absorp}. 
The final absorption structure (F) indicates the local topological features of the absorbing configurations $\pi_\infty$, while the initial absorption structure (I) denotes the features of the initial configurations $\pi_0$ from where potential paths to the absorbing configuration may start. There are four different local topological features (maximum, minimum, neutral, and saddle), which are given for both the cooperative absorbing configuration with $bc(\pi_\infty)=N$ and the defective absorbing configuration with $bc(\pi_\infty)=0$. 
\begin{table}[tb]
\caption{Local absorption structure for the PD and the SD game;  $bc(\pi_\infty)=N$ is for the cooperative absorbing configuration and $bc(\pi_\infty)=0$ for the defective absorbing configuration. Results show the final absorption structure (F) and the initial absorption structure (I). Local topological features of absorbing configurations: maximum (mx), minimum (mn), neutral (nt), saddle (sd)}

\vspace{0.2cm}
\label{tab:absorp}      
Cooperative absorbing \hspace{1.6cm}Defective absorbing

configuration: $bc(\pi_\infty)=N$ \hspace{0.95cm}configuration: $bc(\pi_\infty)=0$

\vspace{0.2cm}
\begin{tabular} {|c|l|l||c|l|l|} \hline
 & PD & SD &  & PD & SD \\ \hline
  BD & F: mn  & F: nt  & BD &F: mx  & F: mx \\ 
&I: sd&I: sd&&I:  sd &I: mn\\ 
 DB & F: mx & F: mx$^a$,mn$^a$,nt  & DB&F: mn& F: mn\\
&I: sd&I: sd&&I:  sd&I: mx, sd\\
 \hline

\end{tabular}
\vspace{0.1cm}

$^a$ Not global for $4\leq N \leq 20$

\end{table}
The results in Tab. \ref{tab:absorp} show some general properties of the local absorption structure for the game settings considered, which in turn can be interpreted as specific for the game landscape proposed by Eqn.    (\ref{eq:bd}) and (\ref{eq:db}). A first general property is that for both games and both absorbing configurations, the absorption structure of BD updating generally differs from DB updating. This may answer the question of why game landscapes that are symmetric with respect to BD and DB for no replacement restrictions yield fixation properties that do differ from BD to DB.  A possible explanation is that while the game landscapes are topologically the same as shown by the landscape measures $\lambda$ and $h_{IC}$ (see Figs. \ref{fig:meas}c,d and \ref{fig:meas}e,f), their absorption structure is not. This suggests the absorption structure to be a determining factor in the relationships between game landscapes and fixation. A second general property is that the absorption structure is inverse complementary for BD and DB, the meaning of which is that if there is a maximum for BD, then DB has a minimum, and vice versa, while a saddle remains a saddle. This follows from the symmetry  properties via the birth and death probabilities (\ref{eq:birth}) and (\ref{eq:death}) as discussed in Sec. \ref{sec:gameland}. There is an exception with the cooperative absorption of the SD game discussed later.   

The local absorption structure also shows differences between the social dilemma games.
A property specific for the PD game 
is that there is no variety over the number of players and coplayers. In other words, within a given game setting changing the number of players and coplayers does not alter the absorption structure. This also applies to  a general absence of variety over different  interaction matrices $A_I(\kappa)$. At least for the interaction matrices tested, the absorption structure is invariant over interaction networks. Also, the maxima/minima of the absorption structure are global.
For the SD game the results are slightly different. For this game setting, we find that the cooperative final absorbing configuration is neutral for BD updating, and a combination of maximal, minimal and neutral for DB updating.   Further analysis shows that these  topological features  vary over $N$, $d$ and $A_I(\kappa)$. For most $N$ and $d$ the absorbing configuration is neutral, while for a few $N$ and $d$ it is either a maximum or a minimum. In these cases, the topological features are fixed for varying interaction networks. In addition, there are $N$ and $d$ for which varying interaction networks give either a mixture of maxima and neutral, or  a mixture of minima and neutral. For the defective absorbing structure, the initial configurations vary over $A_I(\kappa)$, where for most $N$ and $d$ we have saddles, while for some other $N$ and $d$ there is a mix of saddles and maxima.  Also, for the cooperative absorption structure and DB, the maxima/minima are local, not global. 
Numerical evaluation affirms that such a variety of the topological features of the absorption structure over interaction networks can be observed for other game settings as well, particularly those on the line
 $2T=11+P-S$. The SD game with $T=5$ is exactly on this line for the parametrization used ($R=3$, $S=1$, $P=0$). 

\begin{figure*}[t]

\vspace{0.5cm}

\includegraphics[trim = 10mm 65mm 10mm 80mm,clip, width=6cm, height=6cm]{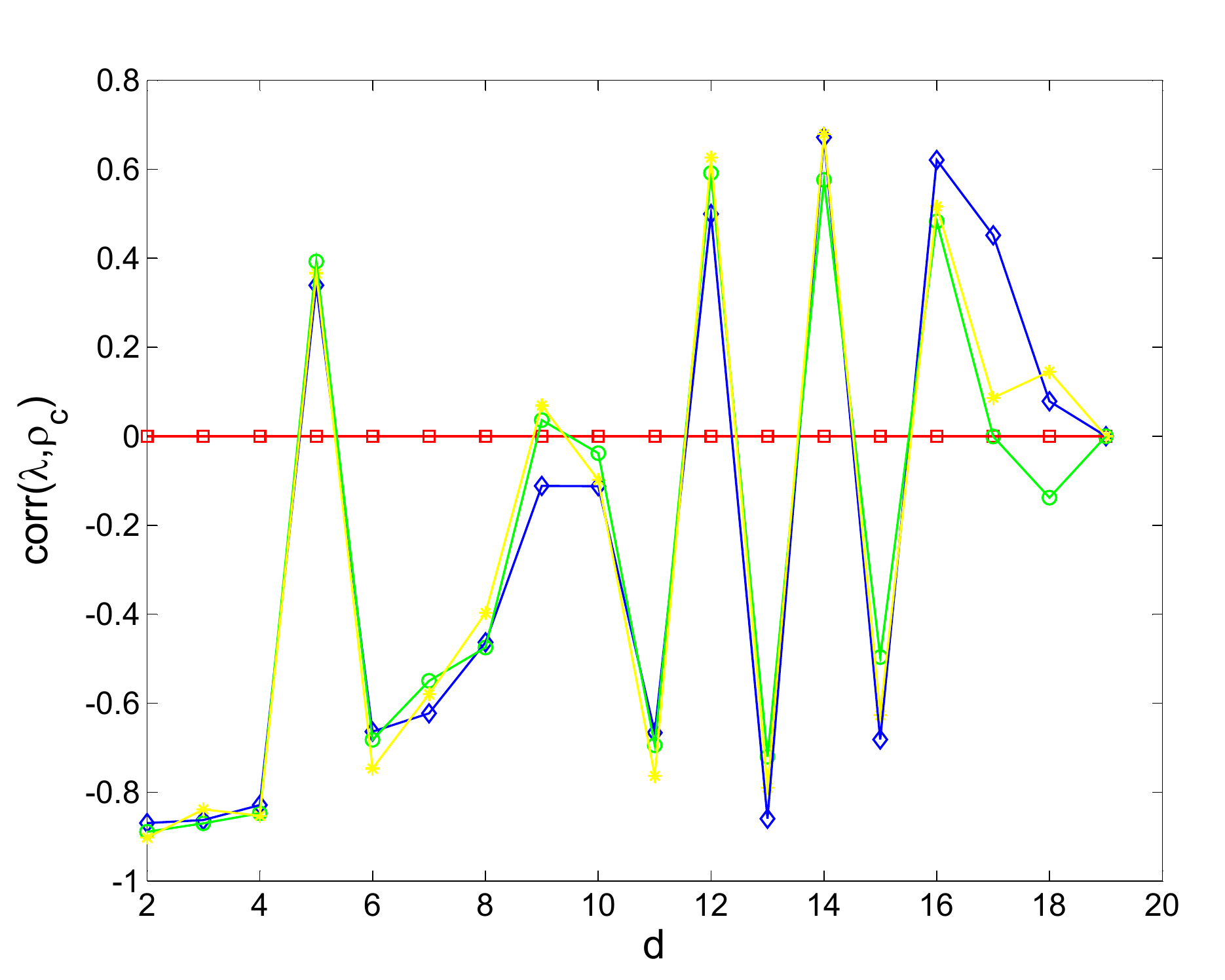} 
\includegraphics[trim = 10mm 65mm 10mm 80mm,clip, width=6cm, height=6cm]{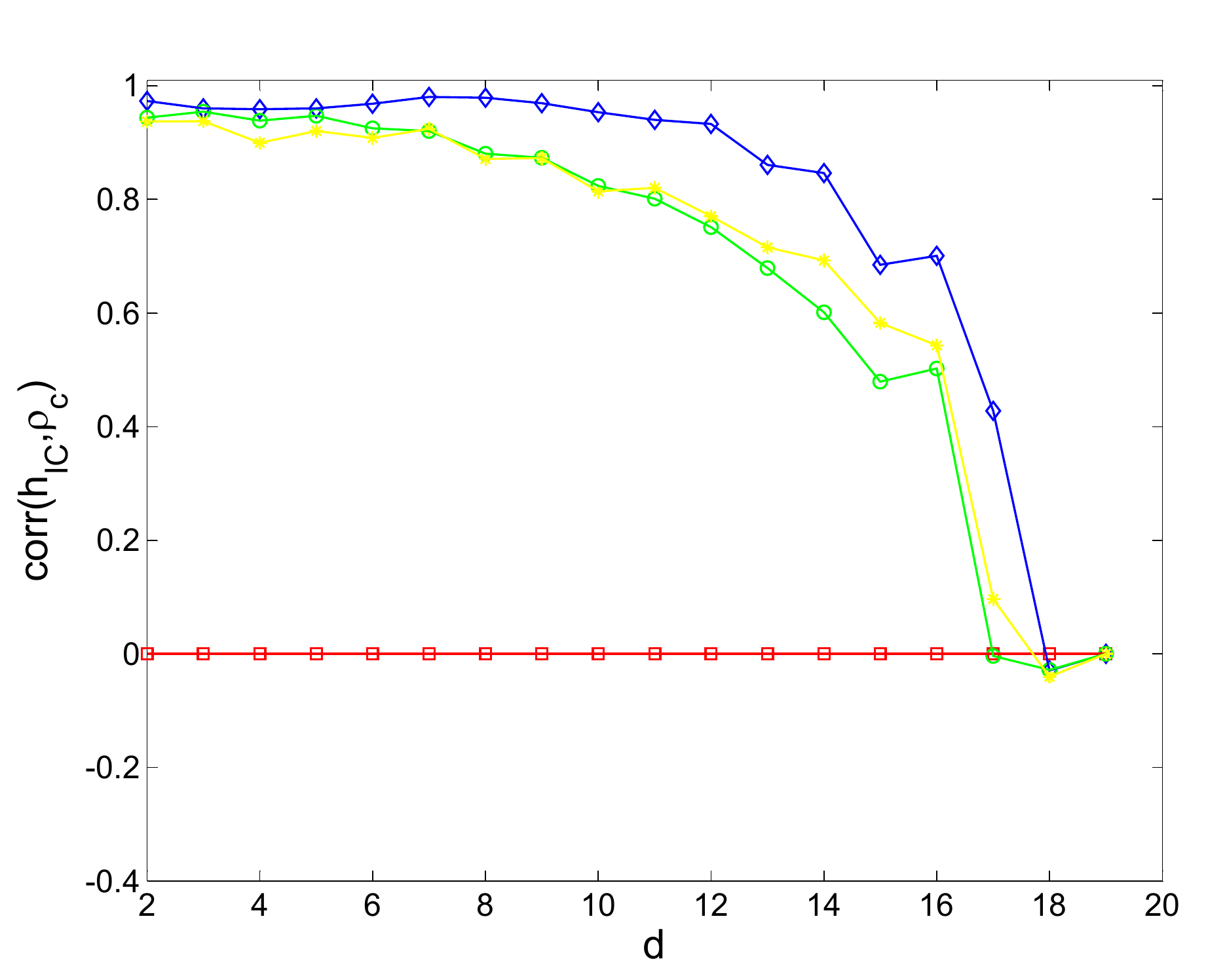} 

\hspace{0.8cm} (a)  \hspace{5.3cm} (b) 
\vspace{0.3cm}

\includegraphics[trim = 10mm 65mm 10mm 80mm,clip, width=6cm, height=6cm]{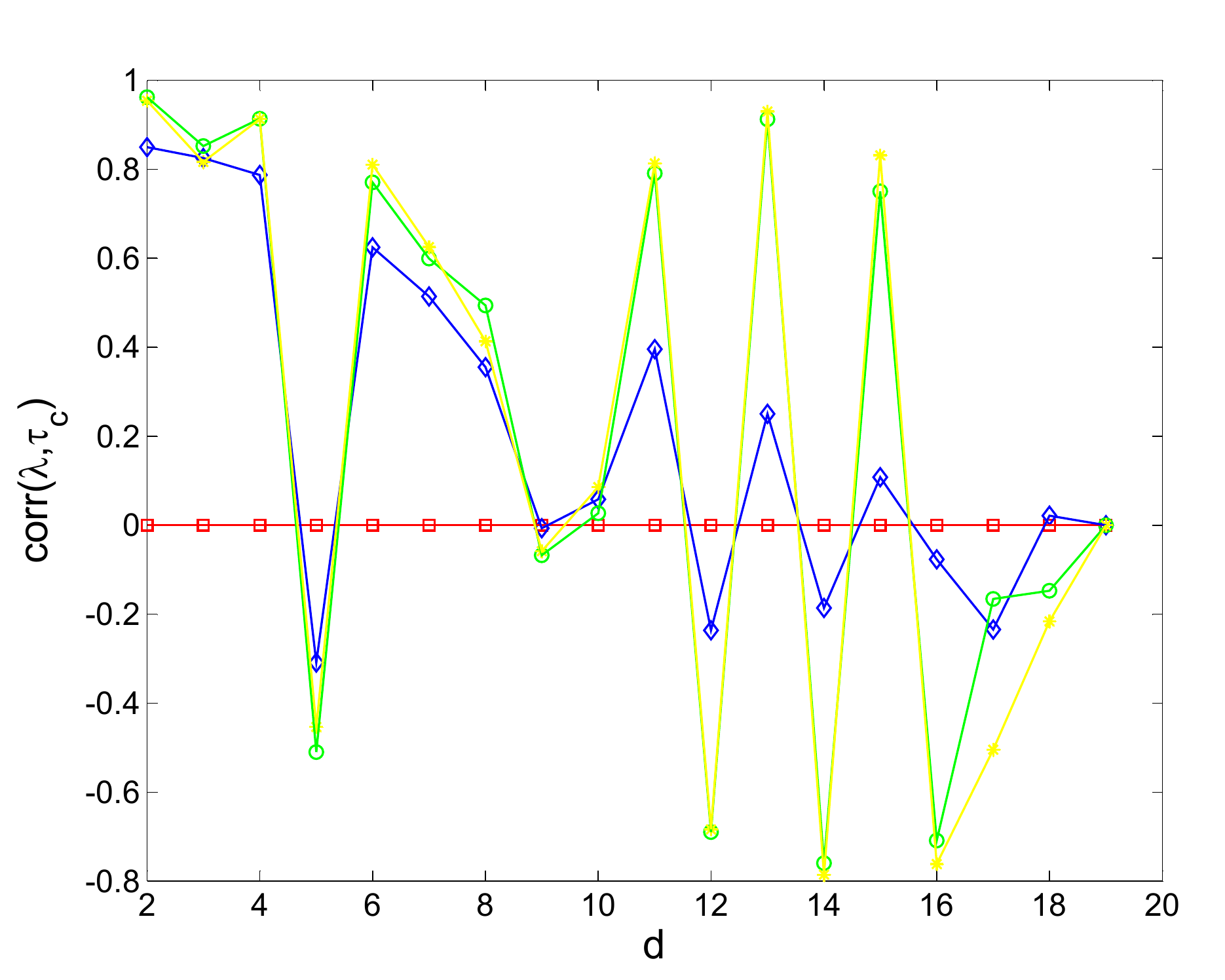} 
\includegraphics[trim = 10mm 65mm 10mm 80mm,clip, width=6cm, height=6cm]{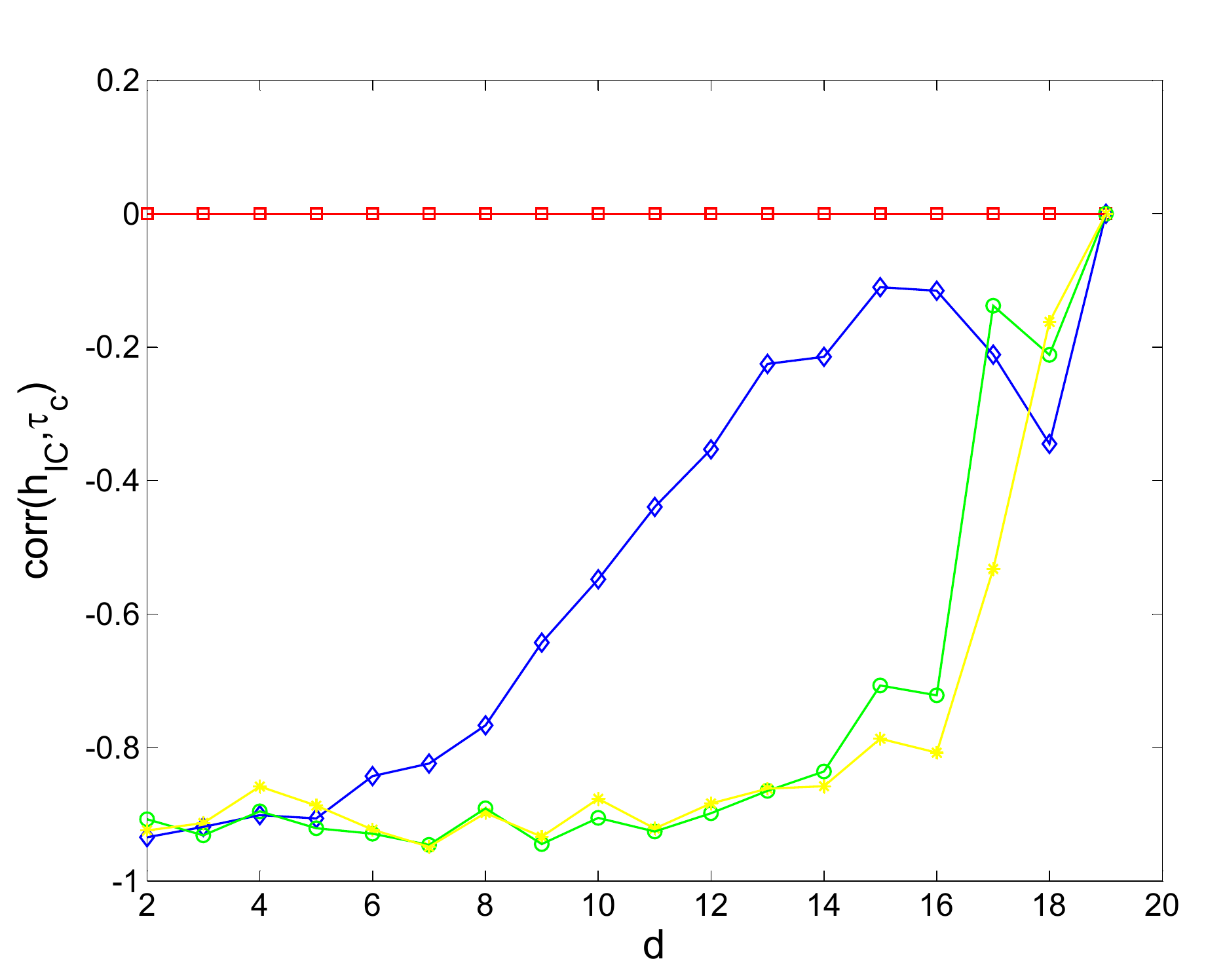} 

\hspace{0.8cm} (c)  \hspace{5.3cm} (d)

\caption{Correlation between fixation properties and landscape measures for the cooperative absorbing configuration over the number of coplayers $d$.  Red markers give the results for PD--BD, green for PD--DB, blue for SD--BD, yellow for SD--DB. Fixation probability $\varrho_c$:   (a)  correlation length $\lambda$,  (b) information content $h_{IC}$.  Fixation time $\tau_c$:     (c)   correlation length $\lambda$,  (d)  information content $h_{IC}$. The information content correlates well to the fixation properties, while the relation to the correlation length is weak. There is no correlation for  PD--BD as the fixation of cooperation is zero for this game and strategy updating.   }
\label{fig:corr_coop}

\end{figure*}

\begin{figure*}[t]

\vspace{0.5cm}

\includegraphics[trim = 10mm 65mm 10mm 80mm,clip, width=6cm, height=6cm]{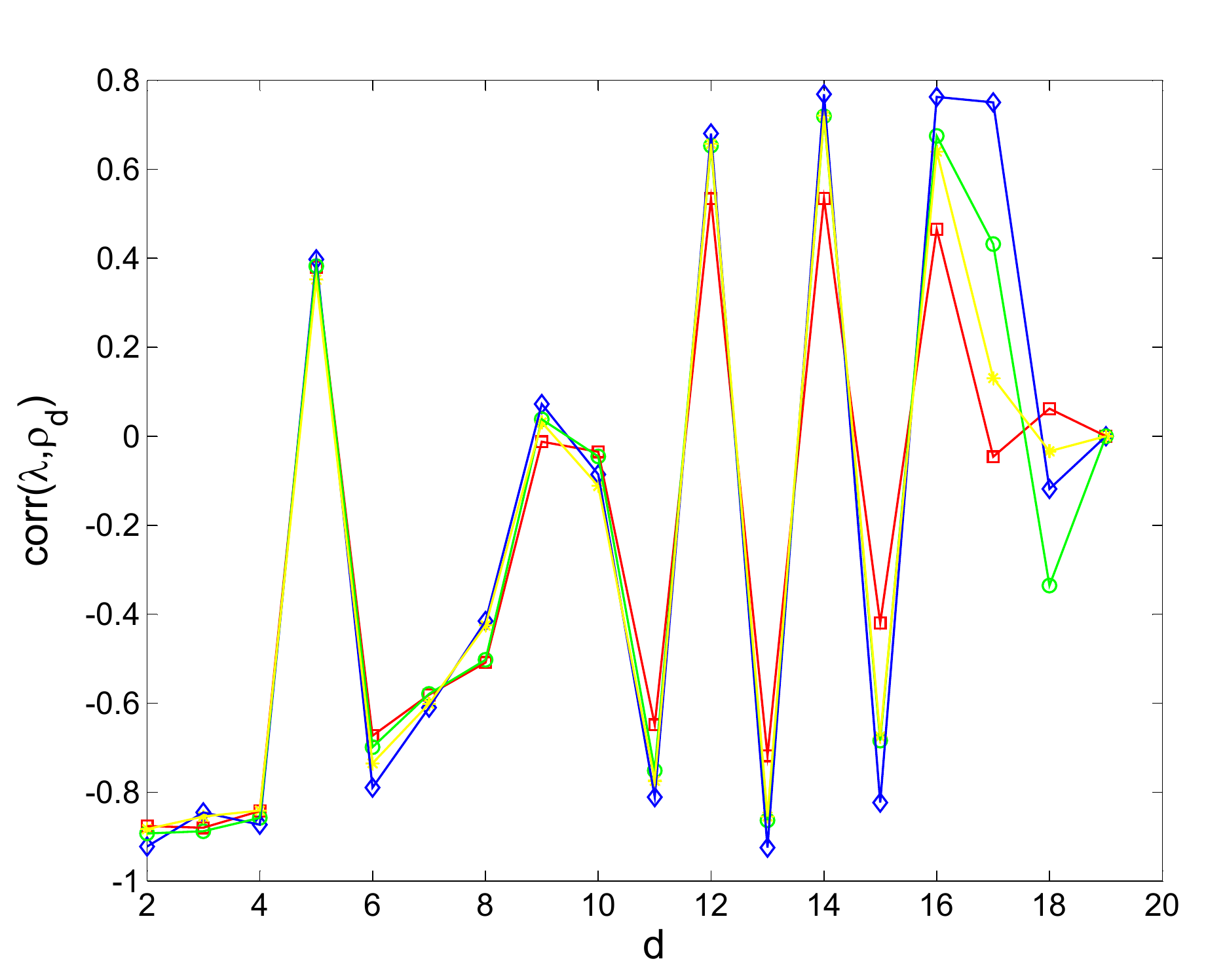} 
\includegraphics[trim = 10mm 65mm 10mm 80mm,clip, width=6cm, height=6cm]{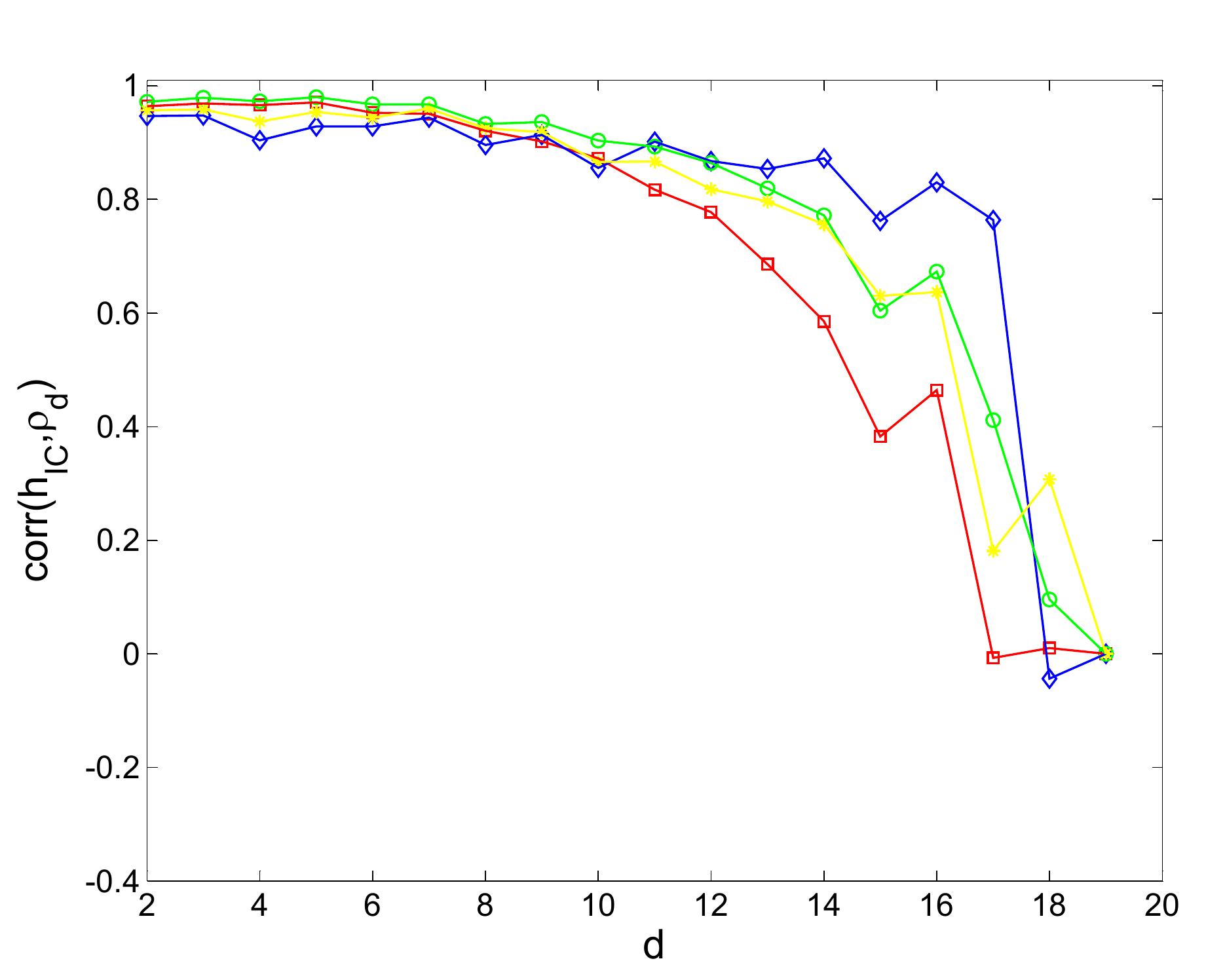} 

\hspace{0.8cm} (a)  \hspace{5.3cm} (b) 
\vspace{0.3cm}

\includegraphics[trim = 10mm 65mm 10mm 80mm,clip, width=6cm, height=6cm]{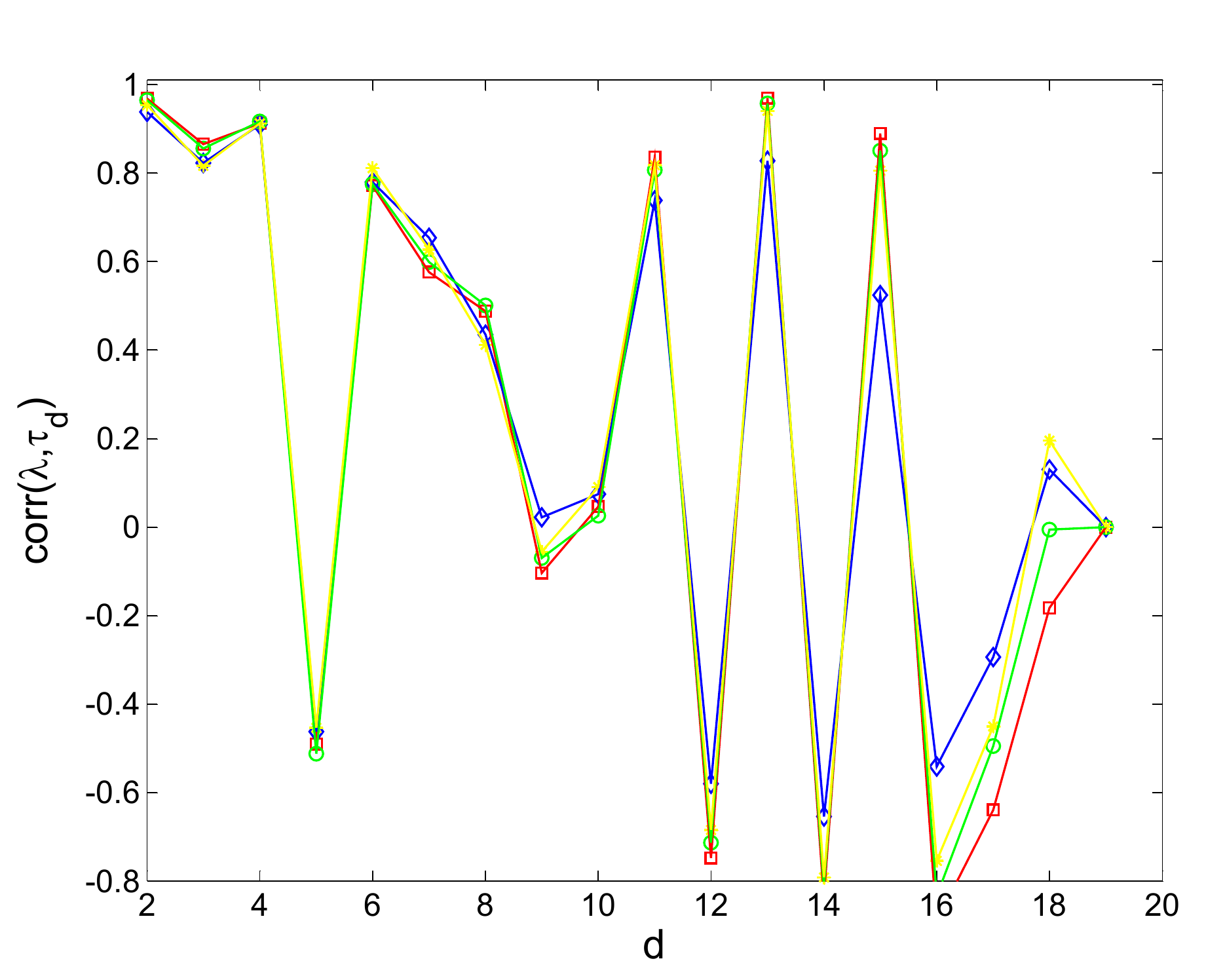} 
\includegraphics[trim = 10mm 65mm 10mm 80mm,clip, width=6cm, height=6cm]{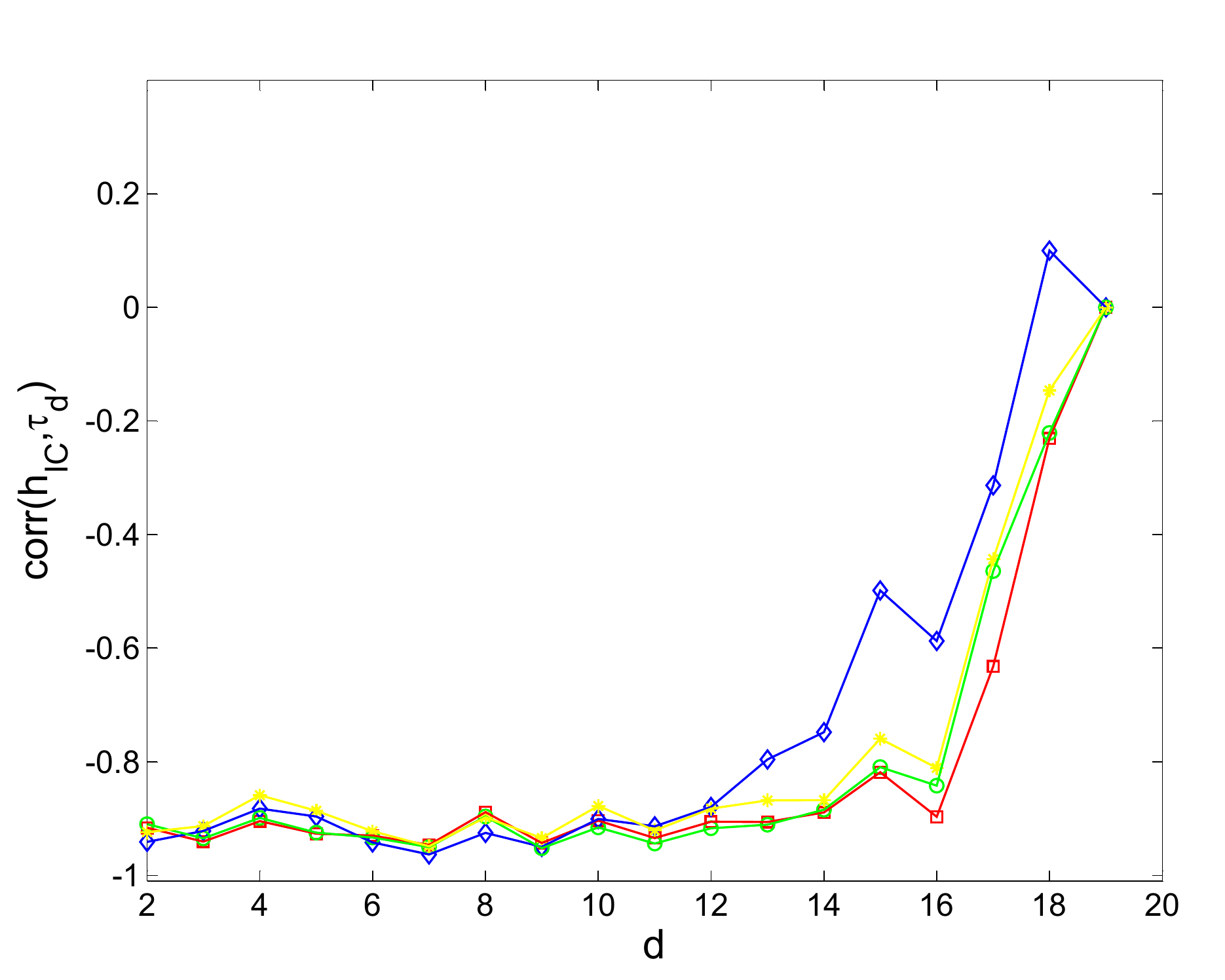} 

\hspace{0.8cm} (c)  \hspace{5.3cm} (d)

\caption{Correlation between fixation properties and landscape measures for the defective absorbing configuration over the number of coplayers $d$.  Red markers give the results for PD--BD, green for PD--DB, blue for SD--BD, yellow for SD--DB. Fixation probability $\varrho_d$:   (a)  correlation length $\lambda$,  (b) information content $h_{IC}$.  Fixation time $\tau_d$:     (c)   correlation length $\lambda$,  (d)  information content $h_{IC}$. As for the cooperative absorbing configuration, the information content correlates well to the fixation properties, while the relation to the correlation length is weak.      }
\label{fig:corr_def}

\end{figure*}

Based on the absorption structure of the game landscapes, the next set of numerical results deals with correlations between landscape measures and fixation properties. As already discussed in Sec. \ref{sec:fix}, relationships between topological structures of the game landscapes and fixation events are likely to be shaped and typified by the absorption structure. The results for the correlations  between landscape measures and fixation properties  are shown in Fig. \ref{fig:corr_coop} for the cooperative absorbing configuration and Fig. \ref{fig:corr_def} for the defective absorbing configuration. The results are for all game settings considered with red markers indicating a PD game with BD updating, green PD game and DB, blue is SD game with BD and yellow SD and DB. The lines connecting the markers are depicted to ease following trends.  
The 
Pearson product--moment correlation coefficient is calculated to aggregate over interaction networks and the number of players $N$ for each number of coplayers $d$. The advantages of such a mode of calculation are that the database for analyzing correlations becomes sufficiently large (it comprised of data for the instance of interaction matrices $A_I(\kappa)$ times the number of players $N$) and that trends over a varying number of players are captured. However, there should be an awareness that the calculation implicitly assumes that fixation properties and landscape measures scale on $N$ in ways compatible with the dependence on $A_I(\kappa)$.

Comparing the results in Fig. \ref{fig:corr_coop} tells us that the correlations between $\lambda$ and either $\varrho_c$ or $\tau_c$ are volatile and hardly evaluable, while for $h_{IC}$ there are clear trends. The same can be said about the defective fixation, see Fig. \ref{fig:corr_def} for  $\varrho_d$ and $\tau_d$.   Further analysis (not shown in a figure) confirms this to remain if the calculation is done for each $N$ and $d$.   Hence, it appears to be justified to conclude that the information content $h_{IC}$ correlates more clearly to fixation properties than the correlation length $\lambda$. Another results (also not shown in a figure for brevity reasons)  shows that the number of local maxima $\#_{LM}$ also correlates poorly to fixation properties.  Apart from the fact that there is no correlation for the PD game, BD updating and the fixation of cooperation as $\varrho_c=0$, there are further results to note.  For $d=19$, the correlations are always zero. This is why: with the experimental setup employed (see Sec. \ref{sec:setup})  only for $N=20$, there can be $d=19$,  with the additionally meaning that such a game is well--mixed with just one instance of the interaction matrix $A_I$. As correlation cannot be based on a single data pair, the correlation must be zero. Furthermore, it can be seen that there is a negative correlation between $h_{IC}$ and $\varrho_c$ (and $\varrho_d$), while the correlation between $h_{IC}$ and $\tau_c$ (and $\tau_d$) is positive. This appears reasonable as the fixation probabilities fall with increasing number of players $N$, yet the fixation times grow. Also, it can be observed that generally the correlations are strongest for small number of players and weaken before they reach zero for $d=19$.

Regarding the shaping and typifying effect of  absorption, the following can be observed comparing the local absorption structure in Tab. \ref{tab:absorp} with the correlations in Figs.   \ref{fig:corr_coop} and \ref{fig:corr_def}. From a topological point of view the correlation should be particularly strong if the absorbing configuration is a maximum and the initial configurations are all minima.  For absorbing and initial configurations being minima and maxima, the opposite should apply. Only for one example the final absorbing configuration is a maximum and the initial configurations are minima for all $N$, $d$ and $A_I(\kappa)$: the defective absorption of the SD game with BD.    For this case, the correlation between $h_{IC}$ and $\varrho_d$ is indeed slightly stronger than for the other cases. However, for the correlation between $h_{IC}$ and $\tau_d$ the opposite is true. In general, it can be noted that the correlations for all settings (except PD--BD for the cooperative absorption) are clearly visible and rather similar. It can be concluded that while there are some hints in the local absorption structure to explain the correlations between landscape measures and fixation properties, the explanatory framework should not be overstretched. Analogous to a landscape analysis that only focuses on selected points in the landscape (for instance the maxima/minima), the local absorption structure only captures a subset of the topological structures that shape coevolutionary game dynamics.  Therefore, extensions toward a global absorption structure seem desirable.

\begin{figure*}[t]

\includegraphics[trim = 10mm 65mm 10mm 80mm,clip, width=6cm, height=6cm]{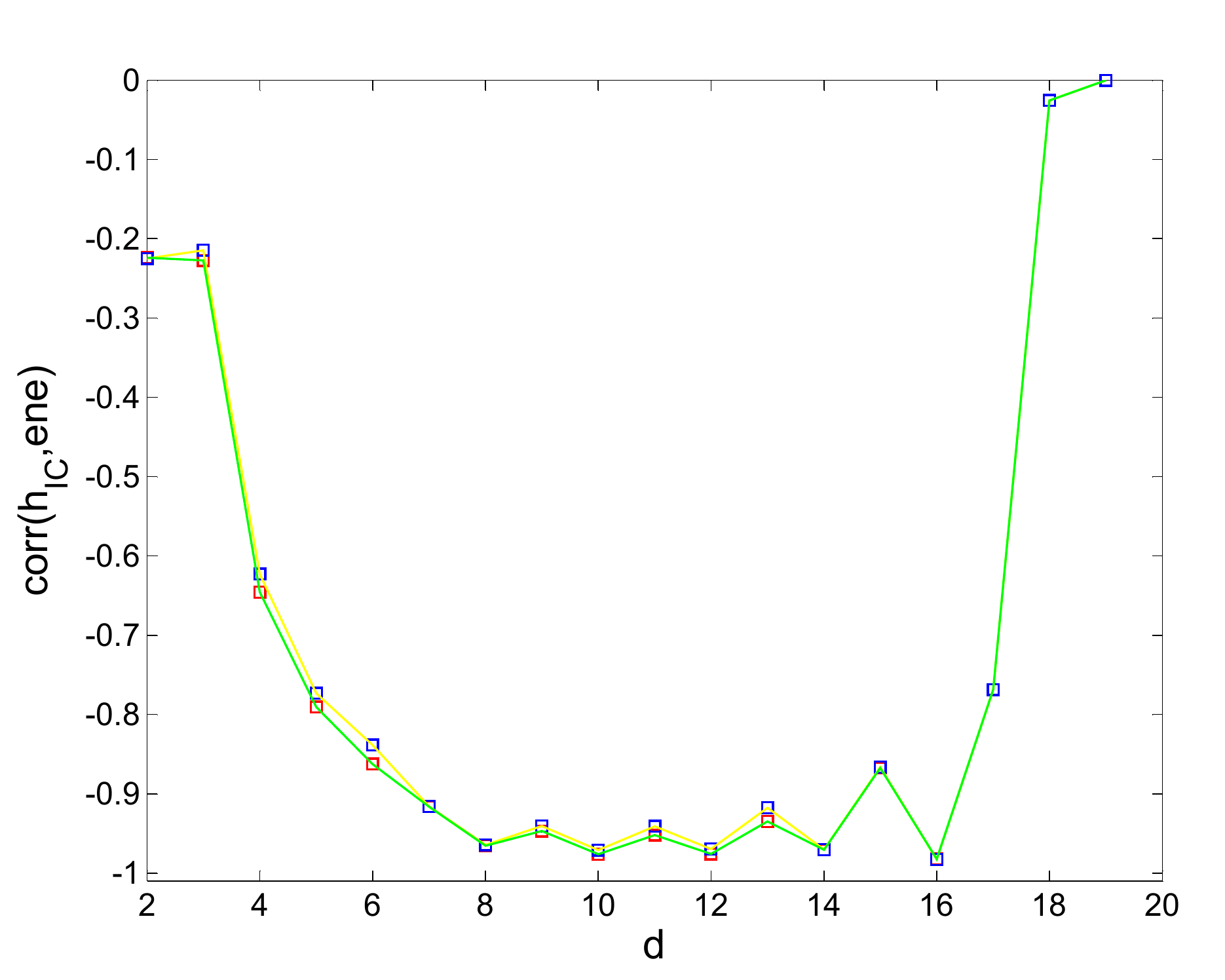} 
\includegraphics[trim = 10mm 65mm 10mm 80mm,clip, width=6cm, height=6cm]{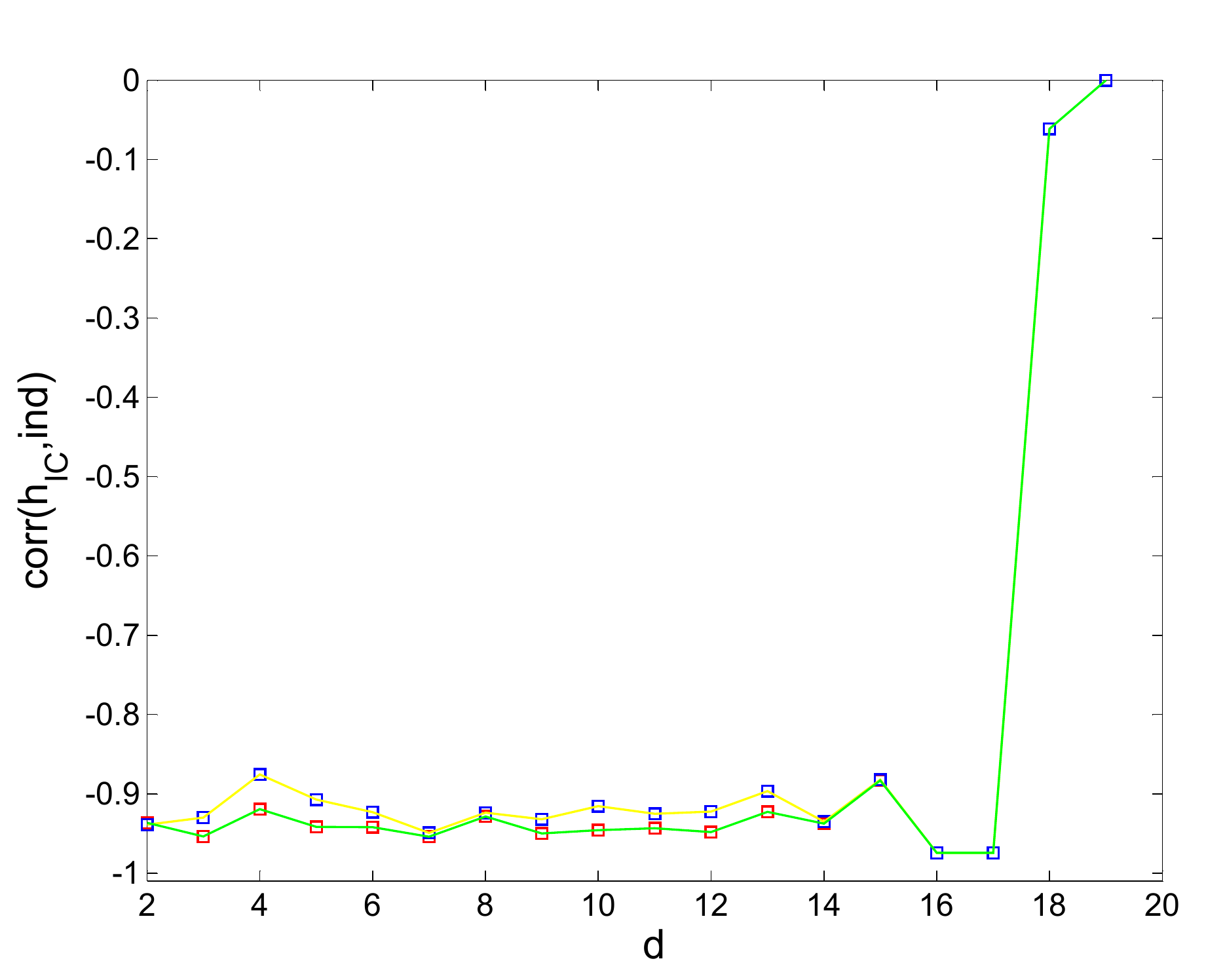} 

\hspace{0.8cm} (a)  \hspace{5.3cm} (b)

\includegraphics[trim = 10mm 65mm 10mm 80mm,clip, width=6cm, height=6cm]{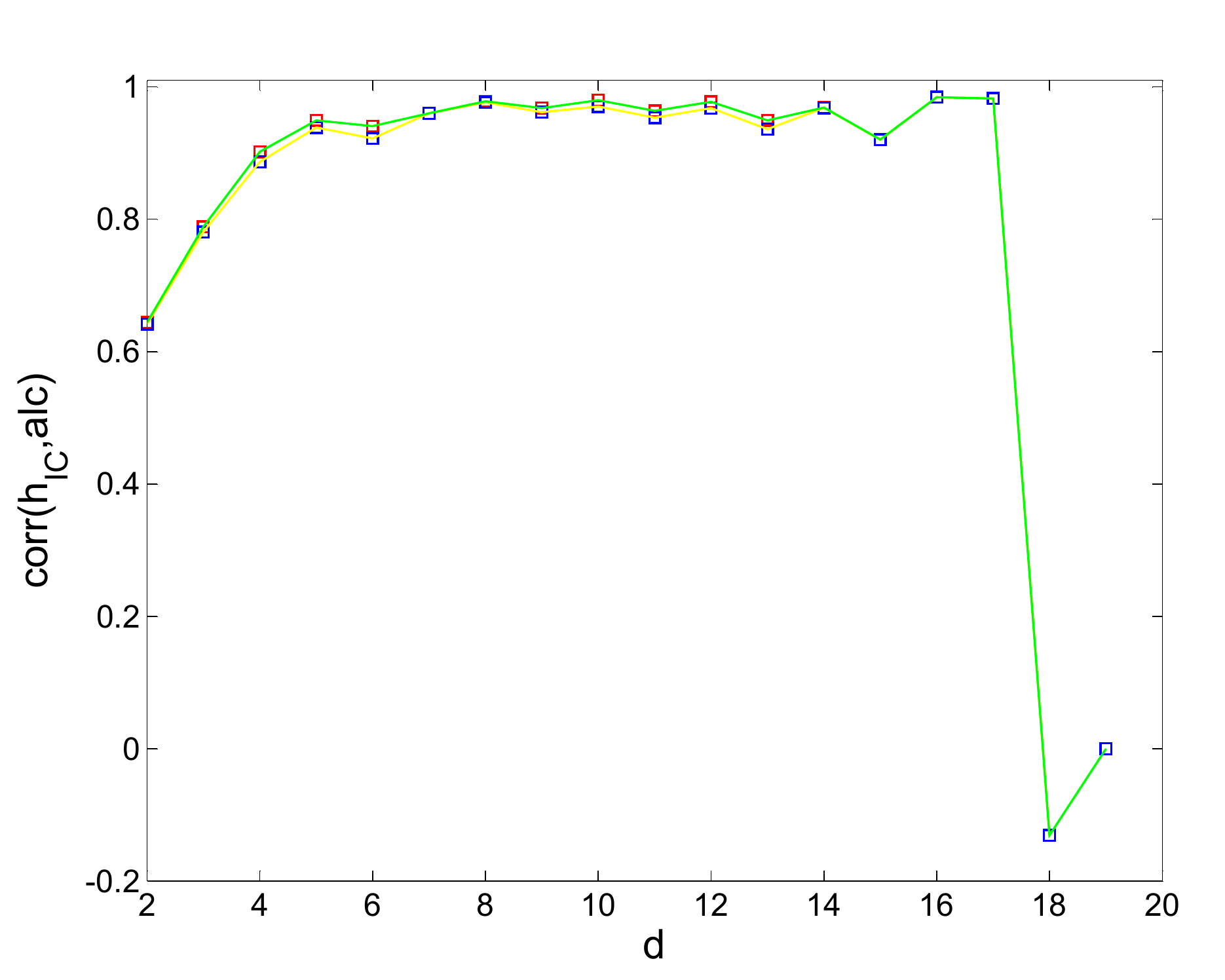} 
\includegraphics[trim = 10mm 65mm 10mm 80mm,clip, width=6cm, height=6cm]{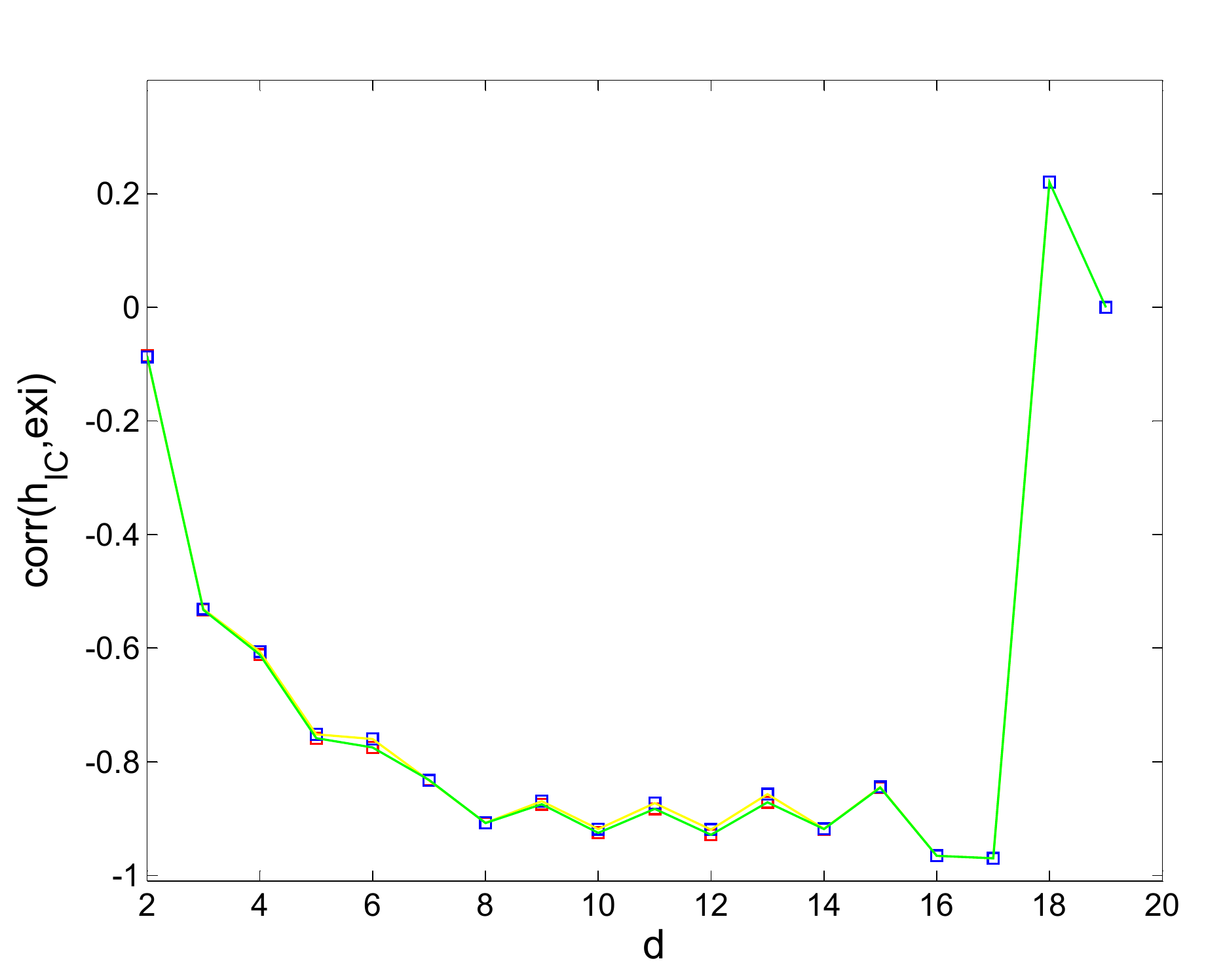} 

\hspace{0.8cm} (c)  \hspace{5.3cm} (d)

\caption{Correlation between matrix measures and information content $h_{IC}$  over the number of coplayers $d$ while no replacement restrictions are imposed.  Red markers give the results for PD--BD, green for PD--DB, blue for SD--BD, yellow for SD--DB.   (a) graph energy $\text{ene}$,  Eq. (\ref{eq:ener}); (b)  independence number $\text{ind}$, Eq. (\ref{eq:ind}); (c) algebraic connectivity $\text{alc}$, Eq. (\ref{eq:conn}); and (d)   expander index $\text{exi}$, Eq. (\ref{eq:expan}). All matrix measures correlate well to the information content and the results are indistinguishable for BD and DB landscapes, reflecting the symmetry properties of these landscapes. }
\label{fig:corr_matrix1}

\end{figure*}

\begin{figure*}

\includegraphics[trim = 10mm 65mm 10mm 80mm,clip, width=6cm, height=6cm]{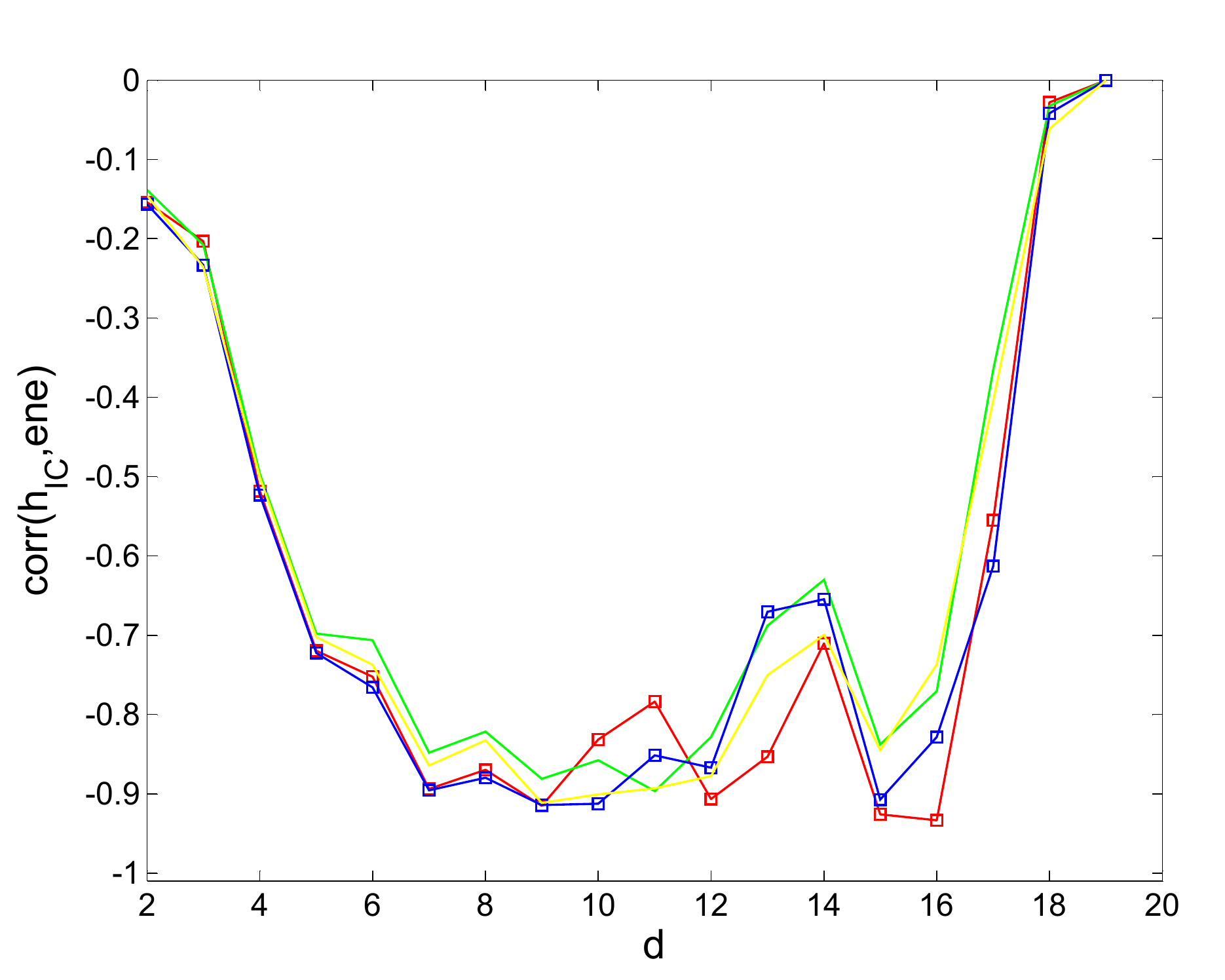} 
\includegraphics[trim = 10mm 65mm 10mm 80mm,clip, width=6cm, height=6cm]{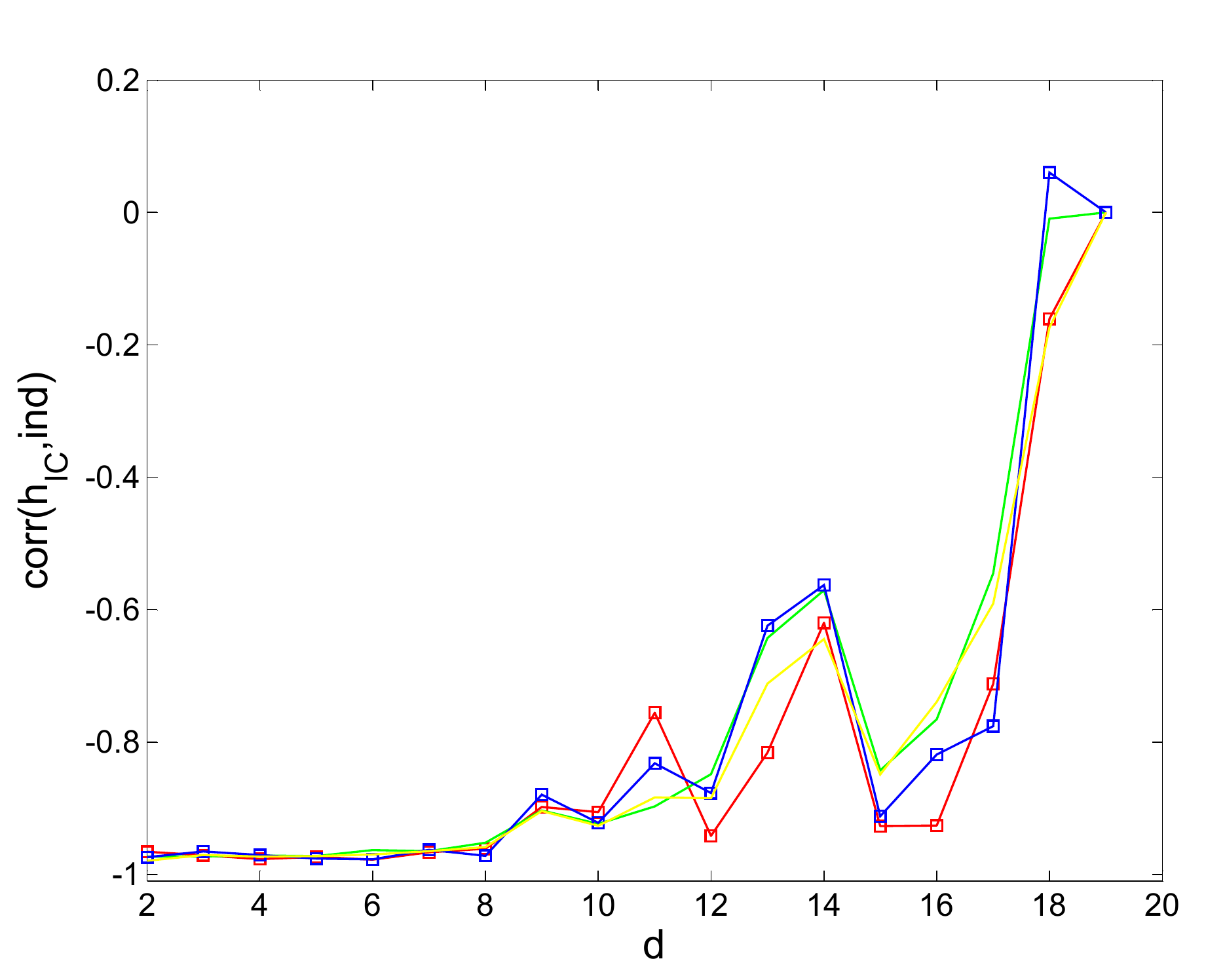} 

\hspace{0.8cm} (a)  \hspace{5.3cm} (b)

\includegraphics[trim = 10mm 65mm 10mm 80mm,clip, width=6cm, height=6cm]{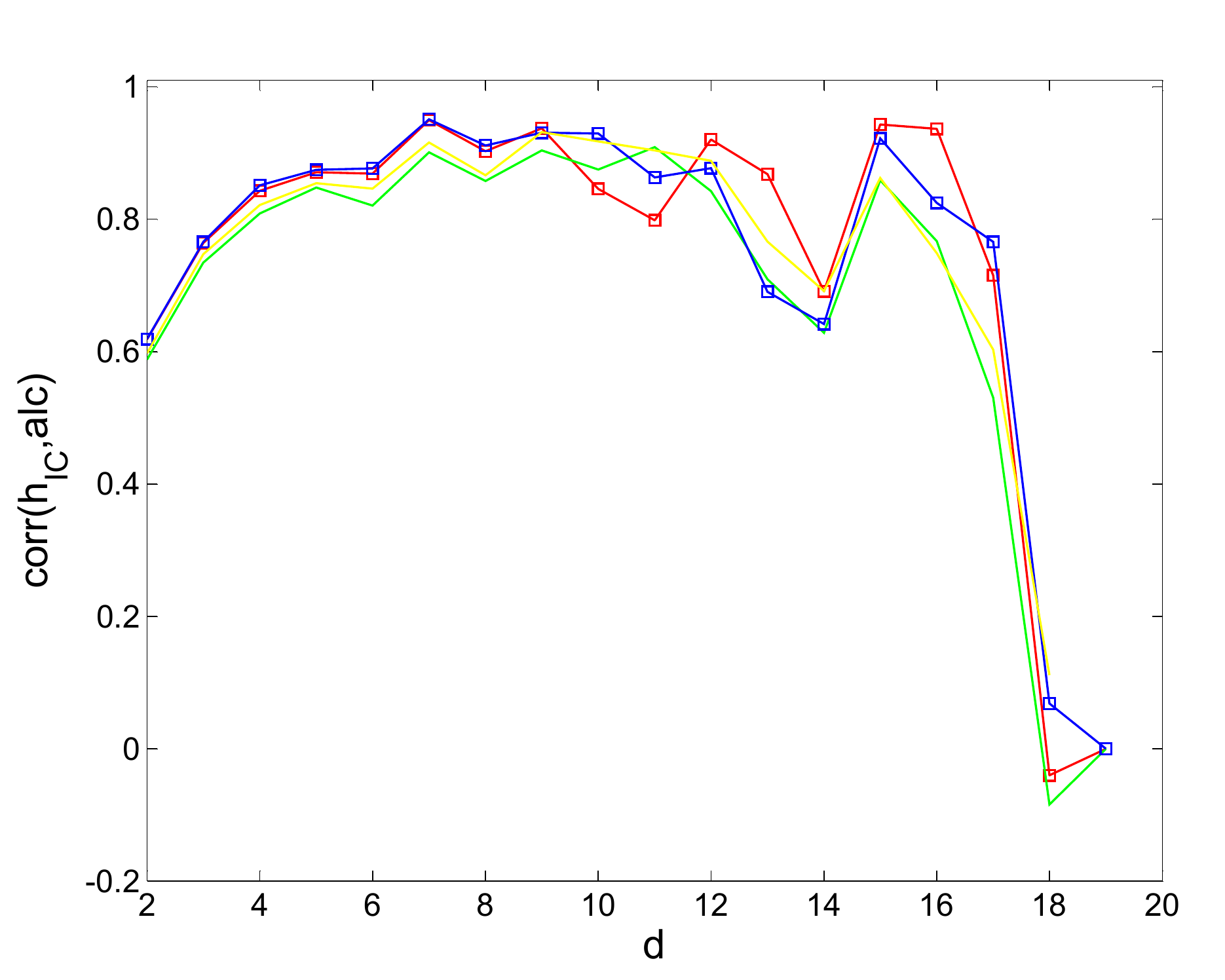} 
\includegraphics[trim = 10mm 65mm 10mm 80mm,clip, width=6cm, height=6cm]{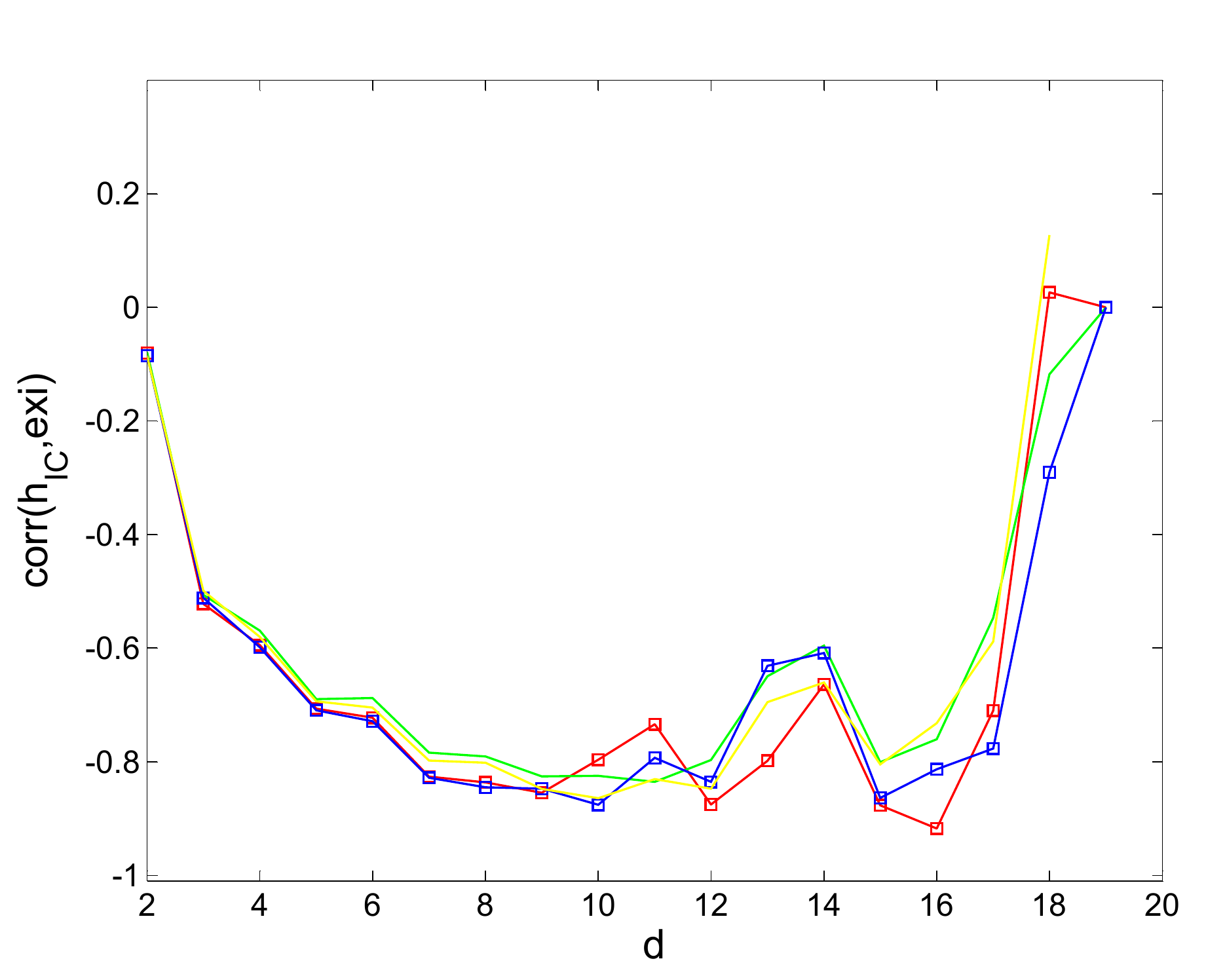} 

\hspace{0.8cm} (c)  \hspace{5.3cm} (d)

\caption{Correlation between matrix measures and information content $h_{IC}$ for replacement restrictions specified by the replacement matrix $W_R$ over the number of coplayers  $d$.  Red markers give the results for PD--BD, green for PD--DB, blue for SD--BD, yellow for SD--DB.   (a) graph energy $\text{ene}$,  Eq. (\ref{eq:ener}); (b)  independence number $\text{ind}$, Eq. (\ref{eq:ind}); (c) algebraic connectivity $\text{alc}$, Eq. (\ref{eq:conn}); and (d)   expander index $\text{exi}$, Eq. (\ref{eq:expan}). Replacement restrictions break the symmetry of BD and DB landscapes and weaken the correlations between matrix measures and the information content. }
\label{fig:corr_matrix3}

\end{figure*}

A last set of experiments reports correlations between landscape measures and network measures of the interaction matrices $A_I(\kappa)$.  Figs. \ref{fig:corr_matrix1} and \ref{fig:corr_matrix3} give the correlations between  information content $h_{IC}$ and either  graph energy,  Eq. (\ref{eq:ener}),  independence number, Eq. (\ref{eq:ind}), algebraic connectivity, Eq. (\ref{eq:conn}), or expander index, Eq. (\ref{eq:expan}). Again, the Pearson coefficient is used  aggregating over interaction networks and the number of players $N$ for each number of coplayers $d$. The same color code for the game settings as for the correlations with fixation properties is used. It can be seen that the information content correlates well to all networks measures, see Fig. \ref{fig:corr_matrix1}.  It is conspicuous that the results are indistinguishable for BD and DB landscapes, which fully reflects the symmetry properties of these landscapes. The symmetry is broken by replacements restrictions. The correlations between the network measures and  $h_{IC}$  reported in Fig. \ref{fig:corr_matrix3} confirm this  as there are differences between BD and DB for replacement restrictions. The correlations, however, are less smooth over $d$ as compared to the results in Fig. \ref{fig:corr_matrix1},  which is most likely due to the additional stochastic nature of replacement restrictions.   Lastly, two more observations can be noted. A first is that the correlations between $h_{IC}$ and the networks measures are negative for graphs energy, independence number and expander index, while they are positive for algebraic connectivity. The main reason is that amongst the network measures studied only algebraic connectivity increases continuously for $d$ getting larger in both mean and variance over instances of the interaction matrices $A_I(\kappa)$. A second is that for the graph energy and the expander index the correlations are weak for both small numbers of coplayers ($d<6$) before they get stronger  to weaken again for larger number of players ($d>16$). For the other two network measures the weakening is only for $d>16$, which is similar to the correlations with fixation properties.

\subsection{Discussion}
The experimental results given above set out relationships between landscape measures and both fixation properties and network properties, and argue that dynamic landscape models of coevolutionary games are viable. In this section, features and implications of such a modeling approach are discussed, the experimental findings of Sec.~\ref{sec:results} are put into context,  and some concluding observations are offered. 

\begin{enumerate}

\item An essential part of the numerical experiments is the study of correlations between landscape measures and both fixation properties and quantifiers of the networks of interaction. A main results is that information content scales well to fixation properties such as fixation probability and fixation time, see Figs.  \ref{fig:corr_coop} and \ref{fig:corr_def}. Particularly ruggedness as measured by the correlation length relates less clearly and much weaker to fixation properties than information content, which is understood to account not specifically for ruggedness, but more for the interplay between smooth, rugged and flat landscape areas. Hence, a conclusion may be that also for game landscapes ruggedness alone is not a good predictor for evolutionary dynamics, as also reported for other types of fitness landscapes,~\cite{mal13}. There are additional entropic landscape measures based on the information content, for instance partial information content, information stability or density--basin information,~\cite{mun15,vassi00}. Thus, it might be interesting to study whether these measures also scale well for game landscapes and may offer further insight into game dynamics. Regarding the correlations between landscape measures and quantifiers of interaction networks, the results are more consistent, see Figs. \ref{fig:corr_matrix1} and \ref{fig:corr_matrix3}. There are clear correlations for all four of the network measures considered, with algebraic connectivity and independence number scaling slightly better than graph energy and expander index. However, it should be noted that 
the correlations established between the landscape measures and network measures are based on the Laplacian or adjacency spectra of the adjacency matrix $A_I$. As these spectra do not uniquely determine the interaction graph, there might be correlations between the graph structure of the interaction network and the game landscape that are not captured.  
The discussion might be extendable  by considering alternatives, for instance generalized graph distance measures as reported by~\cite{gu16}.

\item The experiments studying the effect of different networks of interaction with given $N$ and $d$ on fixation properties and landscape measures only report mean, maximum, and minimum values and their interdependencies. Further statistical analysis, for instance considering variances or higher--order moments, is deliberately omitted. The main argument is that we should  beware of drawing conclusions based on such a  statistical analysis  as it is not clear if it really produces generalizable results.  To illustrate the point, let us  consider $d=2$ coplayers, for which the number of  different graphs can be enumerated exactly by Eq. (\ref{eq:2regular}). The experiments presented are for up to $N=20$ players. Accordingly, at the upper limit of the experimental setup,  we find
$\mathcal{L}_2(20) \approx 1.4 \cdot 10^{17}$. There is no alternative but to conclude that any number of  numerically testable networks of interaction represents just a tiny subset of all interaction networks. At the same time it is far from being clear how well the finite number of graphs generated by the numerical procedures represents all possible different graphs for given $N$ and $d$. 
Thus, it might be possible that some trends are biased by the algorithmic process of 
numerically generating interaction networks $A_I$. Further work is needed to clarify these interdependencies.   
Such a work should go along with categorizing interaction matrices into clusters. These clusters could quantify, on the one hand, similar fixation properties of the coevolutionary games, and on the other, similar spectral properties or similar  generalized graph distance measures of the interaction matrices as mentioned above. Hence, it might be possible to study the effect of network properties of the interaction matrix on coevolutionary game dynamics for a larger number of networks.

\item The experimental results showing landscape measures and their correlations to fixation properties as well as to graph--theoretical quantities of the interaction networks are for the specific algebraic description of the game landscapes  (\ref{eq:bd}) and (\ref{eq:db}). The algebraic form of how to cumulate death and birth probabilities from the player--wise strategy landscapes is definatory and was employed as it fitted well to fixation properties and previous results known about social dilemma game dynamics. It is an open question whether an alternative algebraic form of   (\ref{eq:bd}) and (\ref{eq:db}) can achieve similar or even better results.
Similarly, the results obtained here are specific for the linear relation $f=1+\delta  p$. Hence, it might be interesting to analyze how different payoff--to--fitness relations modify the results, for instance the exponential relations $f=1+\exp{(p)}$ or $f=\exp{(p)}$, as suggested by~\cite{allen14,sha12}.  This may go along with experimental studies of different levels of the intensity of selection $\delta$, which is also opened up by the game landscape approach proposed in this paper. In the case of weak selection, that is for $\delta \rightarrow 0$, the player--wise strategy landscapes lose their distinct topological features, which yields (in the absence of replacement restrictions) a game landscape that is neutral. Consequently, the game dynamics on this landscape would be random drift. For larger or large values of $\delta$ the topological features of the player--wise strategy landscapes become more prominent, and the game landscape is more rugged. From this line of argument it can be conjectured that there is a direct relation between the intensity of selection and the ruggedness of the game landscape, which might be verifiable by future studies.

\item The results of dynamic game landscapes presented are for up to $N=20$ players. Naturally, it would be desirable to extend the experimental setup to a larger population size.  However, if the trends identified in these results remain valid for a larger number of players needs to be studied in further work, as  the maximal number of players used was the upper limit that could be realized within a reasonable time--frame and the computational resources available in this study.  A general feature  surely is that the number of configurations to be analyzed in the landscapes increases exponentially with the number of players, hence setting bounds as to how far such experiments might be extendable. Therefore,  with the computational resources currently available the modeling framework is likely to be confined to a moderate  number of players. However, for an increased number of players there is the framework of replicator dynamics which sufficiently describes  game dynamics for populations becoming large,~\cite{traul05}.

\item Some primary experiments have shown that for replacement restrictions, the correlation between landscape measures on the one hand, and fixations and network properties on the other, cease. Apparently,  the replacement restrictions seriously modify the structure of the game landscapes.
It is another open question if these relationships can be reestablished by taking into account properties of the replacement process, for instance the absorption structure of the restricted network. This may be extended by foregoing the setting that the degree of the replacement matrix $W_R$ matches the degree of the adjacency matrix $A_I$.

\end{enumerate}

\section{Summary and conclusions} \label{sec:con}
Coevolutionary games cast players that update their strategies as well as their networks of interaction. In this study, a
reinterpretation of coevolutionary games as dynamic fitness landscapes is proposed. The dynamic landscapes are based on three major components: (i) a description of strategy updating as a Moran process with definable  probabilities of strategy transitions, (ii) a formulation of updating the interaction network as instances of random regular graphs, and (iii) a linear relation between payoff and fitness.   Using these components, payoff--related fitness landscapes can be defined for each player.  It is further shown that coevolutionary  game dynamics can be expressed by a game landscape derived from these player--wise landscapes by including the strategy updating process. Moreover, different strategy updating processes, such as  death--birth (DB) or birth--death (BD) produce different game landscapes, which can be seen as strategy updating breaking the symmetry of the play--wise landscapes. In numerical experiments it has been demonstrated that landscape measures such as modality, ruggedness and information content allow to differentiate between different game landscapes. 
Fixation probabilities and fixation times have been calculated as well as network measures characterizing the networks of interaction of the coevolutionary games. By correlation analysis it has been shown how the landscape measures relate to both  fixation properties and network measures.

The approach presented is a
technique for analyzing coevolutionary games by landscapes. 
Moreover, the approach is
not restricted to Moran processes as long as strategy transition probabilities can be derived, at least approximately.
Finally, network updating is  currently modeled as a given sequence of random regular graphs, but should be understood as a transition process,
for instance by using
reproducing graphs,~\cite{south10} as a tool to refine the description of transitions between adjacency matrices.

Different settings of the game represented by the numeric values of the payoff matrix and different rules of the strategy updating result into a large variety of coevolutionary game dynamics. A considerable number of works have analyzed and discussed this game dynamics with respect to fixation properties such as fixation probability and fixation time from both a theoretical as well as an experimental point of view,~\cite{allen14,lieb05,nowak06,patt15,sha12}. The results reported here contribute to this discussion by offering a fitness landscape view as an alternative explanatory framework. In other words, by the approach presented coevolutionary games may become amenable to be analyzed by dynamic landscapes.

\newpage

\newsavebox{\smlmat}% Box to store smallmatrix content
\savebox{\smlmat}{$A_I=\left(\begin{smallmatrix}0 &1 & 1& 1\\1 & 0 & 1 & 1 \\1 & 1& 0& 1\\ 1& 1&1&0\end{smallmatrix}\right)$}

\begin{figure*}[t]

\includegraphics[trim = 35mm 65mm 40mm 100mm,clip, width=2.97cm, height=3cm]{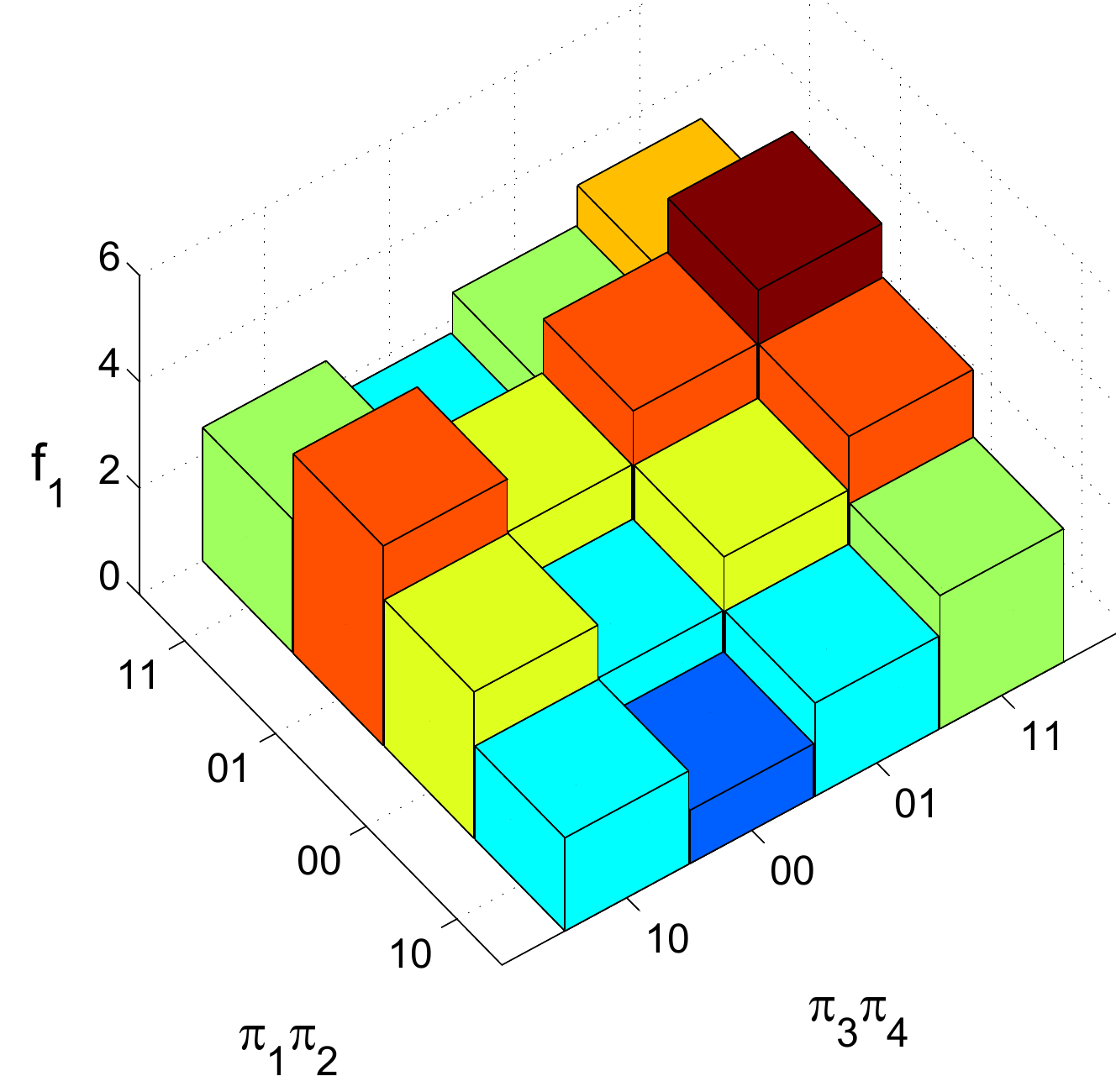} 
\includegraphics[trim = 35mm 65mm 40mm 100mm,clip, width=2.97cm, height=3cm]{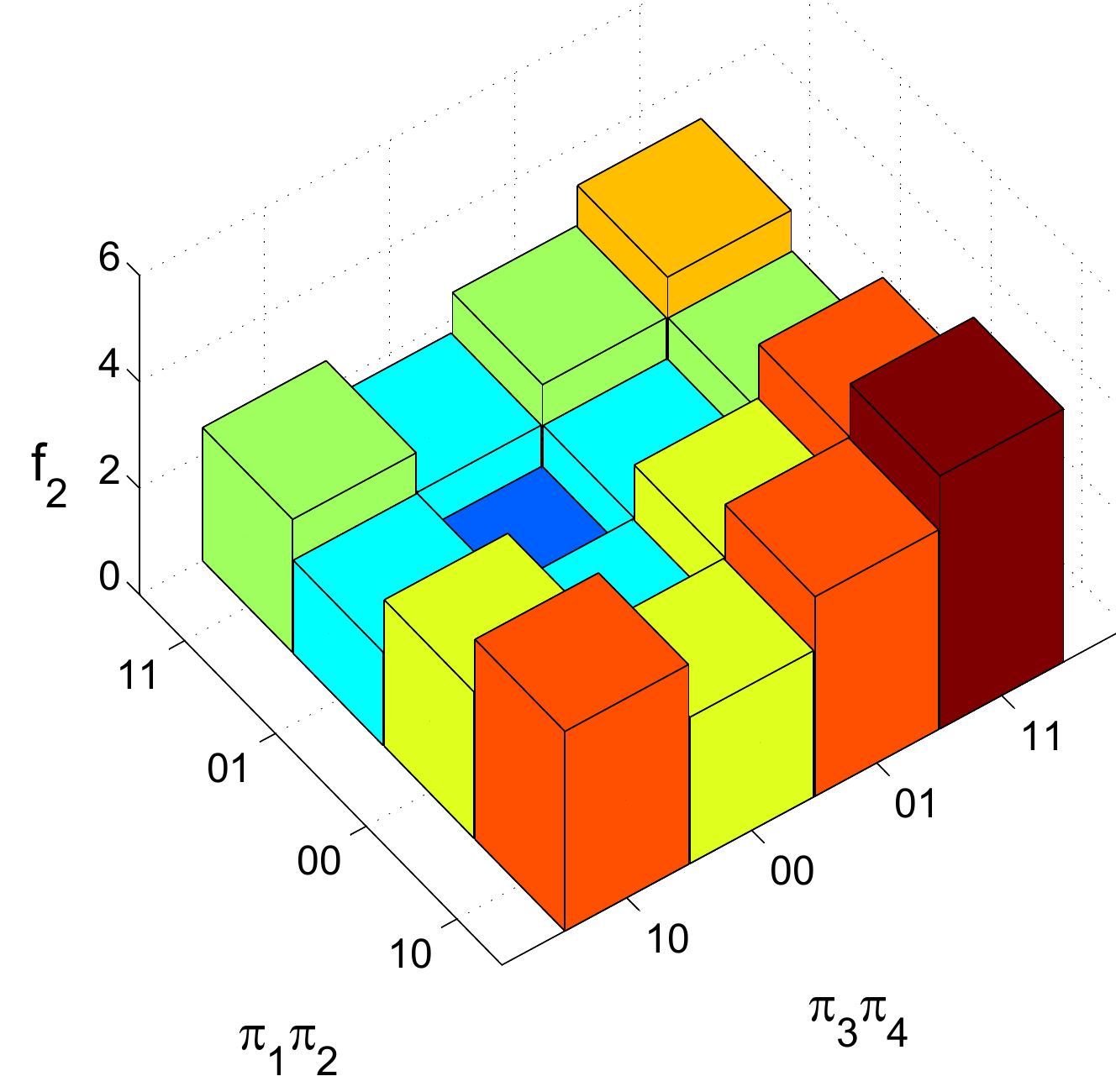} 
\includegraphics[trim = 35mm 65mm 40mm 100mm,clip, width=2.97cm, height=3cm]{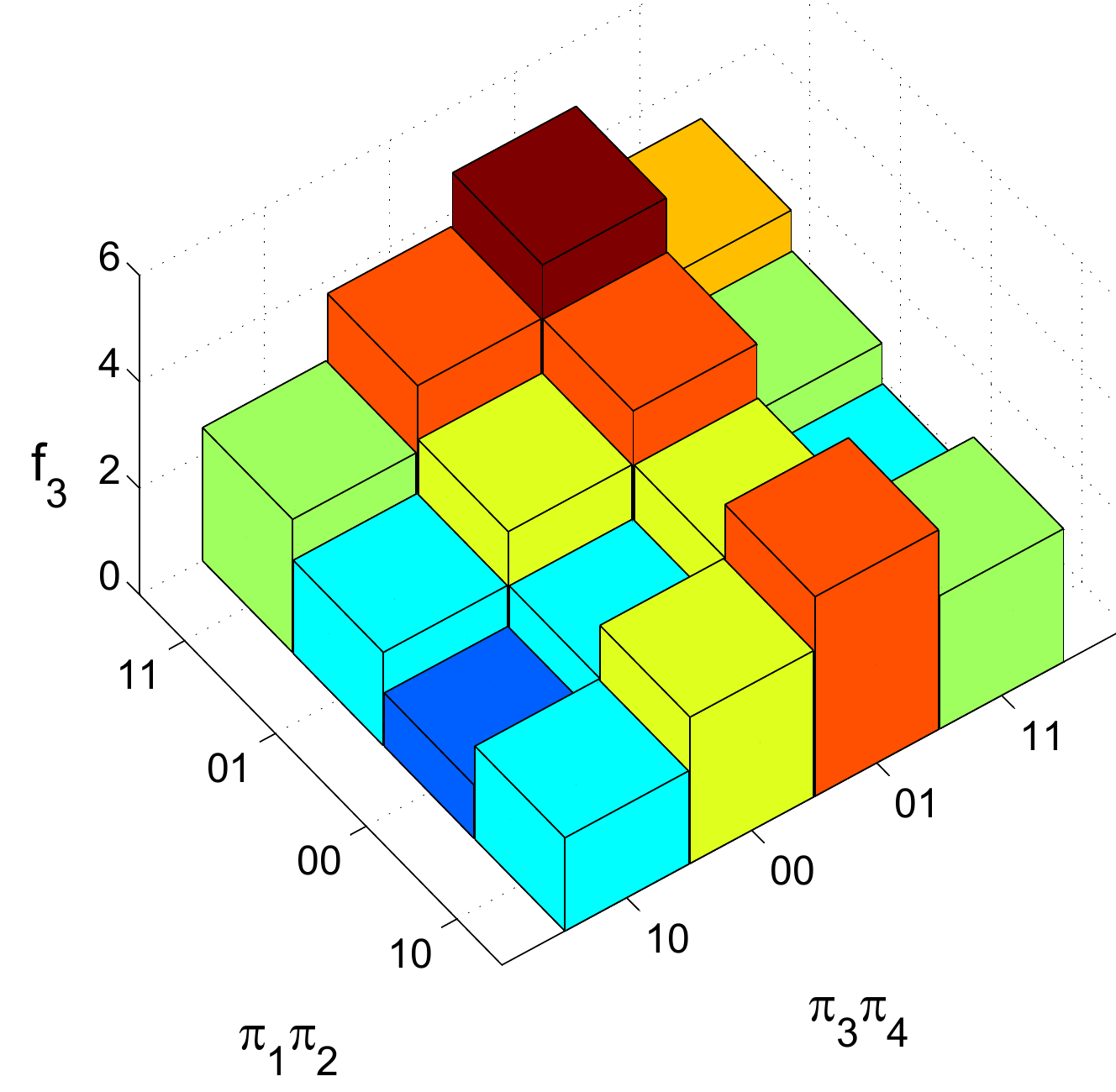} 
\includegraphics[trim = 35mm 65mm 40mm 100mm,clip, width=2.97cm, height=3cm]{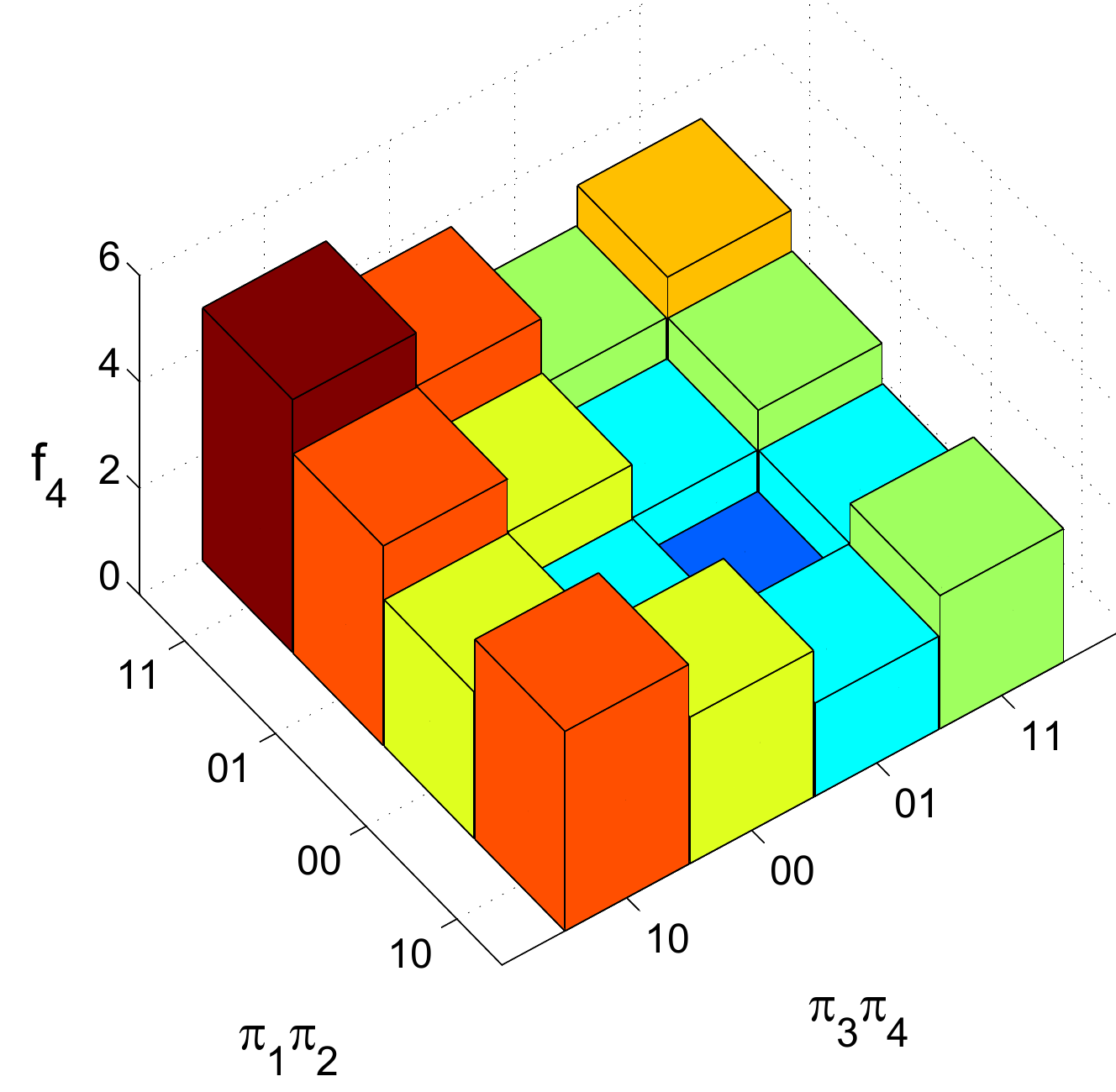}

\caption{Illustration of  two-dimensional strategy landscapes $\Lambda_{\Pi}^i$ for a  PD game with $N=4$, $d=3$ and a complete interaction network with~\usebox{\smlmat}.   Same colors give equal fitness values $f=1+\delta p$ for payoff $p$ with $\delta=0.25$. Each strategy configuration $\pi=(\pi_1 \pi_2 \pi_3 \pi_4)$ has $N=4$ neighbors distanced by Hamming distance $\mathcal{H}_d^1$, while periodic boundary conditions apply. For each player the landscape has one maximum; the player defects, while its three coplayers cooperate. }
\label{fig:n4}
\end{figure*}

\newsavebox{\smlmata}% Box to store smallmatrix content
\savebox{\smlmata}{$A_I(0)=\left(\begin{smallmatrix}0 &0 & 1& 1\\0 & 0 & 1 & 1 \\1 & 1& 0& 0\\ 1& 1&0&0\end{smallmatrix}\right)$}

\newsavebox{\smlmatb}% Box to store smallmatrix content
\savebox{\smlmatb}{$A_I(1)=\left(\begin{smallmatrix}0 &1 & 0& 1\\1 & 0 & 1 & 0 \\0 & 1& 0& 1\\ 1& 0&1&0\end{smallmatrix}\right)$}

\newsavebox{\smlmatc}% Box to store smallmatrix content
\savebox{\smlmatc}{$A_I(2)=\left(\begin{smallmatrix}0 &1 & 1& 0\\1 & 0 & 0 & 1 \\1 & 0& 0& 1\\ 0& 1&1&0\end{smallmatrix}\right)$}

\begin{figure*}[t]

\includegraphics[trim = 35mm 65mm 40mm 100mm,clip, width=2.97cm, height=3cm]{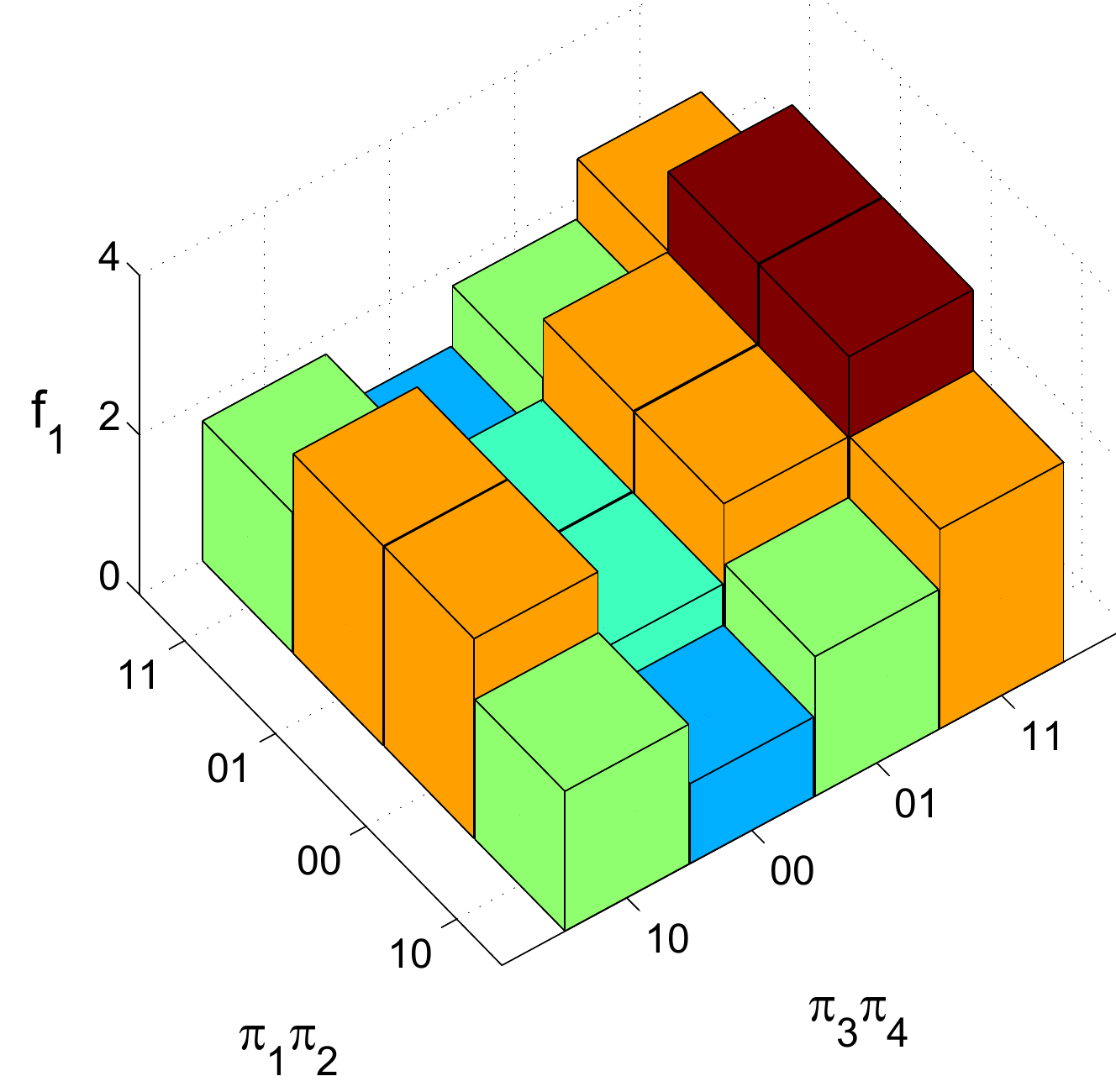} 
\includegraphics[trim = 35mm 65mm 40mm 100mm,clip, width=2.97cm, height=3cm]{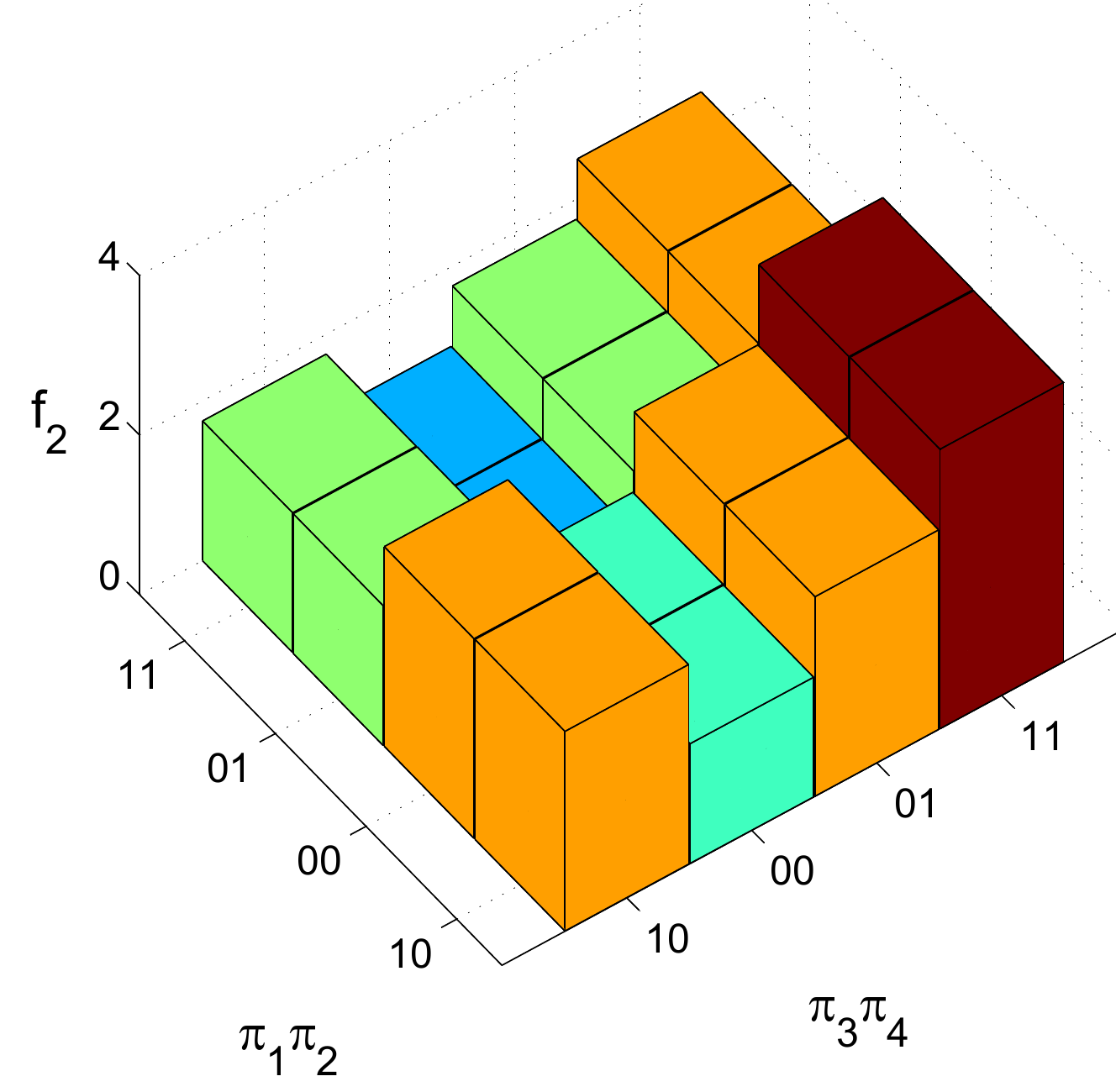} 
\includegraphics[trim = 35mm 65mm 40mm 100mm,clip, width=2.97cm, height=3cm]{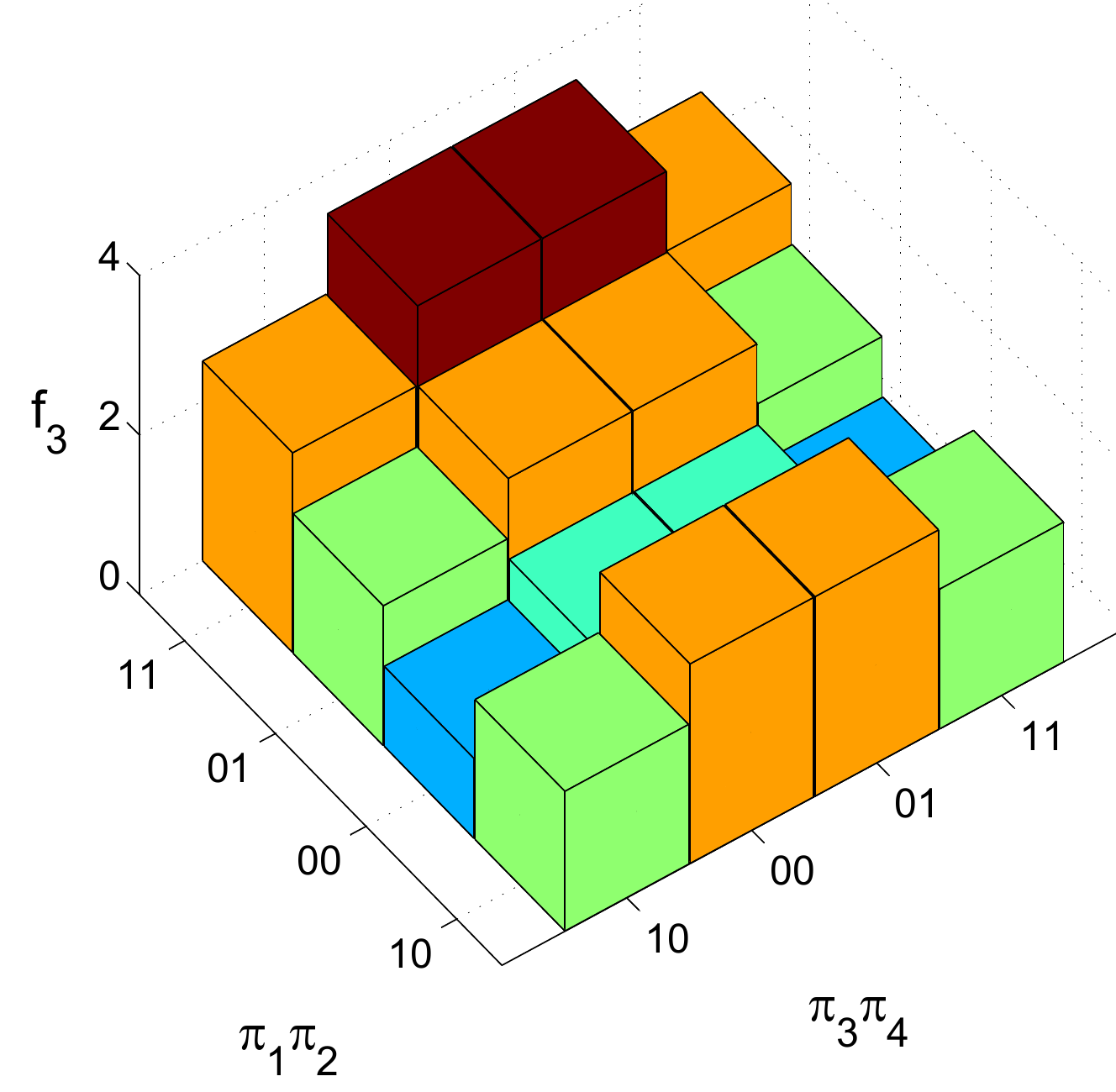} 
\includegraphics[trim = 35mm 65mm 40mm 100mm,clip, width=2.97cm, height=3cm]{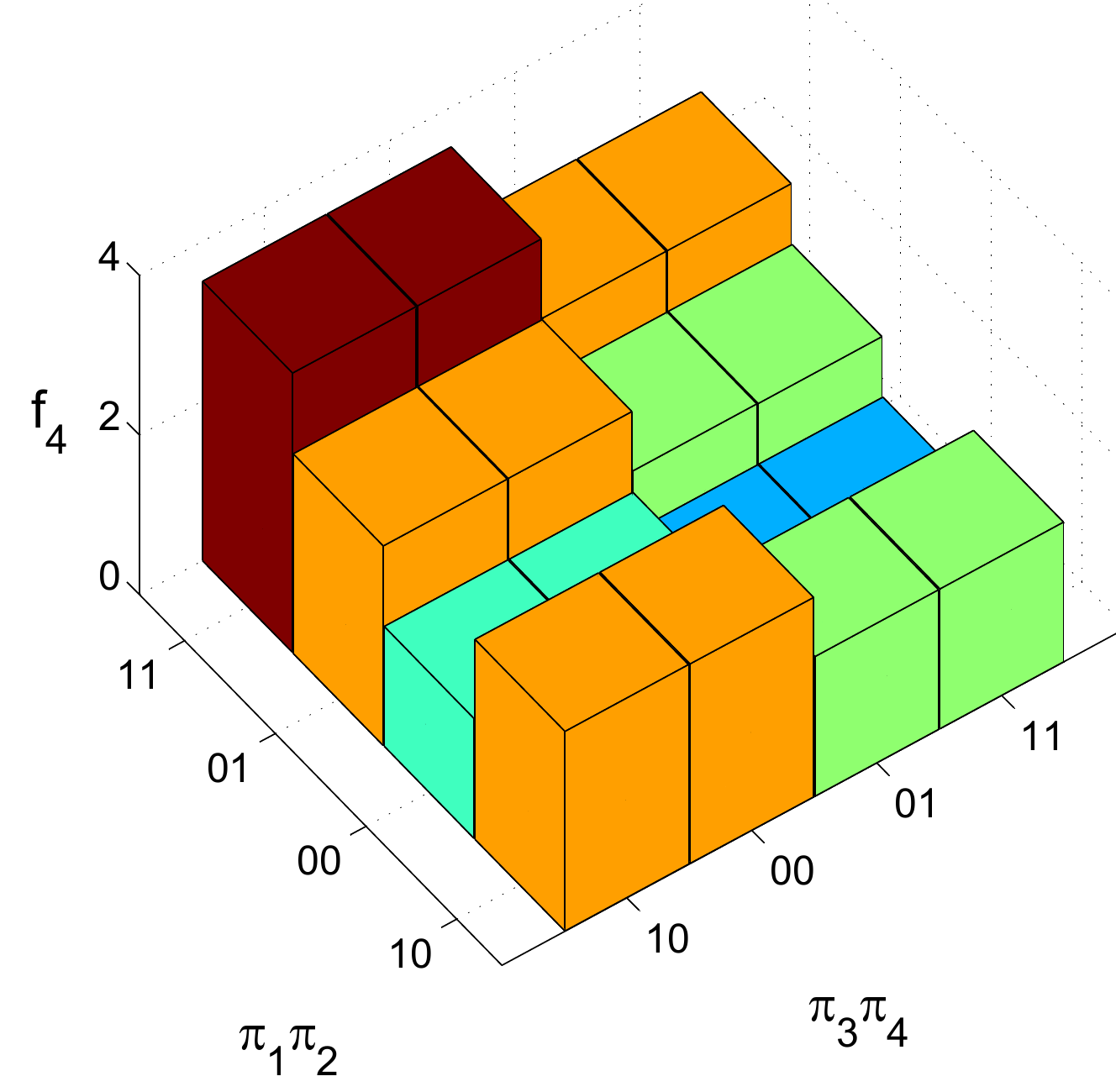} 

\hspace{5.9cm} (a) 

\includegraphics[trim = 35mm 65mm 40mm 100mm,clip, width=2.97cm, height=3cm]{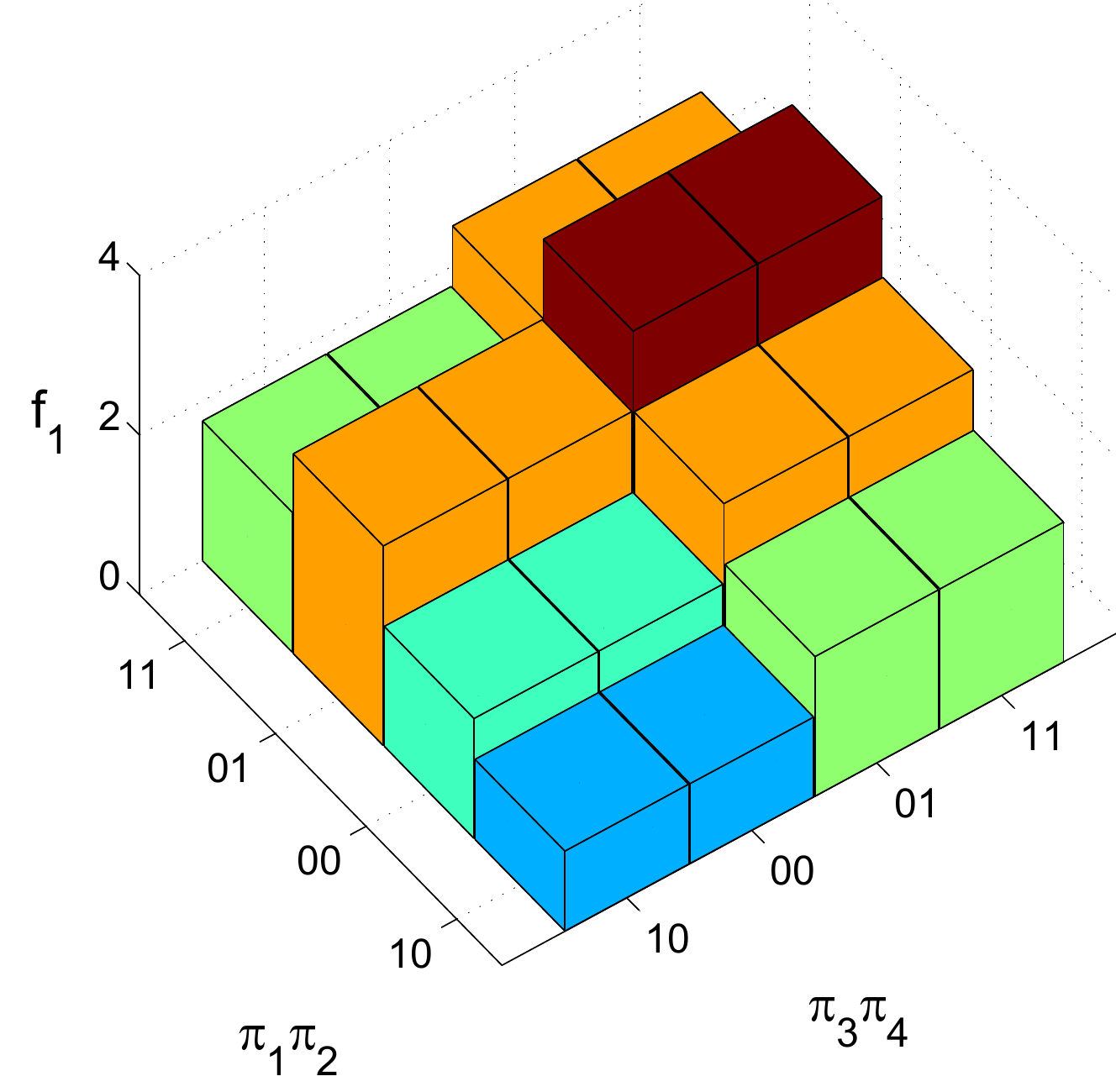} 
\includegraphics[trim = 35mm 65mm 40mm 100mm,clip, width=2.97cm, height=3cm]{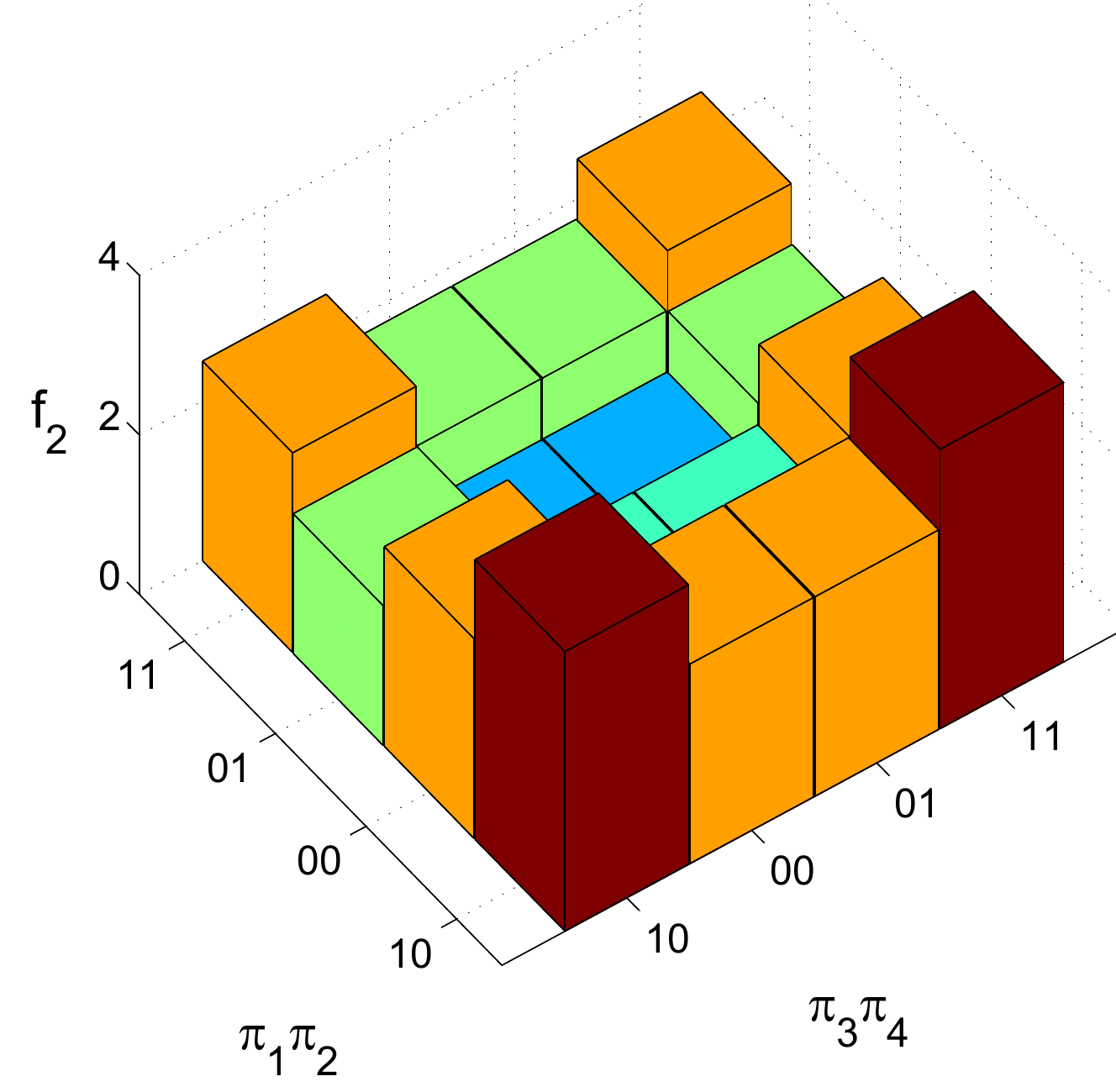} 
\includegraphics[trim = 35mm 65mm 40mm 100mm,clip, width=2.97cm, height=3cm]{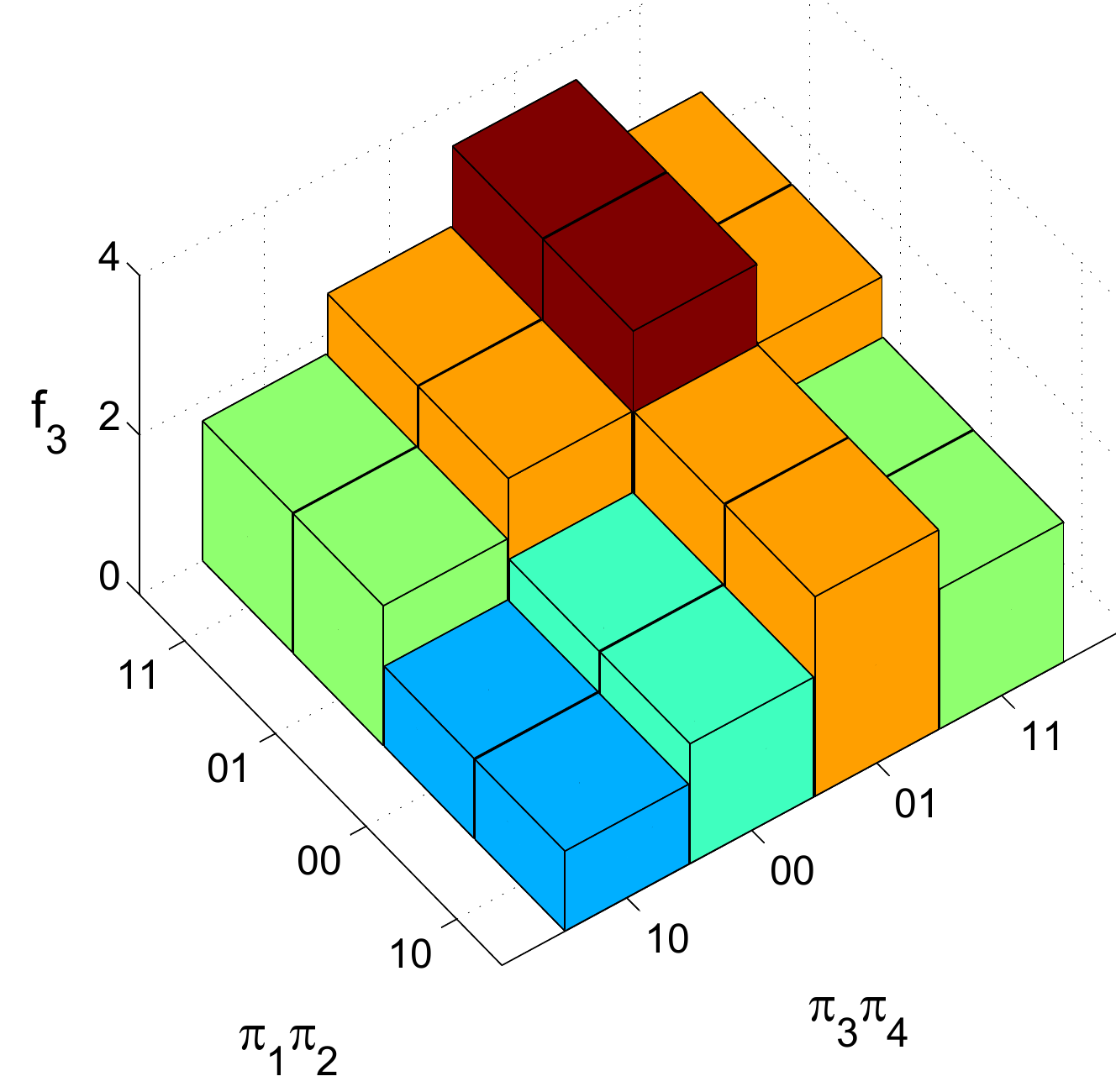} 
\includegraphics[trim = 35mm 65mm 40mm 100mm,clip, width=2.97cm, height=3cm]{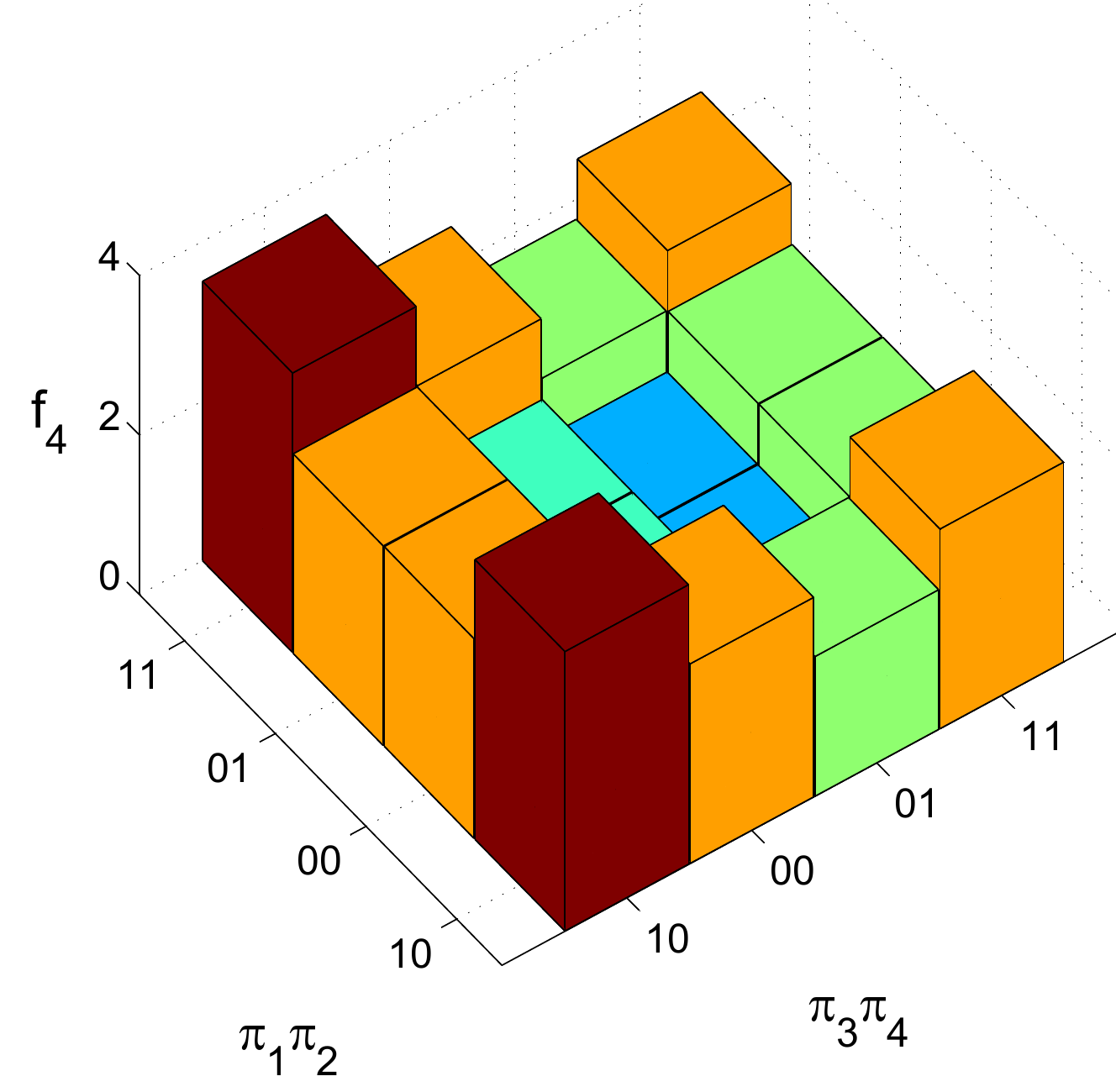} 

\hspace{5.9cm} (b) 

\includegraphics[trim = 35mm 65mm 40mm 100mm,clip, width=2.97cm, height=3cm]{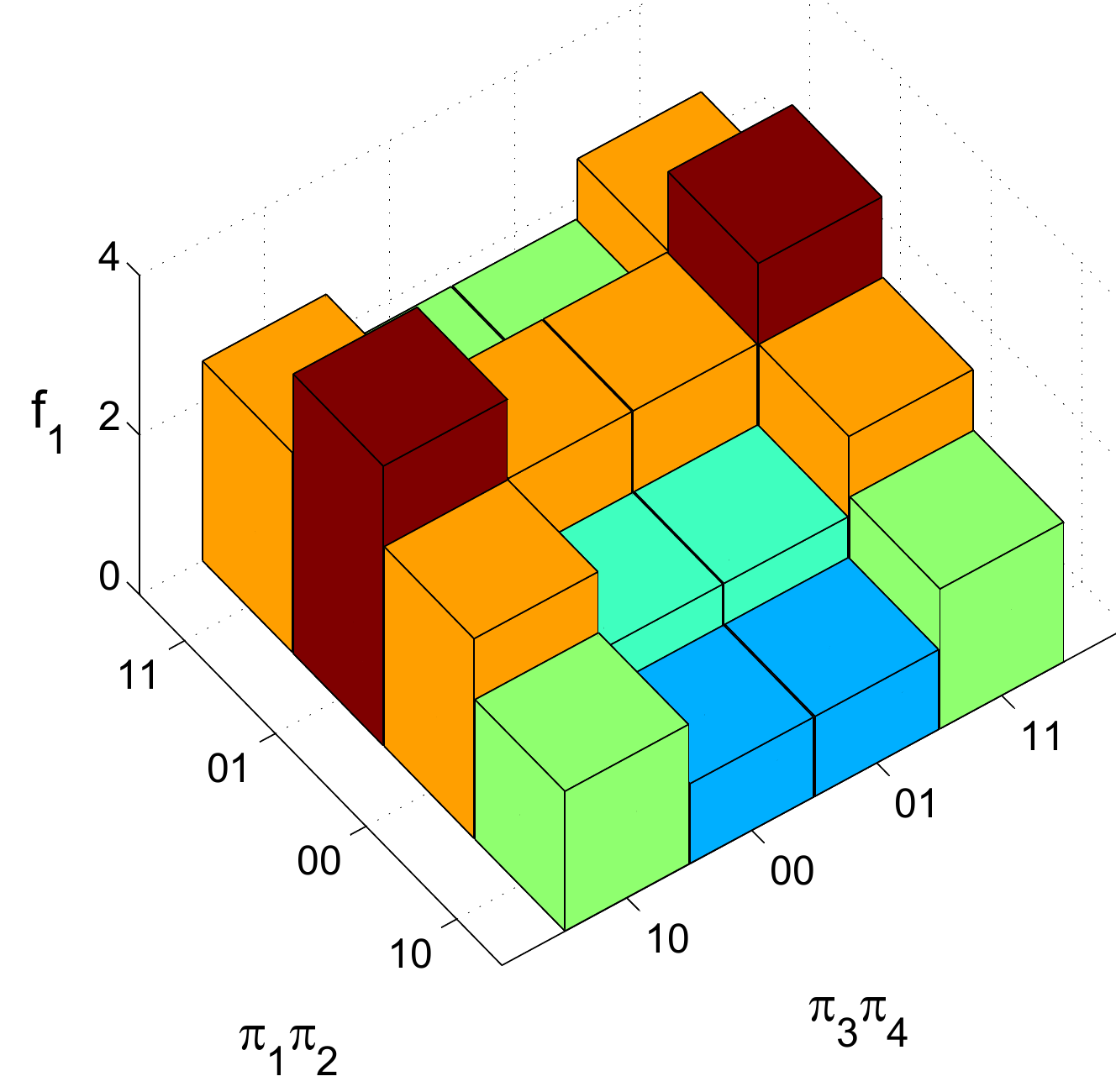} 
\includegraphics[trim = 35mm 65mm 40mm 100mm,clip, width=2.97cm, height=3cm]{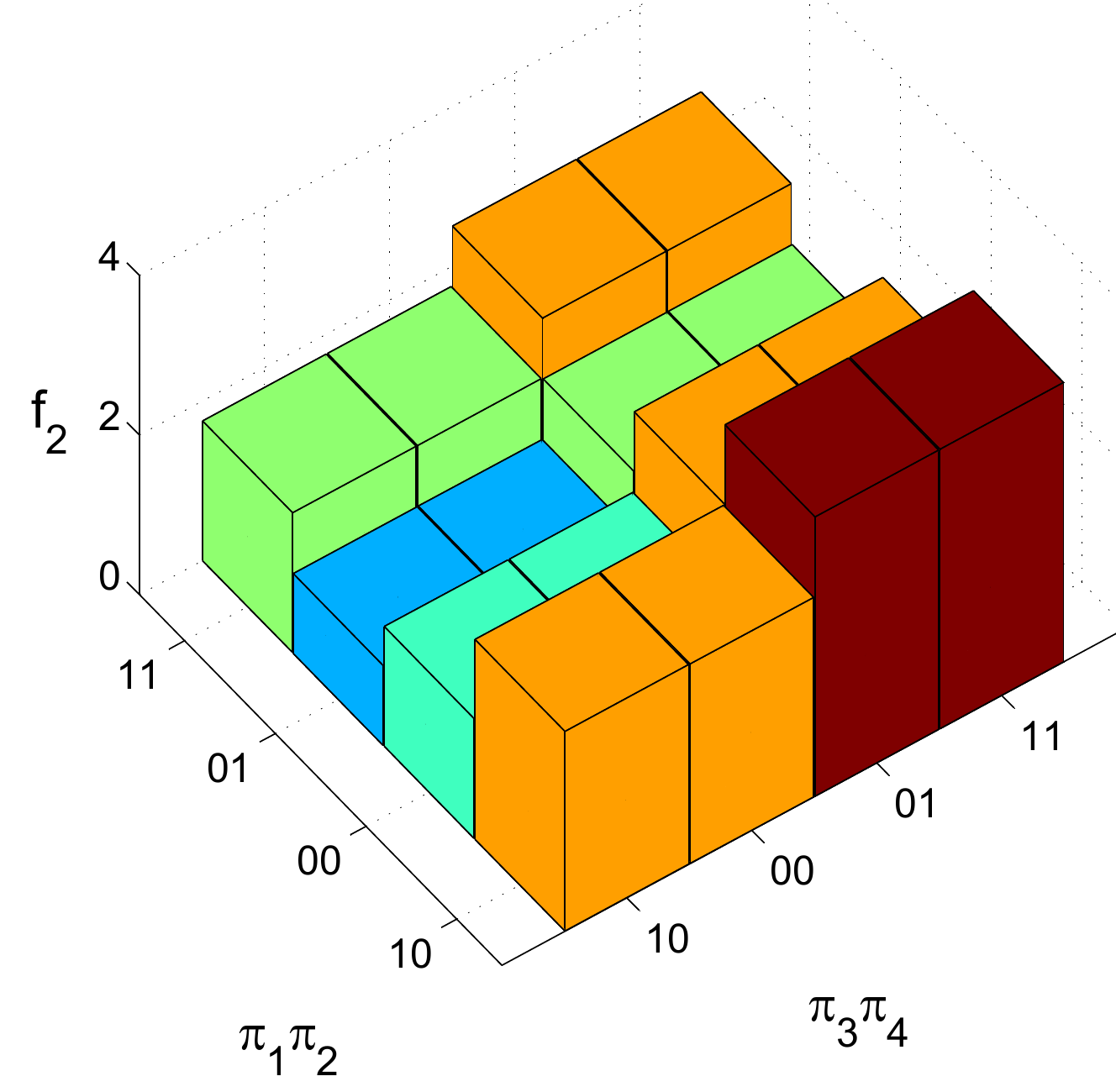} 
\includegraphics[trim = 35mm 65mm 40mm 100mm,clip, width=2.97cm, height=3cm]{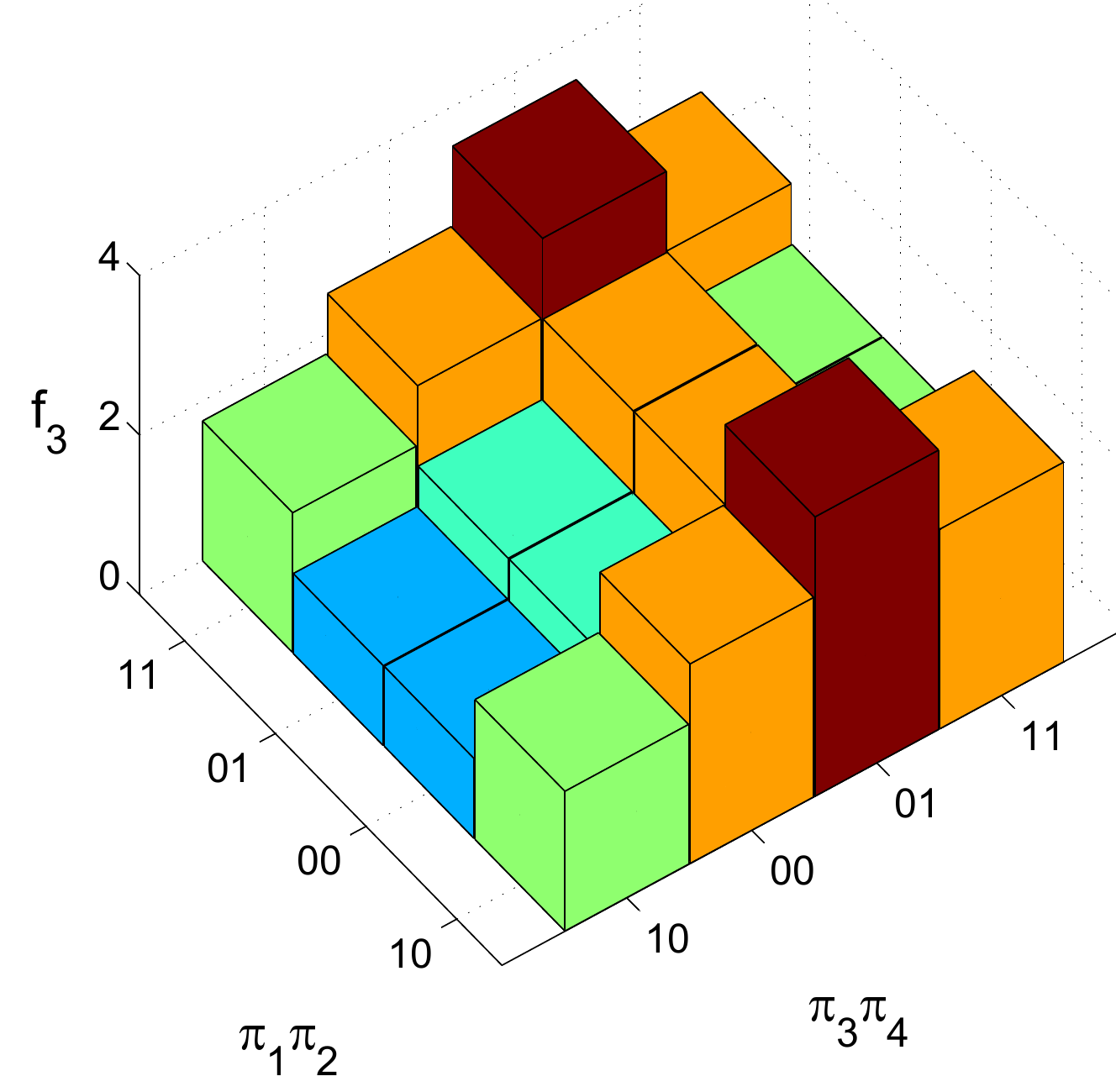} 
\includegraphics[trim = 35mm 65mm 40mm 100mm,clip, width=2.92cm, height=3cm]{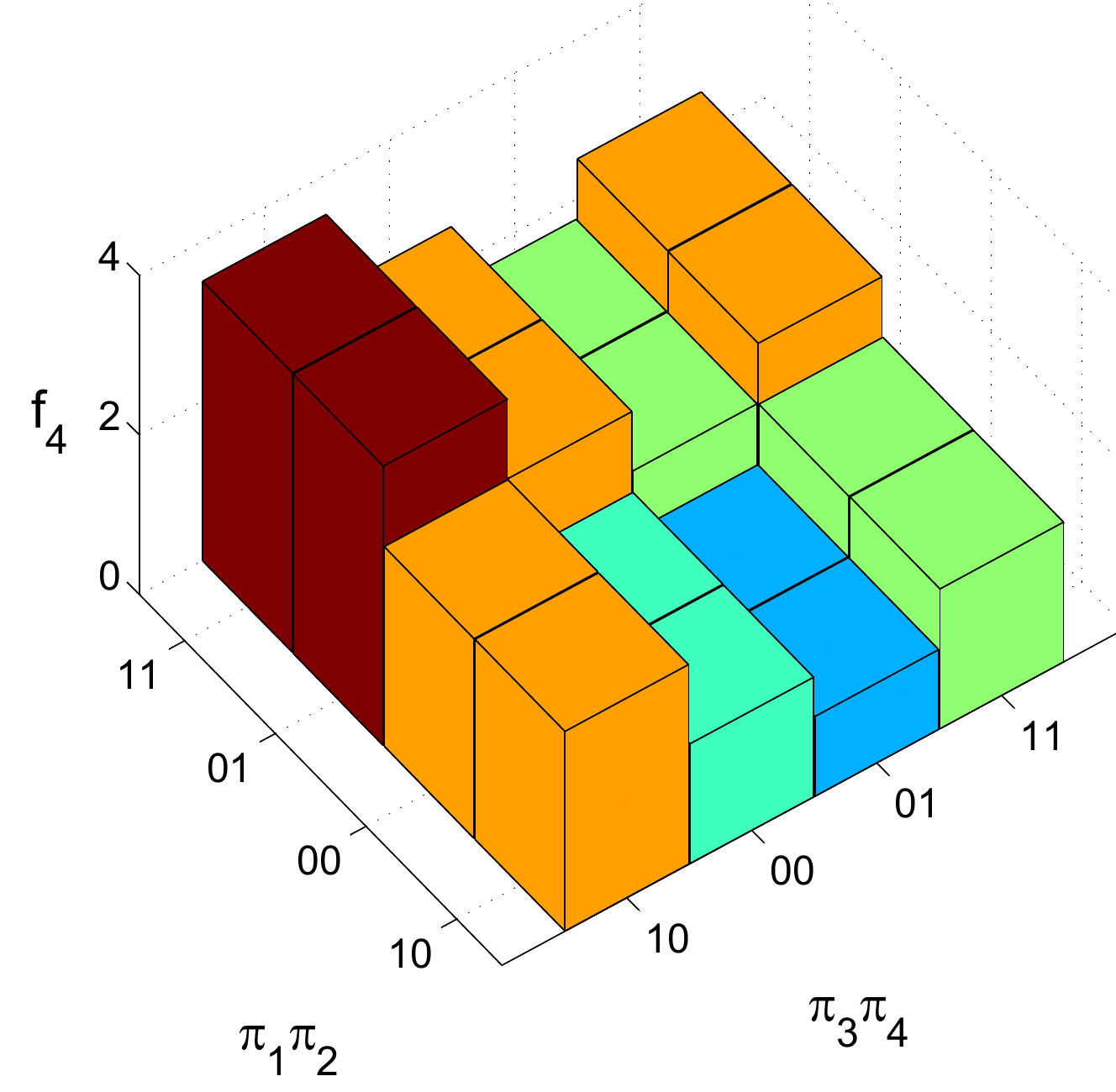} 

\hspace{5.9cm} (c) 

\caption{Illustration of two-dimensional strategy landscapes $\Lambda_{\Pi}^i$ for a  PD game with $N=4$, $d=2$  and  $\mathcal{L}_2(4)=3$ different networks of interaction.   Same colors give equal fitness values $f=1+\delta p$ for payoff $p$ with $\delta=0.25$. (a) ~\usebox{\smlmata} (b) ~\usebox{\smlmatb}  (c) ~\usebox{\smlmatc}. For each player and all three interaction networks the landscape has two maxima; the player defects, while its two coplayers cooperate. There are two maxima as the third player can do either way. }
\label{fig:n5}
\end{figure*}

\clearpage

\clearpage

\end{document}